\documentclass[12pt,a4paper]{article}

\usepackage{a4wide}
\usepackage{fancyhdr}
\usepackage{tikz}
\usetikzlibrary{matrix, decorations.pathreplacing}
\usepackage{hyperref}
\usepackage{fancyref}
\usepackage{graphicx}
\usepackage{subfig}
\usepackage{algorithm}
\usepackage{algpseudocode}
\usepackage{wrapfig}
\usepackage[toc,page]{appendix}
\usepackage{url}
\usepackage{graphicx}
\usepackage{epsfig}
\usepackage{parskip}
\usepackage[utf8]{inputenc}
\usepackage{amsmath}
\usepackage{amssymb}
\usepackage{mathtools}%
\DeclareMathAlphabet{\mathpzc}{T1}{pzc}{m}{it}
\usepackage{bm}
\usepackage{cite}
\usepackage{neuralnetwork}
\usepackage{pdfpages}
\usepackage{tikz}
\usepackage{xcolor}
\usepackage{placeins}
\usepackage[export]{adjustbox}
\usepackage{blindtext}
\usepackage{multicol}

\usepackage[labelfont=bf]{caption}
\usepackage{breqn}

\usepackage{gensymb}
\usepackage{titlesec}
\usepackage{pdfpages}

\usetikzlibrary{calc}
\graphicspath{{./ims/}}

\titleformat{\paragraph}
{\normalfont\normalsize\bfseries}{\theparagraph}{1em}{}
\titlespacing*{\paragraph}
{0pt}{3.25ex plus 1ex minus .2ex}{1.5ex plus .2ex}

\begin{document}

\begin{titlepage}
\vspace*{1cm}
    \begin{tikzpicture}[remember picture,overlay]
    \node[anchor=north west,yshift=-30pt,xshift=30pt]
        at (current page.north west)
        {\includegraphics[height=20mm]{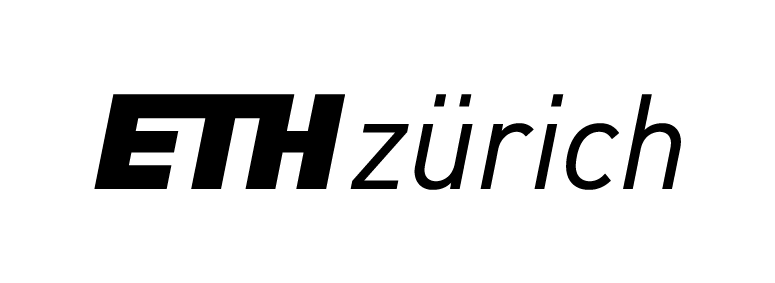}};
    \end{tikzpicture}

\begin{center}
\vspace*{-2cm}
\hrule
\large\text{Institute of fluid dynamics}\\
\vspace*{3cm}
\LARGE\LARGE\textbf{Resistive tearing: numerical exploration of nonmodal effects}
\vspace*{1cm}

	\large\text{MSc Thesis}\\

	\vspace*{8cm}
	\large\textbf{Elias Pratschke}
\end{center}
	
		\vspace*{1cm}
		\begin{center}
		\large\text{(November 2023 - May 2024)}
		\end{center}
		\vspace*{1cm}
	\begin{multicols}{2}
		\flushleft
		\textsc{Supervisors}
		
		Prof. Dr. Patrick Jenny\\
        Prof. Dr. Steven Brunton

		\flushright
		\textsc{Advisors}

        Dr. Daniel W. Meyer-Massetti\\
		Prof. Dr. Alan A. Kaptanoglu\\
		Dr. Christopher Hansen\\		
	
	\end{multicols}
\end{titlepage}

\pagenumbering{roman}

\begin{abstract}
The fluid dynamics community has found substantial success in explaining both the non-modal onset and coherent structure formation in wall-bounded turbulence through examining transient growth and pseudoresonance. Whether similar effects are important in plasmas well-described by magnetohydrodynamics is an open question. Instabilities are extremely common and are challenging to control in plasma physics devices, with resistive instabilities still not being fully understood. In nuclear fusion experiments, the onset of turbulence and the formation of turbulent coherent structures often enhance undesirable heat transport between the core and the edge, and a vast number of studies have attempted to understand these various phenomena via linear and fully nonlinear numerical codes. If there are plasma experiments where instabilities and turbulence onset are dominated by similar nonmodal effects, the tools of non-modal stability and resolvent analysis could prove to be essential to practitioners. 
Towards that goal, in this work we provide a detailed derivation of non-modal stability tools and the resolvent operator for incompressible, resistive magnetohydrodynamics (MHD). 

Concretely, optimal initial conditions that maximize transient growth are computed, along with bounds on linear transient growth, and both are numerically verified with a nonlinear spectral solver to explore whether these linear effects can be reproduced in nonlinear simulations. The potential of such initial conditions to initiate the tearing mode in a spectrally stable system is explored. We found the common approach in nonmodal hydrodynamic stability of using pseudospectral methods to be infeasible for many parameter combinations due to numerical ill-conditioning. Further, transient growth in the Harris current sheet proved to be highly norm-dependent, with the observed differences being of an order of magnitude. Nonlinear simulations showed that transient growth occurs to some extent in all norms, both in the limit of infinitesimal as well as finite perturbations. No case was observed of transient growth being strong enough to trigger the full plasmoid instability.\\

\end{abstract}

\newpage

\tableofcontents
\addcontentsline{toc}{section}{Contents}
\listoffigures

\newpage
\pagenumbering{arabic}
\section{\label{sec:Intro}Introduction}

 Plasma is known as the ``fourth state of matter'' and abundant not only in our universe, but also on planet earth. A plasma is an ionized gas, consisting of a mix of free electrons, ions and neutrals. Because of the presence of charged particles, a plasma is conductive. There are many scientific and engineering disciplines interested in the behavior of plasma, among them diverse fields such as extreme ultraviolet lithography, fusion power and solar physics \cite{bellan2008fundamentals}. 

Engineering problems can often be reduced to controlling and understanding complex systems. Fluids and plasmas are among the most difficult systems to quantify and manipulate. Due to the fact that a plasma is conductive, two of the most difficult disciplines in physics have to be combined: Fluid- and electrodynamics. A long history of deriving and trying to understand simplified systems has been established, with equations of motion being easily derived, but usually challenging or impossible to solve. Each of these simplified models typically requires a tailored numerical or analytical approach to gain understanding from them. Even for these simplified, canonical systems, fundamental questions such as stability are still active areas of research. Unfortunately, many promising technologies rely on a very detailed understanding of plasma, the most prominent being creating energy with fusion reactions.

Thermonuclear fusion requires plasma to be confined inside the reactor at very high temperatures and densities for a sufficient amount of time to generate power. This is known as the fusion `triple product' \cite{Lawson_1957}. This confinement is one of the most difficult control problems in engineering. A huge variety of instabilities can disrupt the confinement, and many of them are severely limiting design options and operation regimes \cite{deVries_2011} \cite{Schuller_1995}. Much of the engineering knowledge about plasma stems from analytic and numerical studies on the ideal magnetohydrodynamic (MHD) equations, which assume a continuum, single-fluid plasma with infinite conductivity and no viscosity, the equations being a significant simplification compared to using kinetic equations for each atomic species \cite{galtier2016introduction}\cite{freidberg2014ideal}. The ideal MHD equations have many nice properties that make them amenable to both analytical and numerical studies, although 3D-simulations of realistic geometries are still at the frontier of possible computations \cite{Jardin}. The equation linking electric field $\mathbf{E}$, velocity $\mathbf{v}$, magnetic field $\mathbf{B}$ and the current density $\mathbf{J}$ is known as Ohm's law:

\begin{align}
    \mathbf{E} + \mathbf{v} \times \frac{\mathbf{B}}{c} = \eta\bm J.
    \label{eq:ohm_1}
\end{align}

 For an ideal plasma, the right-hand side of this equation is zero. One of the most important properties of an ideal MHD plasma is the following: It can be shown that an ideal plasma has the intrinsic property that the magnetic and velocity fields move together, and this coupling can never be broken \cite{freidberg2014ideal}. It is known as Alfvèns theorem or the ``Frozen flux'' theorem and is a direct consequence of using the ideal Ohm's law \cite{Alfven_1942}. Allowing for a small amount of resistivity relaxes the coupling of magnetic and velocity fields in some regions of the domain. The magnetic field can now diffuse independently of the plasma flow. This gives rise to the possibility of breaking and reconnecting magnetic field lines and - in consequence - topological changes in the magnetic field. Under certain circumstances, this can allow the plasma to reconfigure itself into a state of lower magnetic energy that would have been prohibited in ideal MHD \cite{schnack2009lectures}. This process is known as magnetic reconnection and was explored via linear stability analysis in the seminal work of Furth, Kileen and Rosenbluth \cite{furth_1963}. 

These resistive instabilities have been primarily observed in reactors, in the form of sawtooth crashes, and astrophysical flows, as solar flares. They occur on a wide variety of timescales, which are between the fastest, ideal, instabilities occurring at the level of the Alfvèn time $\tau_{a}$ and the slow diffusive phenomena $\tau_{r}$. One of the first research efforts into modelling reconnection was in fact due to solar flares \cite{Sweet_1958} \cite{parker_1963}, by Sweet and Parker. Still, much faster phenomena were being observed than were described by the newly established Sweet-Parker theory. Patschek theory \cite{patschek_1964} was significantly better at modeling these phenomena, although it was seen to be insufficient in many simulations, with the impulsive onset still not being explained. All these classical models have in common that increasing the Lundquist number leads to lower predicted growth rates. For astrophysical Lundquist numbers, these models would predict that a solar flare is an event that occurs over a timescale of years. Also, both Sweet-Parker and Patschek models are fundamentally steady state. Work by Loureiro \cite{Louriero} and Bhattacharjee \cite{2009Bhattacharjee} has shown reconnection to occur on faster timescales than predicted by the classical models, with the mechanism behind these fast dynamics having possibly large implications on other areas of plasma research, such as the structure and formation of MHD turbulence. This has become known as the plasmoid instability and has mainly been explained as a nonlinear phenomenon. The linearized system corresponding to the plasmoid instability is the well-known tearing mode, which features a linear growth rate that, again, decreases with the Lundquist number and therefore can only explain the nonlinear behavior in an unsatisfactory way.

In contrast to the linearized ideal MHD force operator, the linearized resistive MHD equations are non-normal \cite{borba_1994}. It is possible that non-modal effects could help explain some of the discrepancies between established theory, numerical results and observations. While linear stability analysis fails to predict the rapid onset of the plasmoid instability, transient growth could very well extend the validity of the linear tearing mode. Recent work \cite{2009Bhattacharjee} seems to indicate a sensitivity of the onset of the plasmoid instability to noise. This is an indication that non-modal effects could play a role, as they are generally sensitive to the choice of initial condition. Compared to the extensive amount of research done on non-modal stability in the fluid dynamics community, the field is still relatively unexplored in plasma physics. Some work has been done with respect to MHD turbulence \cite{Landreman_2015} \cite{friedmann_2014}, and on the linear amplification of pseudomodes \cite{Squire_2014} \cite{mactaggart_2018} \cite{LOVERSO2024100042}. 

More specifically, Squire \& Bhattacharjee \cite{Squire_2014} focus on computing linear transient growth for the Magnetorotational instability and found transient effects to be significantly more relevant than eigenmode dynamics at short time horizons. Landremann provides evidence that gyrokinetic turbulence and transient linear amplification are linked and can lead to subcritical turbulence, similar to the classic results in hydrodynamic flows \cite{trefethen_stabiltiy}, although Landremann puts a greater emphasis on self-sustained processes in turbulence. MacTaggart \cite{mactaggart_2018}\cite{MacTaggart2020} looks at the transient growth of the resistive tearing mode, using two different norms, although a focus is put on finding scaling laws at spectrally unstable wavenumbers.

While adding more terms to the ideal MHD equations clearly leads to a more holistic picture of the physics, the terms come with an increase in the separation of scales and therefore an increased difficulty in simulations. Resistive and viscous terms in particular fundamentally change the nature of the MHD equations. Apart from the non-orthogonal eigenfunction basis that can be caused by diffusion operators, the limit of infinitely small resistivity is singular, similar to the inviscid limit of the Navier-Stokes equations \cite{Mcmillan_2004}. Low diffusion coefficients lead to very thin boundary layers and require simulations to resolve very small structures with high gradients, again in analogy to hydrodynamics. These boundary layers are known for their high demands on grid generation and raw compute power. These difficulties lead to a general preference for using simpler systems, such as ideal MHD, whenever possible.

In the fluid dynamics community, experiments have played a vital role in validating mathematical models and have often led to the discovery of fundamental flaws in said models, with a very prominent example being the hydrodynamic flow in a pipe, for which modal stability analysis predicts stability up to a certain Reynolds number \cite{Orszag_Kells_1980}. A lot of effort has been put into computing this critical Reynolds number, although experiments were repeatedly contradicting the results of these computations and showing sub-critical transition and turbulence. This discrepancy led to completely new approaches to hydrodynamic stability analysis, building on the pseudospectrum of an operator, instead of the most unstable eigenvalues. One of the first comprehensive works was the seminal paper by Trefethen et al. \cite{trefethen_stabiltiy}, followed by efforts to model transition to turbulence \cite{Reddy_Henningson_1993} \cite{Henningson_bypass_1993} and finally a series of successful reduced-order models of self-sustaining processes in wall-bounded turbulence \cite{McKEON_SHARMA_2010} \cite{Gomez_Blackburn_Rudman_Sharma_McKeon_2016}. In contrast to fluid mechanics, plasma experiments are costly to construct and measurements are difficult to obtain. Therefore, plasma practitioners must put sustained effort into the development of numerical and mathematical modeling. 
 


\subsection{\label{sec:motivation}Motivation - Non modal stability analysis: Going beyond eigenvalues}

 Linearizing equations of motion is often the very first step to gaining any understanding of a nonlinear system. If all eigenvalues have a negative real part, the system is stable, as all initial conditions lead to solutions that decay to zero as $t \rightarrow \infty$. This concept is known as asymptotic stability \cite{chicone}. In many engineering disciplines, a large emphasis is placed on computing asymptotic stability via eigenvalues. This works well if one can make three very strong assumptions: The eigenvalues can be calculated exactly (i.e., without round-off error), $A$ is known exactly (i.e., no modeling or discretization error), and we only care about the behavior of the system at $t\rightarrow \infty$ 
\cite{TrefethenEmbree+2005}. 

In practice, these assumptions are rarely fulfilled: Eigenvalues are computed with iterations that have to be stopped at some point \cite{trefethen1997numerical}, entries of system matrices contain uncertainties, be it from discretization error or heuristic parameters, and short-term behavior can sometimes be more important than the behavior at infinity. The first two issues could, in theory, be improved with arbitrary-precision data types and better measurements or discretization. In general, the issue that eigenvalues tell us nothing about short-term and transient behavior cannot be fixed. However, if $\mathbf{A}$ is normal, the transient behavior is equivalent to the behavior at infinity, i.e., for a stable system, decay starts instantly, and the eigenvalue analysis will give useful results. For non-normal system matrices $\bm A$, this is not true, and the system may exhibit arbitrarily large transient growth before eventually reaching a state of decay as predicted by an eigenvalue analysis \cite{TrefethenEmbree+2005}. If $\mathbf{A}$ represents a linearization of a nonlinear system, transient growth can be large enough to make $\mathbf{A}$ no longer a valid approximation. This can lead to instability, even though $\mathbf{A}$ is formally spectrally stable. The discipline that seeks to augment linear stability analysis with these non-normal effects is called non-modal stability analysis and is mainly concerned with both transient growth and pseudoresonances\cite{Farrell_1996}.

A well-studied example is fluid flow through a pipe, which is known to exhibit turbulence well before the linearized governing equations (Orr-Sommerfeld) become spectrally unstable \cite{trefethen_stabiltiy}, \cite{Reddy_Henningson_1993}. Even in the case where both experiment and Orr-Sommerfeld agree on stability, the transition to turbulence is orders of magnitude faster than predicted by the unstable eigenmode. Other examples where eigenvalue analysis can provide misleading results are:  viscoresistive magnetohydrodynamics \cite{borba_1994} \cite{Dorsselaer_2003}, numerical linear algebra \cite{Greenbaum1996AnyNC} and many other disciplines. 

\subsection{\label{sec:objectives}Objectives}
\begin{figure}
    \centering
    \includegraphics[width=0.5\linewidth]{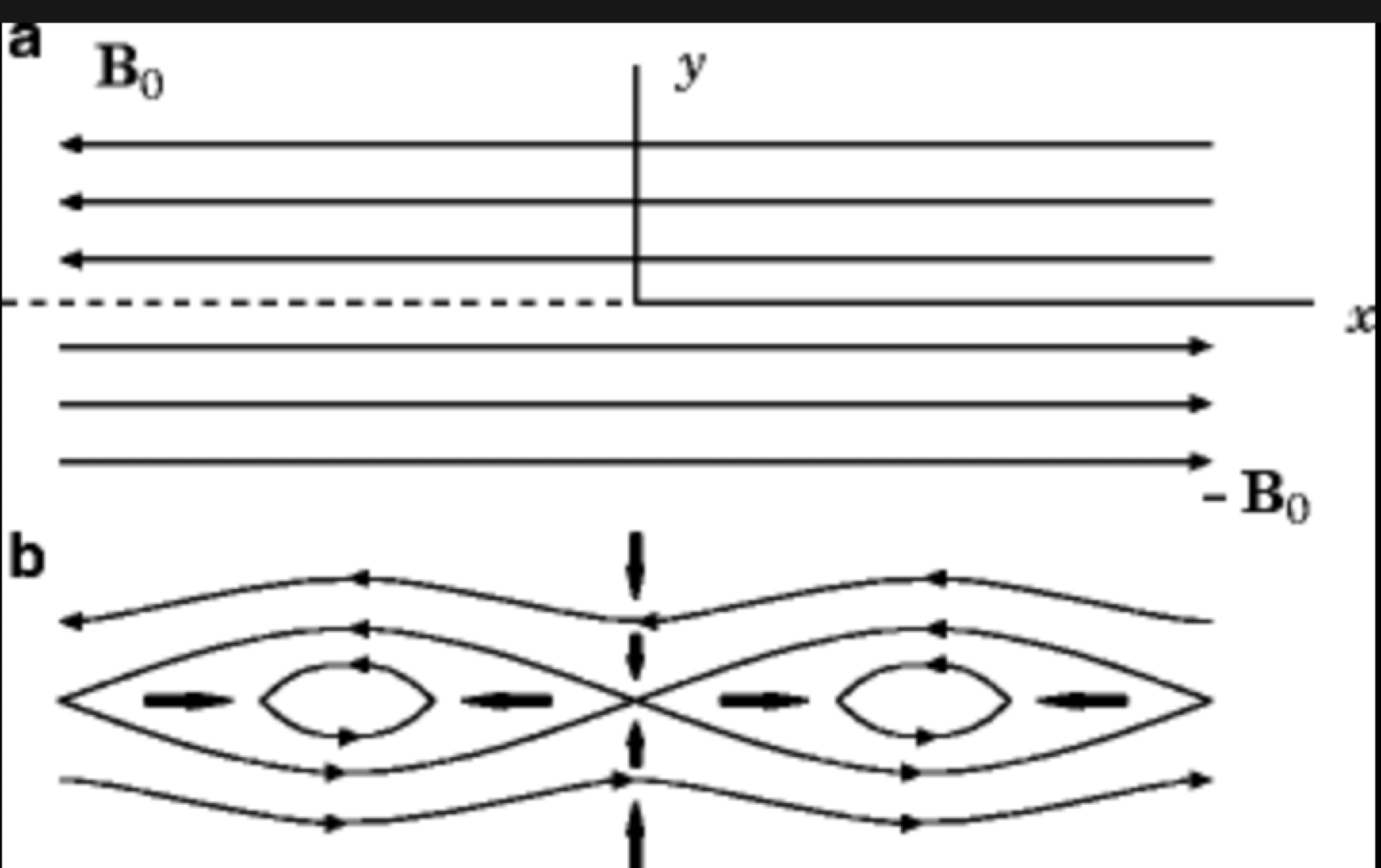}
    \caption{Sketch of the Harris sheet equilibrium (a) and the island state produced by the tearing instability (b). Illustration adapted from \cite{battacharjee1995}}
    \label{fig:sketch}
\end{figure}

Understanding resistive instabilities is paramount for extending the operating range and safety margins of fusion reactors. Because relatively little is known about the onset of resistive instabilities, including nonmodal effects, we seek to gain a more in-depth understanding of these phenomena. Specifically, in this work we focus on the onset of the resistive tearing instability in a slab geometry \cite{battacharjee1995}, which can be seen as a simple, prototypical case of a resistive instability. We use the incompressible, viscoresistive MHD equations for this study, as they are the simplest system that allows for resitive instabilities and magnetic reconnection. Due to the fundamental difference in the mathematical properties compared to the ideal MHD equations, this work is of particular interest. A sketch of the problem can be seen in Fig. \ref{fig:sketch}. 

We seek to gain an understanding of the nonmodal stability and of possible amplification mechanisms at the root of the tearing instability. Firstly, we investigated the influence and effects of transient growth on the linear and nonlinear evolution. Secondly, we want to obtain a quantitative sense of how numerical and physical parameters influence this transient growth. Lastly, we want to investigate how initial conditions, designed to maximize transient growth, are able to influence nonlinear resistive MHD simulations and whether they have the potential to change the stability region of the tearing instability. This builds strongly on the work of MacTaggart (\cite{mactaggart_2018}, \cite{MacTaggart2020}), with the emphasis being on extending the work to different parameter regimes and verifying linear results using a nonlinear solver. We found that transient growth occurs in both norms considered, although the growth in the energy norm was found to be more modest than in the $L_2$ norm. We also found calculations in the energy norm to be very sensitive w.r.t. the exact numerical algorithm and parameters used. Transient growth can be reproduced with a nonlinear solver, although it is not significant enough to trigger a full, nonlinear instability.

\subsection{\label{sec:structure}Structure of thesis}
In the first section, the mathematical and physical foundations of the work are specified. The reduced set of incompressible and resistive MHD equations is derived. The concepts of pseudopsectra, pseudo- and resolvent modes, non-normality, and the necessary numerical methods are introduced, along with some useful bounds from non-modal stability theory. Only the essential aspects are covered due to reasons of brevity. After deriving a reduced MHD system and a linearized version of it, our results are presented for this system. These are segmented into two broad categories: Results from the linearized system and results concerning the nonlinear simulations. Finally, an overview of possible approaches to build further on the foundation provided by this thesis is given.

\section{\label{sec:methodology}Theory}

\subsection{\label{subsec:equations}Equations and assumptions}

The equations we shall use to model the tearing mode instability are called the visco-resistive, incompressible MHD equations. These equations and transport coefficients can be derived by taking moments of the Boltzmann equation, followed by a series of approximations on the length and time scales involved in the dynamics. More specifically, these equations hold for length scales significantly larger than the Debye length, the ion and electron  Larmor radii and skin depths. Similarly, these equations only hold for velocities less than or equal to the ion thermal speed, and for frequencies smaller than the plasma, gyro, collisional, and equilibration frequencies~\cite{freidberg2014ideal}. In contrast to equations such as the Euler or Navier-Stokes equations, this means that there is a fundamental limit for the various scales at which MHD is valid, prohibiting the practice of infinitely resolved grids for DNS-like accuracy. 

The incompressible viscoresistive MHD equations are defined as follows:
\begin{align}
\label{eq:MHD_equations}
    \nabla \cdot \mathbf{u} &= 0, \quad \nabla\cdot\bm B = 0,\\ \notag
    \rho \left[  \partial_{t}\mathbf{u} +  \mathbf{u} \nabla \mathbf{u} \right] &= 
    - \nabla P + \mathbf{J} \times \mathbf{B} + \nu \nabla^{2}\mathbf{u}, \\\notag
    \partial_{t} {\mathbf{B}} &= - \nabla \times  \mathbf{E}, \\\notag
    \mu_{0}\mathbf{J} &= \nabla \times \mathbf{B},\\ \notag
\mathbf{E} + \mathbf{u}\times \mathbf{B} & = \eta \mathbf{J} \label{eq:ohm},
\end{align}

where $\mathbf{u}$ denotes the velocity field, $\mathbf{J}$ is the current density, $\mathbf{B}$ the magnetic field, $P$ is a scalar pressure, and $\mathbf{E}$ is the electric field. The coefficients are the kinematic viscosity $\nu$, density $\rho$, conductivity $\eta$ and permeability $\mu_{0}$.

The two differences from ideal MHD are the resistive term in Ohm's law and the diffusion term in the momentum equation. 

\subsection{\label{subsec:plasma_equilibria}Equilibrium and instability}
An equilibrium state of the MHD equations is one in which all forces balance, and the system is steady-state. A static equilibrium requires that the velocity is not just steady-state, but also zero, while a dynamic equilibrium permits a steady-state flow of plasma. We will focus on static equilibria. In this case, the equations that must hold are the following:
\begin{align}
    \nabla \cdot \mathbf{u} &= 0, \quad \nabla\cdot\bm B = 0,\\ \notag
    0 &= - \nabla P + \mathbf{J} \times \mathbf{B} \label{eq:equi1}, \\ \notag
    \nabla\times\bm B &= \mu_0\bm J. \notag
\end{align}
Plasma equilibrium can only be achieved via external sources. This is a consequence of the virial theorem, which states that ``a magnetized fluid cannot be in MHD equilibrium under forces generated by its own internal currents''~\cite{freidberg2014ideal}. Typically, these external fields are applied using coils to produce an external magnetic field. Nonetheless, much of plasma physics concerns itself with astrophysical conditions where no coils or walls are present. Any process that allows a plasma equilibrium to reconfigure itself to a state of lower energy is called an instability, although sometimes this word is reserved for linear, exponentially fast instabilities. Instabilities can occur spontaneously and destroy the previous equilibrium.

Resistive terms in the MHD equations allow the plasma to break field lines to reconfigure itself to an even lower energy, after all ideal MHD mechanisms (such as pressure-driven and current-driven instabilities) have been exhausted. Since diffusion is a much slower process than ideal MHD processes, resistive instabilities are typically slower \cite{bellan2008fundamentals}.

The equilibrium we use in this thesis is static, i.e., the equilibrium velocity is zero. It consists of a magnetic shear layer, with the magnetic field reversing direction in a narrow layer at $y=0$. We chose a profile corresponding to the Harris sheet equilibrium, which is characterized by a hyperbolic tangent function. A diagram of the configuration can be seen in Fig. \ref{fig:sketch}. The figure also shows a sketch of the magnetic islands generated by the tearing instability.

\subsection{\label{sec:stability_derivation}Linear stability of current sheet}

Since the onset of many instabilities can be described via the growth of infinitesimal perturbations, we shall derive an analytical approximation of the linear growth rates of the resistive tearing mode.

After linearizing at the static equilibrium described by a Harris current sheet, the equations that govern the evolution of the infinitesimal perturbations to first order are these:

\begin{align}
    \nabla \cdot \mathbf{B_{1}} &= 0,
    \nabla \cdot \mathbf{u_{1}} = 0,\\ \notag
    \partial_{t} \mathbf{u_{1}} &= - \nabla P_{1} + \frac{1}{\mu_{0}} \left[ (\nabla \times \mathbf{B_{1}}) \times \mathbf{B_{0}} + (\nabla \times \mathbf{B_{0}}) \times \mathbf{B_{1}} \right] ,\\ \notag
     &= -\nabla \left(P_{1} + \frac{\mathbf{B_{0}} \mathbf{B_{1}}}{\mu_{0}} \right) + \frac{1}{\mu_{0}} \left[ (\mathbf{B_{0}} \cdot \nabla) \mathbf{B_{1}} + (\mathbf{B_{1}\cdot \nabla}) \mathbf{B_{0}} \right], \\ \notag
    \partial_{t} {\mathbf{B_{1}}} &= - \nabla \times \left[ \frac{\eta}{\mu_{0}} \nabla \times (\mathbf{B_{0}} + \mathbf{B_{1}}) -  \mathbf{u_{1}} \times \mathbf{B_{0}}\right], \\ \notag
    &= \frac{\eta}{\mu_{0}} \nabla^{2}(\mathbf{B_{0}}+\mathbf{B_{1}}) + \left[(\mathbf{B_{0}\cdot \nabla})\mathbf{u_{1}} + (\mathbf{u_{1}}\cdot \nabla)\mathbf{B_{0}}\right]
\end{align}

Inserting our values for the equilibrium $B_{0}$, the equations simplify to:

\begin{align}
    \partial_{t}\mathbf{B_{1}} &= \begin{bmatrix} -\partial_{y}(u_{1y}B_{0x})\\
    \partial_{x}(u_{1y}B_{0x})\\
    \partial_{x}(u_{1z}B_{0x})\end{bmatrix} + \frac{1}{S} 
    \begin{bmatrix} \nabla^{2}(B_{0x}+B_{1x})\\
    \nabla^{2}B_{1y}\\
    \nabla^{2}B_{1z}\end{bmatrix} \\
    \partial_{t}\nabla^{2}u_{1y} &= B_{0x}\partial{x}\nabla^{2}B_{1y} - B_{0x}^{\prime\prime} \partial_{x}B_{1y} \notag
\end{align}

The two equations above form a coupled system that we can use to obtain solutions for both the y-components of the velocity and mag. field perturbation. The x-components can be obtained from the solenoidal constraints for both velocity and magnetic field. For an analytical estimate, the equations are further simplified. Note that inserting our flux function formulation and integrating the induction law yields the same equation we derived for the magnetic flux function, verifying the derivation. The y-component of the induction equation (with the assumption that $\partial_{xx}B_{y}$ is very small) produces a relation between $B_{y}$ and $u_{y}$:

\begin{align}
\label{eq:goldston_derivation_1}
    \omega B_{y} &= -k_{x} B_{0x} u_{y} + \frac{i\eta}{\mu_{0}} \partial_{yy} B_{y}
\end{align}
 This equation also showcases the fact that for a zero resistivity, the magnetic and velocity fields are intrinsically linked. Only in the area where the first right-hand side term becomes small can the magnetic field lines diffuse. Next, we will separately look at regions inside and outside the diffusive layer, match solutions, and obtain an approximate linear growth rate. The curl of the momentum equation gives us a second equation relating $B_{y}$ and $u_{y}$
 \begin{align}
     -\frac{\omega\mu_{0}}{k_{x}} \left[ \partial_{y}(\rho \partial_{y}u_{y}) - k_{x}^{2}\rho u_{y} \right] &= \partial_{y}\left[B_{x0}^{2} \partial_{y} (\frac{B_{y}}{B_{x0}}) \right] - k_{x}^{2} B_{x0} B_{y}.
 \end{align}
 Outside the resistive layer, we can obtain a condition for $B_{y}$:
\begin{align}
    \partial_{y}\left[B_{x0}^{2} \partial_{y} (\frac{B_{y}}{B_{x0}}) \right] - k_{x}^{2} B_{x0} B_{y} &= 0.
\end{align}
 We conclude that $B_{y}$ approaches a finite, constant value at $y=0$. Using the incompressibility constraint of $\mathbf{B_{1}}$:
 \begin{align}
     \partial_{y} B_{y} + ik_{x}B_{x0} = 0.
 \end{align}
 we can deduce that a discontinuity in $B_{x0}$ implies a discontinuity in $B_{y}$. The normalized $\Delta^{\prime}$ quantifies the discontinuity in $B_y$,
\begin{align}
    \Delta^{\prime} &= \frac{1}{B_{y}} \left(\partial_{y}B_{y}\vert_{x=0+} - \partial_{y}B_{y}\vert_{x=0-} \right).
\end{align}
Inside the resistive layer, we assume that $B_{x0} = B_{x0}^{\prime} \cdot x$, i.e. a first order Taylor approximation and $\overline{B_{y}}$ is the constant part of $B_{y}$ and get, after inserting into Eq. \ref{eq:goldston_derivation_1}:
\begin{align}
    \omega \overline{B_{y}} + k_{x}B_{x0}^{\prime} y u_{y} &= \frac{i\eta}{\mu_{0}} \partial_{yy}B_{y}.
\end{align}
The momentum equation is simplified by assuming the y derivatives are much larger than the x derivatives leading to:
\begin{align}
    -\omega \rho_{0} \mu_{0} \partial_{yy}u_{y} &= k_{x}B_{y0}^{\prime}\partial_{yy}B_{y}\\
    \gamma \eta \rho_{0} \partial_{yy} u_{y} &= k_{x} B_{x0}^{\prime}y (i\gamma \overline{B_{y}} + k_{x} B_{x0}^{\prime}yu_{y} ).
\end{align}
Where $\omega=i\gamma$ in anticipation of pure growth. We can solve this equation numerically to obtain $u_{y}$. Matching the solution inside the resistive layer to the outer region, we get the growth rate:
\begin{align}
\label{eq:gamma}
  \gamma = \frac{0.55 (\Delta^{\prime}a)^{\frac{4}{5}}}{\tau_{A}^{\frac{2}{5}} \tau_{R}^{\frac{3}{5}}}.
\end{align}
The full derivation of the growth rate was not included in this thesis for brevity, and can be found in \cite{goldston2020introduction}. We can see that stability is determined by the sign of $\Delta^{\prime}$. Solving the equations of the outer region, we get an expression for it:
\begin{align}
\label{eq:delta_prime}
    \Delta^{\prime} &= \frac{2k_{x}a [\mathrm{exp}(-2k_{x}a) - 2k_{x}a + 1]}{\mathrm{exp}(-2k_{x}a) + 2k_{x} -1 }.
\end{align}
The expression is larger than zero for small wavenumbers, or large $L_x$. This is a current sheet that is sufficiently thin to exhibit resistive tearing. The expression Eq. \ref{eq:gamma} also highlights the fact that the tearing mode grows at an intermediate rate, between the Alfvèn time, and the resistive timescale. The above derivation is following \cite{goldston2020introduction}

\subsection{\label{subsec::nonlinear_sim_theory} Evolution of the tearing mode}

The nonlinear dynamics of the tearing instability are obtained from solving Eq. \ref{eq:MHD_equations}. We start out with a magnetic shear layer that consists of a field that smoothly changes the orientation of the field lines in a relatively narrow shear layer, which contains a single neutral line where the magnetic field is identically zero. This initial condition is perturbed and results in the tearing instability, given the current sheet is thin enough, as seen in Sec. \ref{sec:stability_derivation}. The nonlinear evolution and saturation of the instability depend largely on the parameter $L_{x} = \frac{2\pi}{k_{x}}$. 

\begin{figure}%
    \centering
    \subfloat{{\includegraphics[width=0.4\linewidth]{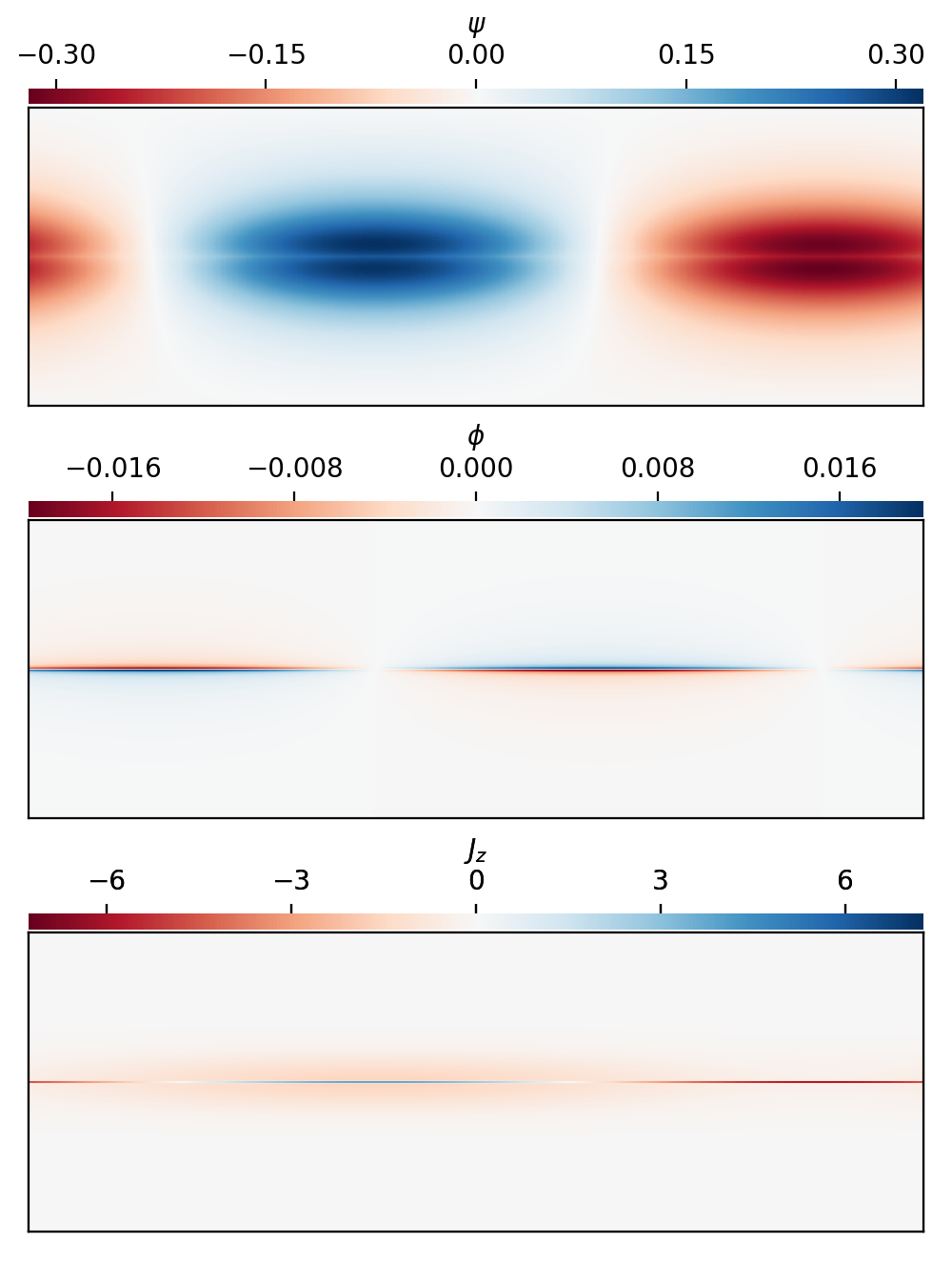} }}%
    \qquad
    \subfloat{{\includegraphics[width=0.4\linewidth]{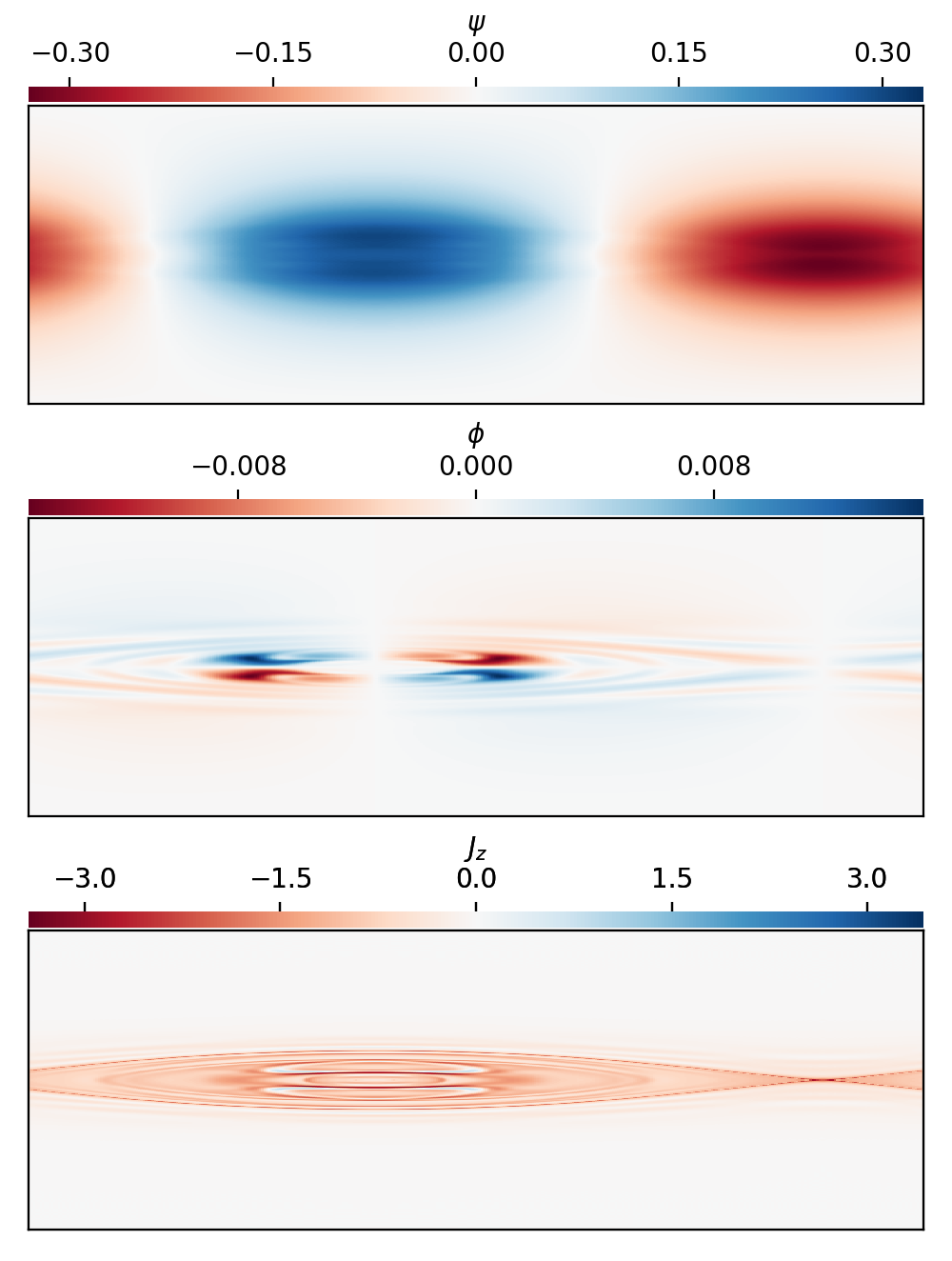} }}%
    \caption{Evolution of a tearing mode simulation initialized with a large eigenvector. It can be seen that the system saturates in a state that includes one magnetic island. The left plot shows the initial condition at $t=0$}%
    \label{fig:saturation}%
\end{figure}

For smaller values of $L_{x}$ the nonlinear evolution is comparatively simple. The instability settles into an algebraic nonlinear growth phase \cite{Rutherford_1973}, after an initial phase of exponential growth, and eventually reaches a saturated state with one magnetic island \cite{RDP}. The case of a single magnetic island can be seen in Fig. \ref{fig:saturation}, which was produced using a fairly large eigenvector as an initial condition at a periodic wavenumber of $k_x=0.7$ and a correspondingly large $L_x = \frac{2\pi}{k_x}$. 

\begin{figure}%
    \centering
    \subfloat{\includegraphics[width=0.4\linewidth]{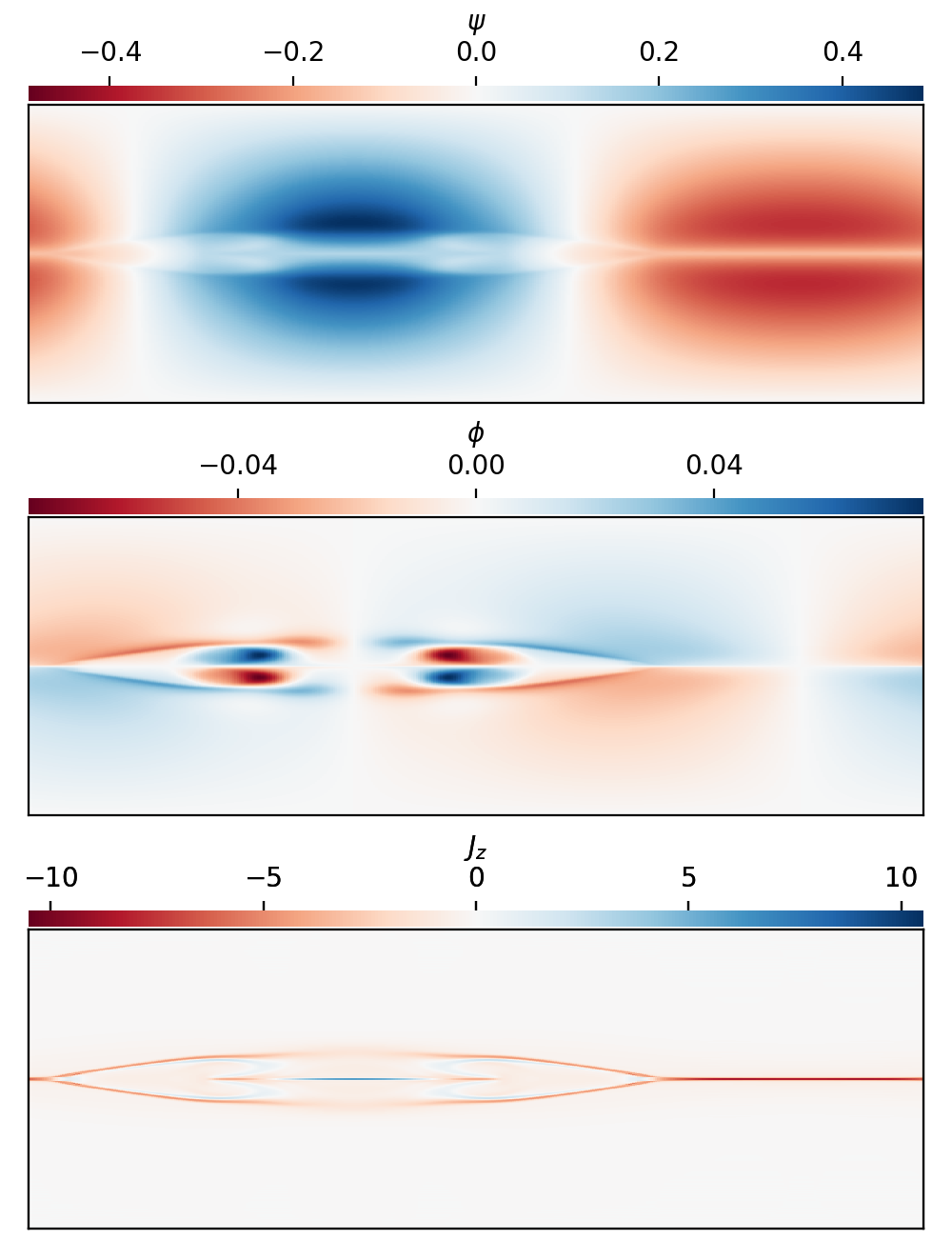} }%
    \qquad
    \subfloat{\includegraphics[width=0.4\linewidth]{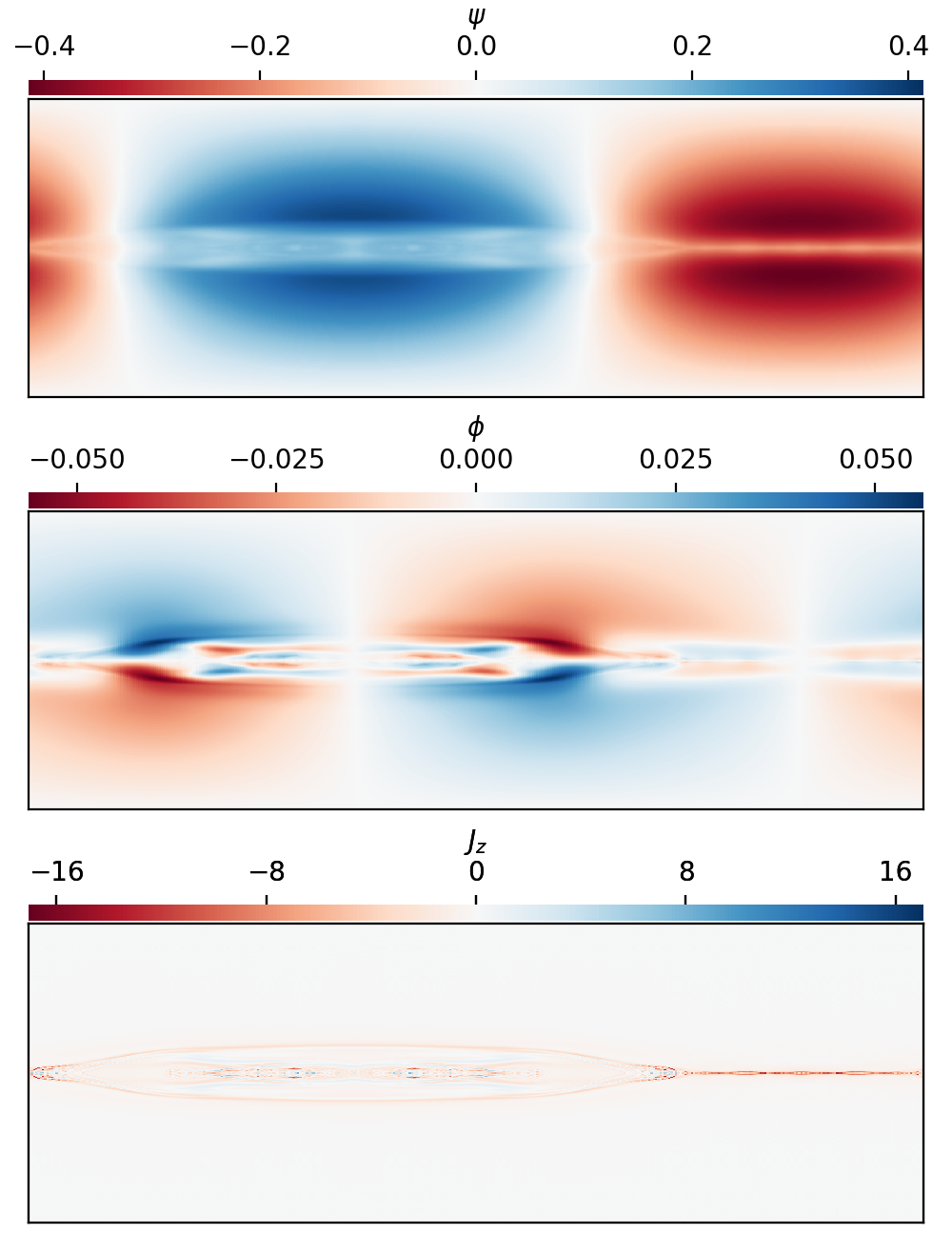} }%
    \caption{Evolution of a tearing mode simulation initialized with a large eigenvector. The left plot shows the system at time $t=17$, the right one at $t=39.5$}%
    \label{fig:secondary}%
\end{figure}

For larger $L_x$, after the formation of a single magnetic island, a secondary instability can be triggered, and many secondary plasmoids are generated, with Fig. \ref{fig:secondary} showing the first phase of this secondary instability. We can see that the primary island is short enough in x-direction that the current sheet in between the island ends is elongated enough to form a secondary plasmoid. This plot was produced again using a large eigenvector perturbation at a wavenumber of $k_x=0.2$. Both cases were computed at a Lundquist number of $S=10^4$ and a small viscosity, corresponding to $Re=10^6$. The work of Parker \cite{RDP} goes into great depth on how these secondary instabilities depend on physical parameters and provides a detailed description of the dynamics of the secondary instability.

As seen in the linear analysis, a stability threshold exists, for which a small \textit{modal} perturbation, i.e., in the shape of the most unstable eigenvector, will not lead to a magnetic island formation, this is the case for sufficiently small $L_x$, or equivalently, large $k_x > 1$.

\subsection{\label{sec:streamfunction} Formulation in incompressible basis}

Because our flow configuration does not require the full 3-dimensional, compressible MHD equations, we will simplify Eq. \ref{eq:MHD_equations} before continuing. Since we do not account for compressibility in this study and are looking at a two-dimensional flow field, we can rewrite the above equations using a stream function $\phi$ and a vector potential $\psi$. This allows us to project the dynamics onto an incompressible set of vector fields and subsequently, to get rid of any pressure terms in the equations. We look at equilibria with zero velocity. We can decompose the fields around an equilibrium or mean state of these equations by adding a perturbation to the fields $f = f_{0} + f_{1}$. We define: 

\begin{align}
\label{eq:streamfunction_def}
    \mathbf{u} &= \mathbf{e_{z}} \times \nabla \phi + u_{1z}(x, y) \mathbf{e_{z}} 
    =      \begin{pmatrix}
            -\partial_{y} \phi\\
            \partial_{x} \phi\\
            u_{1z}
        \end{pmatrix}, \\ 
    \mathbf{B} &=B_{0}(y)\mathbf{e_{x}} + \mathbf{e_{z}} \times \nabla \psi + B_{1z}(x, y) \mathbf{e_{z}} = \begin{pmatrix}
            B_{0} -\partial_{y} \psi\\
            \partial_{x} \psi\\
            B_{1z}
        \end{pmatrix},  \notag
\end{align}

where $\psi, \phi$ are scalar functions. The equilibrium magnetic field at time $t=0$ is given by $B_{0} = B_{0}\mathrm{tanh}(\frac{y}{\lambda_{B}})$. The density profile is assumed to be constant. The dimensions of the rectangular box are given by $-L_{x}/2 \leq x \leq L_{x}/2$ and $-d \leq y \leq d$. We will assume $\lambda_{\rho} = \lambda_{B} = 1$. The parameter $d$ is chosen to be of order 10, such that the boundaries have a small effect on the resistive layer while not leading to an overly large computational domain. The effect of wall distance is investigated at a later point in the thesis. In general, this setup is comparable to the GEM challenge, designed to benchmark nonlinear MHD solvers \cite{GEM}.

We will follow a similar procedure to \cite{RDP} in deriving the reduced MHD equations. First, we look at the induction law and integrate it either in y- or x-direction to obtain an expression for $E_{z}$ depending on $\psi$ and a constant of integration, which is set to zero for now:
\begin{align}
    E_{z} = \partial_{t}\psi + E_{const} = \partial_{t}\psi \label{eq:constant_of_integration}.
\end{align}
We also get from Amperes law:
\begin{align}
        \mu_{0}\mathbf{J} &= 
        \begin{pmatrix}
            \partial_{y} B_{1z}\\
            -\partial_{x}B_{1z}\\
            \nabla^{2} \psi - \partial{y}B_{0}
        \end{pmatrix}.
\end{align}
We look at the z-component of Ohm's law [\ref{eq:ohm}]:
\begin{align}
\label{eq:ohms_law_zcomponent}
    E_{z} + (u_{x}B_{y} - u_{y}B_{x}) &= \eta J_{z}, \\ \notag
    \partial_{t}\psi + (\partial_{x} \phi \cdot \partial_{y}\psi -\partial_{y}\phi \cdot \partial_{x}\psi) - u_{y}B_{0} &= \frac{\eta}{\mu_{0}} (\nabla^{2}\psi -  \partial_{y}B_{0}), \\\notag
    \partial_{t}\psi -\partial_{x}\phi B_{0} - \frac{\eta}{\mu_{0}}\nabla^{2}\psi =\frac{\eta}{\mu_{0}}\partial_{y}B_{0} &+ (\partial_{y}\phi \cdot \partial_{x}\psi - \partial_{x} \phi \cdot \partial_{y}\psi), \\\notag
    \partial_{t}\psi -\partial_{x}\phi B_{0} - \frac{\eta}{\mu_{0}}\nabla^{2}\psi =\frac{\eta}{\mu_{0}}\partial_{y}B_{0} &+ \{\psi, \phi\},
\end{align}
where curly braces indicate the Poisson bracket:
\begin{align}
 \{f, g\} = \partial_{x}f\cdot\partial_{y}g - \partial_{x}g\cdot\partial_{y}f,
\end{align}
 The next task is to obtain an equation for the evolution of $\phi$. In order to do this, we take the curl of the momentum equation and look at the z-component, which contains new information. Taking the curl eliminated all terms depending on the gradient of a scalar field, such as pressure.
\begin{align}
\label{eq:momentum_work_1}
    \mathbf{e_{z}} \cdot\nabla \times \mathbf{u} = \mathbf{e_{z}} \cdot \nabla \times (\mathbf{e_{z}} \times \nabla \phi) &= \nabla^{2}\phi, \\ \notag
    \partial_{t} \nabla^{2}\phi - \mathbf{u}\cdot\nabla(\nabla^{2}\phi) + (\nabla \times \mathbf{u})\cdot\nabla \mathbf_{u} \cdot\mathbf{e}_{z}&= \frac{(\nabla\times(\mathbf{J}\times\mathbf{B}))\cdot \mathbf{e_{z}}}{\rho}, \\ \notag
    &= \frac{1}{\rho}[(\mathbf{B}\cdot\nabla)J_{z} - (\mathbf{J}\cdot\nabla)B_{z}]. \notag
\end{align}
Looking at the individual terms in Eq.\eqref{eq:momentum_work_1} and inserting information from Eq.\eqref{eq:streamfunction_def} and Amperes law we can simplify:
\begin{align}
(\nabla \times \mathbf{u})\cdot\nabla \mathbf_{u} \cdot\mathbf{e}_{z} &= \partial_{y}u_{z}\cdot\partial_{x}u_{z} - \partial_{x}u_{z}\cdot\partial_{y}u_{z} = 0,\\ \notag
\mathbf{u} \cdot \nabla (\nabla^{2}\phi) &= -\partial_{y}\phi \cdot \partial_{x}\nabla^{2}\phi + \partial_{x}\phi\cdot \partial_{y}\nabla^{2}\phi = \{\phi, \nabla^{2}\phi\},\\ \notag
(\mathbf{B}\cdot\nabla)J_{z} &= (B_{0} - \partial_{y}\psi)\cdot\partial_{x}[\nabla^{2}\psi - \partial_{y}B_{0}] + \partial_{x}\psi \cdot \partial_{y}[\nabla^{2}\psi - \partial_{y}B_{0}],  \\\notag
&= \{\psi, \nabla^{2}\psi\} + B_{0} \partial_{x} \nabla^{2} \psi - B_{0}^{\prime\prime}\partial_{x}\psi ,\\\notag
(\mathbf{J}\cdot\nabla)B_{z} &= (\partial_{y}B_{z}\cdot\partial_{x}B_{z} - \partial_{x}B_{z}\cdot\partial_{y}B_{z}),\\\notag
    &=0.
\end{align}
The viscous dissipation term can be written as:
\begin{align}
    \nabla \times \nabla^{2}(\mathbf{u}) \cdot \mathbf{e_{z}} &= \nabla^{4}\phi.
\end{align}
More compactly with Poisson brackets:
\begin{align}
    \rho \partial_{t} \nabla^{2}\phi + \partial_{x}\psi B_{0}^{\prime\prime} -B_{0}\partial_{x}{\nabla^{2}\psi} -\nu \nabla^{4}\phi &= \rho \{\nabla^{2}\phi, \phi\} + \{\psi, \nabla^{2}\psi \}.
\end{align}
We have two equations from which we can obtain $\phi$ and $\psi$. For the out-of-plane velocity and magnetic field terms, we get separate equations by looking at the corresponding component of the curl of the momentum equation  and the equation obtained by combining Ohm's law and the induction law. The curl of the momentum equation can be simplified by applying the chain rule and integrating either in y or x:
\begin{align}
    \partial_{t}B_{z} &= - \nabla \times (\eta \mathbf{J} - \mathbf{u}\times\mathbf{B})\cdot\mathbf{e_{z}},\\ \notag
    &= \frac{\eta}{\mu_{0}} \nabla^{2} B_{z} + B_{0} \partial_{x}{u_{z}} + \{\psi, u_{z}\} + \{B_{z}, \phi\}, \\ \notag
    \rho\partial_{t}u_{1z} &= \{\phi, u_{z}\} + \frac{1}{\mu_{0}} \{B_{z}, \psi\} - \frac{1}{\mu_{0}} B_{0}\partial_{x}B_{z}.
\end{align}
We normalize the variables as follows: $\hat{x}=\frac{x}{\lambda_{B}}, \hat{y}=\frac{y}{\lambda_{B}}, \hat{\psi}=\frac{\psi}{\lambda_{B}B_{0}}, \hat{\phi}=\frac{\phi}{\lambda_{B}^{2} / \tau_{A}}, \hat{\eta}=\frac{\eta}{\eta_{0}}, \hat{t}=\frac{t}{\tau_{a}}$, where the Alfven time is taken as a $\tau_{A} = \frac{a}{v_{A}} = \frac{a(\mu_{0}\rho)^{0.5}}{B_{0}}$, where $a$ is a characteristic length and $v_{A}$ is the Alfven velocity. The resistive diffusion time is set to $\tau_{R} = \frac{a^{2}\mu_{0}}{\eta}$, where $\eta$ is the resistivity. The Lundquist number $S$ is the ration between Alfvèn and resistive timescales: $S=\frac{\tau_{R}}{\tau_{A}}$.

and obtain the full nonlinear system as follows:
\begin{align}
\label{eq:perturbation_equation_full}
\partial_{\hat{t}}\hat{\psi} -\partial_{\hat{x}}\hat{\phi} B_{0} - \frac{1}{S}(\hat{\nabla}^{2}\hat{\psi}) &= \frac{1}{S}(\partial_{\hat{y}}B_{0}) +\{\hat{\psi}, \hat{\phi} \},\\\notag
\partial_{\hat{t}}\hat{\nabla}^{2}\hat{\phi} + \partial_{\hat{x}}\hat{\psi} B_{0}^{\prime\prime} - B_{0}\partial_{\hat{x}}{\hat{\nabla}^{2}\hat{\psi}} - \frac{1}{Re}\hat{\nabla}^{4}\phi &= \{ \hat{\nabla}^{2}\hat{\phi}, \hat{\phi}\} + \{\hat{\psi}, \hat{\nabla}^{2}\hat{\psi}\}, \\ \notag
\partial_{\hat{t}}\hat{u}_{z} - B_{0}\partial_{\hat{x}}\hat{B}_{z} &= \{\hat{\phi}, \hat{u}_{z}\} + \{\hat{B}_{z}, \hat{\psi} \}, \\ \notag
\partial_{\hat{t}}\hat{B_{z}} -\frac{1}{S} \hat{\nabla}^{2} \hat{B}_{z} - B_{0} \partial_{\hat{x}}\hat{u}_{z} &= \{\hat{\psi}, \hat{u}_{z}\} + \{\hat{B}_{z}, \hat{\phi}\}.
\end{align}
All variables are normalized according to the above formulas, with the hats being dropped for notational simplicity. The above equations are valid for any size of perturbation. If we want to look at the evolution of the full nonlinear system, we can simply define the state of interest as the perturbation and drop all terms related to the equilibrium. We obtain the following system, known as the reduced MHD equations:
\begin{align}
\label{eq:reduced_MHD_2d}
    \partial_{t}\psi - \frac{1}{S}(\nabla^{2}\psi) &= \{\psi, \phi \},\\ \notag
\partial_{t}\nabla^{2}\phi - \frac{1}{Re}\nabla^{4}\phi &= \{ \nabla^{2}\phi, \phi\} + \{\psi, \nabla^{2}\psi\}.\notag
\end{align}
In order to stop the background field in x-direction from diffusing, it is common to modify the equations with an external field and a non-uniform resistivity as follows: $\eta(x) = E_{const}/J_{z0}(x) = \mathrm{cosh}(x)^{2}$ and $E_{const} = 1$. We can add an external field since a constant of integration arises during the derivation of Eq. \ref{eq:constant_of_integration}. Adding these terms, we get the following set of equations:

\begin{align}
    \partial_{t}\psi - \frac{1}{S}\left(\eta(y) \nabla^{2}\psi\right) &= \{\psi, \phi\} - \frac{1}{S}E_{const}. 
\label{eq:reduced_anti_diss}
\end{align}

Adding these terms has the very simple effect of mitigating the diffusion of the background field, while allowing changes to be made to it. Inserting $\phi = 0$, we can easily check that the resistivity leads to an exact cancellation of the diffusion term in conjunction with the external field. We can rearrange the evolution equations for perturbations in the following way, where $\mathbf{\hat{F}}$ is the Fourier transform of the vector of nonlinear terms:
\begin{align}
(-i\omega \cdot \rho \mathcal{M} - \mathcal{L}) \begin{bmatrix}
        \psi \\
        \phi \\
        \end{bmatrix} = \mathbf{\hat{F^{}}} , \quad
\mathcal{M} = \begin{bmatrix}
            \mathbf{1} & 0 & \\
            0 & \nabla^{2}\\
        \end{bmatrix}, \\\mathcal{L} = \begin{bmatrix}
            -\frac{1}{S}\nabla^{2} & -B_{0}\partial_{x}  \\ \notag
            \partial_{x}(B_{0}^{\prime\prime}-B_{0} \nabla^{2}) & -\frac{1}{Re}\nabla^{4}   \\ 
        \end{bmatrix}.
\end{align}
Rearranging and absorbing the constant factor $\rho$, we obtain,
\begin{align}
     \begin{bmatrix}
        \psi \\
        \phi 
        \end{bmatrix} = \mathcal{H}(\omega) \cdot \mathbf{\hat{G}} , \quad \mathbf{\hat{G}} = \mathcal{M}^{-1} \cdot \mathbf{\hat{F}}, \\  \label{eq:resolvent_operator}
        \mathcal{H}(\omega) = (i\omega\cdot\mathbf{1} - \mathcal{M}^{-1}\mathcal{L})^{-1}.
\end{align}

The four scalar fields that represent our solution are sought using the following structure: $f(x, y, t) = a(y)e^{i\left(k_{x}x - i\omega t\right)}$. This reflects the fact that we assume our solution is periodic in x-direction. 

At the boundaries in y-direction, no-penetration and perfect conductivity conditions are prescribed. 
\begin{align}
\label{sec:u_BCs}
    u_{y}\vert_{y = \mathrm{\pm L_{y}}} = \partial_{x} \phi \vert_{y = \mathrm{\pm L_{y}}} &= 0 \\\notag
    u_{x}\vert_{y = \mathrm{\pm L_{y}}} = \partial_{y} \phi \vert_{y = \mathrm{\pm L_{y}}} &= 0 \\\notag
    B_{y}\vert_{y = \mathrm{\pm L_{y}}} = \partial_{x} \psi \vert_{y = \mathrm{\pm L_{y}}} &= 0.
\end{align}
Which is equivalent to:
\begin{align}
    \phi\vert_{y = \mathrm{\pm L_{y}}} = 0, \quad \partial_y\phi\vert_{y = \mathrm{\pm L_{y}}} = 0, \quad \notag
    \psi\vert_{y = \mathrm{\pm L_{y}}}  = 0.
\end{align}

For our numerical implementation, this suggests using Chebyshev polynomials for the discretization in the y-direction. 

\subsection{Linearized system}
While the equations derived above are valid for any size of perturbation, if we are only interested in small perturbations, we can linearize system \ref{eq:perturbation_equation_full} around an equilibrium. We only consider the equations essential for the 2d dynamics of the tearing plane. The matrices of the linear system are rewritten using the periodicity assumption:
\begin{align}
\mathcal{M} &= \begin{bmatrix}
            \partial_{yy}-k_{x}^{2} & 0 \\
            0&\mathbf{1}\\
        \end{bmatrix},\\
    \mathcal{L} &= \begin{bmatrix}
     \frac{1}{Re}(k_{x}^{2} -     \partial_{yy})^{2} & ik_{x}[B_{0}(\partial_{yy} + k_{x}^{2}) + B_{0}^{\prime\prime}] \\
            -iB_{0}k_{x}&-\frac{1}{S}(\partial_{yy}-k_{x}^{2}) \\
        \end{bmatrix}
\end{align}
We will concentrate our efforts on the linear 2-d system that is structurally somewhat similar to the Orr-Sommerfeldt-Squire equations, after inserting the Fourier ansatz for $\phi, \psi$:
\begin{align}
-i\omega \mathcal{M} \begin{bmatrix}
    \phi \\
    \psi
\end{bmatrix} &= \mathcal{L} \begin{bmatrix}
    \phi \\ \psi
\end{bmatrix}, \hspace{0.2in}
\mathcal{M} = \begin{bmatrix}
          \partial_{yy}-k_{x}^{2} & 0 \\
            0 & I \\
        \end{bmatrix},\\
    \mathcal{L} &= \begin{bmatrix}
          \frac{1}{Re}(k_{x}^{2} - \partial_{yy})^{2} & ik_{x}[B_{0}(\partial_{yy} - k_{x}^{2}) - B_{0}^{\prime\prime}] \\
          iB_{0}k_{x} & \frac{1}{S}(\partial_{yy}-k_{x}^{2}) \\ \notag
        \end{bmatrix}, \\ \notag
        -i\omega \mathcal{M} \mathbf{v} &= \mathcal{L} \mathbf{v}. \notag
\end{align}

The prefactor of $-i$ is chosen to adhere to the eigenvalue convention of the hydrodynamic stability and plasma literature, where most often, a positive imaginary real part implies linear instability. The linearized system we derived is nearly identical with the one used in \cite{mactaggart_2018}, with the only difference being the fact that we consider $\psi$ and $\phi$ rather than primitive variables $u_y, B_y$. In the case of a periodic domain, this simply amounts to a multiplication with the periodic wavenumber $k_{x}$, which cancels out. This means that the only difference is how we interpret the output of the system; the fundamental structure is the same for our case and the one in \cite{mactaggart_2018}. We could compute the fluctuating quantities in y-direction using the solenoidal constraints and then solve a linear boundary value problem to obtain the corresponding stream functions. This procedure is not necessary in our case since we would only obtain precise knowledge about a scaling factor, and we will rescale the stream function anyway to achieve a desired perturbation norm.

\subsection{\label{sec:tranditional_stab}Traditional methods of stability analysis}

Typically, one of the main reasons to linearize a system is to perform stability analysis. Traditional linear stability analysis is concerned with the behavior of perturbations to a system in the asymptotic limit ($t\rightarrow\infty$). Typically, we are concerned with some sort of measure of energy, denoted $E_V$ here. Asymptotic stability is defined as follows:
\begin{align}
    \lim_{t\to\infty} \frac{E_{V}(t)}{E_{V}(0)} \rightarrow 0 \label{eq:stability}.
\end{align}
 In many cases, this limit can depend strongly on the magnitude of $E_{V}(0)$. Intuitively, this makes sense if we consider for example a linearization which is only valid in a certain region of phase space, in this case the magnitude of the initial condition could lie outside the domain of validity and the linear stability analysis would fail. This leads to the notion of conditional stability. This is the case if there exists a threshold $\delta > 0$ such that Eq. \ref{eq:stability} holds given $E_{V}(0) < \delta$. If $\delta = \infty$, we call the solution globally stable, meaning no matter how large this initial condition, decay to zero will always be observed. Note, that this is still asymptotic, i.e. the energy could grow to any finite value before decaying. A more strict definition of stability requires that the perturbation energy decreases for all $t$. This is called monotonic stability:
\begin{align}
    \frac{\mathrm{d}E_{V}(t)}{\mathrm{dt}} < 0, \hspace{0.2in} \forall t > 0.
\end{align}
The first three concepts have one thing in common: They are only concerned with long time behavior, and  disregard any transient behavior for $t < \infty$. They are also the most heavily used criteria for stability analysis. Sometimes, this analysis does not provide an adequate picture. We shall explain why and when this is the case, and present mathematical tools to alleviate these shortcomings. 

\subsection{\label{sec:non_normal_theory}Non-normal operators}

Transient growth can provide a fuller picture of linear stability but requires an analysis of non-normal linear operators. The previously mentioned asymptotic stability concepts are perfectly valid when one is dealing with a normal operator. Unfortunately, many operators in fluid dynamics and plasma physics are non-normal, leading to the issue of asymptotic behavior not reflecting finite-time behavior. In magnetohydrodynamic models other than ideal MHD, non-normality is ubiquitous and arises through many physical channels: finite flows, finite viscosity and resistivity, and many two-fluid effects. Non-normality is also observed in linearized gyro-kinetic~\cite{Landreman_2015} and kinetic plasma models. A normal operator, represented by a matrix, fulfills the following conditions:
\begin{align}
    \bm A^{H} \bm A &= \bm A \bm A^{H} \\
    \exists \bm P \hspace{0.1in} s.t \hspace{0.1in} \bm A &= \bm P \boldsymbol \Lambda \bm P^{*},   \hspace{0.2in} \bm P^{*} \bm P = \bm P \bm P^{*} = \bm I. \notag
\end{align}
We shall demonstrate some core principles of non-normal growth on the following toy problem of the form $\mathbf{x^{\prime}} = \bm A \mathbf{x}$:
\begin{align}
\label{eq:toy_problem}
     \frac{\mathrm{d}}{\mathrm{d}t} \begin{bmatrix}
        v \\
        b \\
        \end{bmatrix} =  \begin{bmatrix}
            -\frac{1}{R} & 1 \\
            0 & -\frac{2}{S}
        \end{bmatrix} \begin{bmatrix}
            v \\
            b
        \end{bmatrix}
\end{align}
The eigenvalues and -vectors of this system are:
\begin{align}
    \sigma_{1} &= -\frac{1}{R} \hspace{0.3in} \mathbf{v_{1}} = \begin{bmatrix}
        1 \\
        0
    \end{bmatrix} \\
    \sigma_{2} &= -\frac{2}{S} \hspace{0.3in} \mathbf{v_{2}} = \begin{bmatrix}
        -\frac{R \cdot S}{2R - S} \\
        1
    \end{bmatrix} \notag
\end{align}

Using these quantities, we can write the explicit solution of the differential equation as follows, in eigenvector basis:

\begin{align}
    \begin{bmatrix}
        v \\
        b \\
        \end{bmatrix} = k_{1} \cdot \begin{bmatrix}
        1 \\
        0
    \end{bmatrix} \cdot e^{-\frac{1}{R} t} + k_{2} \cdot \begin{bmatrix}
        -\frac{R \cdot S}{2R - S} \\
        1
    \end{bmatrix} \cdot e^{-\frac{2}{S} t}
\end{align}

Where $k_{1}, k_{2}$ are chosen such that initial conditions are satisfied. The solution of the system can be written as $\mathbf{x}(t) = e^{At}\mathbf{x_{0}}$. The symbolic form of the matrix exponential is the following:
\begin{align}
    e^{\bm At} = \begin{bmatrix}
        \mathrm{exp}(-1/R) & -\mathrm{exp}(-2/S)/(2/S - 1/R) + \mathrm{exp}(-1/R)/(2/S - 1/R) \\
        0 & \mathrm{exp}(-2/S)
    \end{bmatrix}. \label{eq:matrix_exponential}
\end{align}
Looking at the two eigenvectors we use to construct a solution reveals an interesting fact: The angle between them depends on the parameters $S, R$. Explicitly it is:
\begin{align}
    cos(\theta) = \frac{-S}{\sqrt{S^{2} + 1}}.
\end{align}
This expression clearly tends to zero, as $S$ tends to infinity, meaning the two eigenvectors become less linearly independent as $S\rightarrow\infty$. From the full solution, it is also clear that if $S=R \rightarrow 0$, the transient growth will also shrink due to the equations becoming less coupled. 

\begin{figure}[htbp]
  \centering
  \begin{minipage}[b]{0.49\textwidth}
    \centering
    \includegraphics[width=0.99\textwidth]{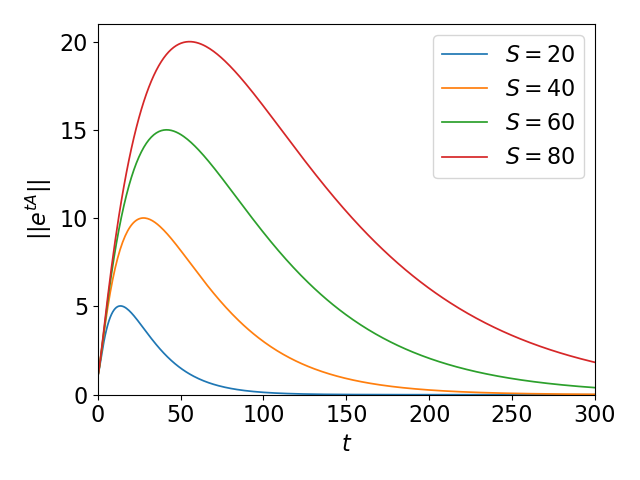}
    \caption{2-norm of the matrix exponential \ref{eq:matrix_exponential} for different times and different values of the parameter $S$. Increasing the value of $S$ leads to stronger transient growth. This matrix exponential is the upper bound of possible transient growth.}
    \label{fig:toy_growth}
  \end{minipage}
  \hfill
  \begin{minipage}[b]{0.5\textwidth}
    \centering
    \includegraphics[width=0.99\textwidth]{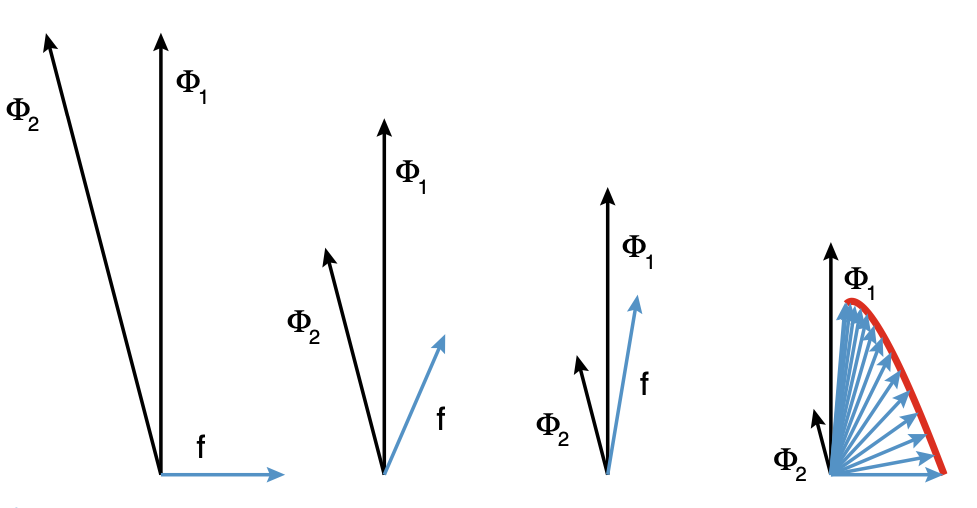}
    \caption{Graphical illustration of the mechanism behind transient growth in a 2 dimensional system representation. The blue arrow represents the solution, with $\Phi_1, \Phi_2$ being the non-orthogonal eigenvectors of the system. Figure reproduced from Schmidt~\cite{schmidt_nonmodal}.}
    \label{fig:toy_growth_arrows}
  \end{minipage}
\end{figure}

A graphical illustration of this unintuitive concept is given in Fig. \ref{fig:toy_growth_arrows}. Clearly, $\Phi_1$ and $\Phi_2$ are not orthogonal. Assuming that one of these eigenfunctions decays faster than the other, in this case $\Phi_2$, we see that instead of making the distance of the solution to the origin smaller, it contains a larger dose of $\Phi_1$, which is still fairly large. Once the solution contains mostly $\Phi_1$, it will start decaying at the rate of $\Phi_1$. Depending on how extreme the difference in decay rates is and, more importantly, how non-orthogonal the functions are, this effect can have significant implications for transient growth. 


Operators with this property of non-orthogonal eigenfunctions, or matrices with non-orthogonal eigenvectors, are called \textit{non-normal}. In the context of stability, it is obviously important to know whether the system at hand in non-normal, and therefore needs special attention, or whether computing eigenvalues suffices. In order to determine if nonmodal effects are important, one makes use of a generalization of the spectrum, the $\epsilon$-pseudospectrum, of an operator.

\subsection{\label{sec:pseudospectra_theory}Pseudospectra and pseudomodes of matrices}

While the all the concepts previously discussed apply to infinite dimensional operators as well as finite dimensional ones, in engineering we are typically concerned with finite dimensional ones, i.e. matrices. While it is true that PDEs are infinite-dimensional, most established computation routines rely on converting the operator into finite dimensional representations, that - ideally - converge to the infinite-dimensional one for relatively small, finite dimensions. In any calculation involving a computer, we end up wanting to know more about a matrix $A$, which is often square, s.t. $\bm A \in \mathbb{C}^{n \times n}$. Every square matrix has a set of complex numbers associated with it called the eigenvalue spectrum:

\begin{align}
    \sigma(A) &= \{ \omega \in \mathbb{C} \vert \hspace{0.05in} {\bm A}\mathbf{v} = \lambda\mathbf{v} \}
\end{align}

In other words, this set consists of complex numbers for which the matrix resolvent $\bm R = (\lambda \bm I - \bm A)^{-1}$ is not well-defined. A matrix $\bm A$ with $n$ distinct eigenvalues has the same amount of eigenvectors that form a complete set. In this case, we can factorize our matrix as $\bm A = \bm V \boldsymbol \Lambda \bm V^{-1}$ where $\boldsymbol \Lambda$ is diagonal and contains the eigenvalues of $\bm A$. Such a matrix is called nondefective. These eigenvectors represent a change of coordinates in the eigenbasis. This change of coordinates makes many problems, such as computing the matrix exponential, trivial.

For a general linear operator (not necessarily a matrix), such as a differential or integral operator, similar notions exist. There is a resolvent operator that exists at all points that do not belong to the spectrum, with the spectrum being a closed set in the complex plane. In contrast to the matrix case, not every point in the spectrum has to be an eigenvalue. The branch of mathematics that deals with these concepts for general linear operators is called spectral theory and requires a more thorough and rigorous treatment than is possible in this thesis.

A concern that arises naturally, at the very latest, once eigenvalues are computed with, inevitably, finite precision, is how sensitive these computed eigenvalues are, i.e., how a slight change in the matrix entries affects the eigenvalues.

A much more robust picture in this regard is painted by looking at the pseudospectrum of a matrix, which generalizes the notion of its spectrum. Instead of asking, ``where does the resolvent not exist'', i.e., asking for points at which it is singular and therefore infinite, we want points where it simply becomes larger than a certain threshold. More formally, we can write the first definition of a pseudospectrum:

\begin{align}
   \sigma_{\epsilon}({\bm A}) = \{ z \in \mathbb{C} \vert \hspace{0.05in} \vert\vert (\omega {\bm I} - {\bm A})^{-1} \vert\vert \geq \frac{1}{\epsilon}\}, \quad \epsilon > 0 \label{eq:og_ps_definition}
\end{align}

This is simply the subset of the complex plane that is bounded by a level curve of the resolvent norm with a value of $\epsilon^{-1}$. In the case of a normal matrix, these level curves would be formed by spheres with eigenvalues at their centers. The pseudospectra for different values of $\epsilon$ are nested sets. The points that are contained by all $\epsilon$-pseudospectra form the spectrum of the matrix.

This can be shown to be conceptually equivalent to looking at the spectrum of a system perturbed in its coefficients with magnitude $\epsilon$, leading to the second, equivalent definition:

\begin{align}
    \sigma_{\epsilon}({\bm A}) = \{ z \in \mathbb{C} \vert \hspace{0.05in} z \in \sigma({\bm A} + {\bm E}),  \vert\vert {\bm E} \vert\vert < \epsilon \}, \quad \epsilon > 0
\end{align}

Intuitively, this means the pseudopectrum gives us the information about where the eigenvalues of a matrix can go if we perturb the matrix with a certain magnitude. This sensitivity-based interpretation is especially interesting in engineering applications, where matrices can contain experimentally determined parameters or numerical approximations. In that case, the $\epsilon$-pseudospectrum shows how errors in discretization or modeling can impact the results of an eigenvalue analysis. This definition implies that pseudospectra could be found by using a sampling (sampling $\bm E$, that is) procedure and simply plotting the union of all sampled spectra. A third definition is:

\begin{align}
    \sigma_{\epsilon}({\bm A}) = \{ z \in \mathbb{C} \vert \hspace{0.05in} \vert\vert (z - {\bm A}) \mathbf{v} \vert\vert < \epsilon, \hspace{0.05in}  \vert\vert \mathbf{v} \vert\vert = 1 \}, \quad \epsilon > 0
\end{align}

This definition gives rise to the concept of a pseudomode $\mathbf{v}$. If we choose the norm $\vert\vert \cdot \vert\vert$ as the 2-norm, the norm of a matrix is its largest singular value, and the inverse of its norm is the inverse of the smallest singular value, as follows:

\begin{align}
    \vert\vert(\bm I z-{\bm A})^{-1}\vert\vert_{2} = \frac{1}{S_{min}(\bm I z-{\bm A})} \label{eq:singular_value_identitiy}
\end{align}

We therefore get another definition of the pseudospectrum that only holds if we choose the 2-norm after combining Eq. [\ref{eq:og_ps_definition}] and [\ref{eq:singular_value_identitiy}]:

\begin{align}
    \sigma_{\epsilon}({\bm A}) = \{ \bm I z \in \mathbb{C} \vert \hspace{0.05in} S_{min}(\bm I z-{\bm A}) < \epsilon \}, \epsilon > 0 \label{eq:sminPS}
\end{align}

This definition is what we will use to actually compute the pseudospectra, since a number of algorithms exist to efficiently compute either the largest or smallest singular values of even fairly large matrices, whereas computing the full spectrum can quickly become expensive. Because we are mostly interested in physical systems, $A$ will often be transformed by a similarity transform, such that $\vert\vert WAW^{-1}\vert\vert_{2}$ is representative of a physical norm, also implying that $\vert\vert W \mathbf{u}\vert\vert$ is a representation of a physical norm of the state vector $\mathbf{u}$.

\begin{figure}%
    \centering
    \subfloat{{\includegraphics[width=0.45\linewidth]{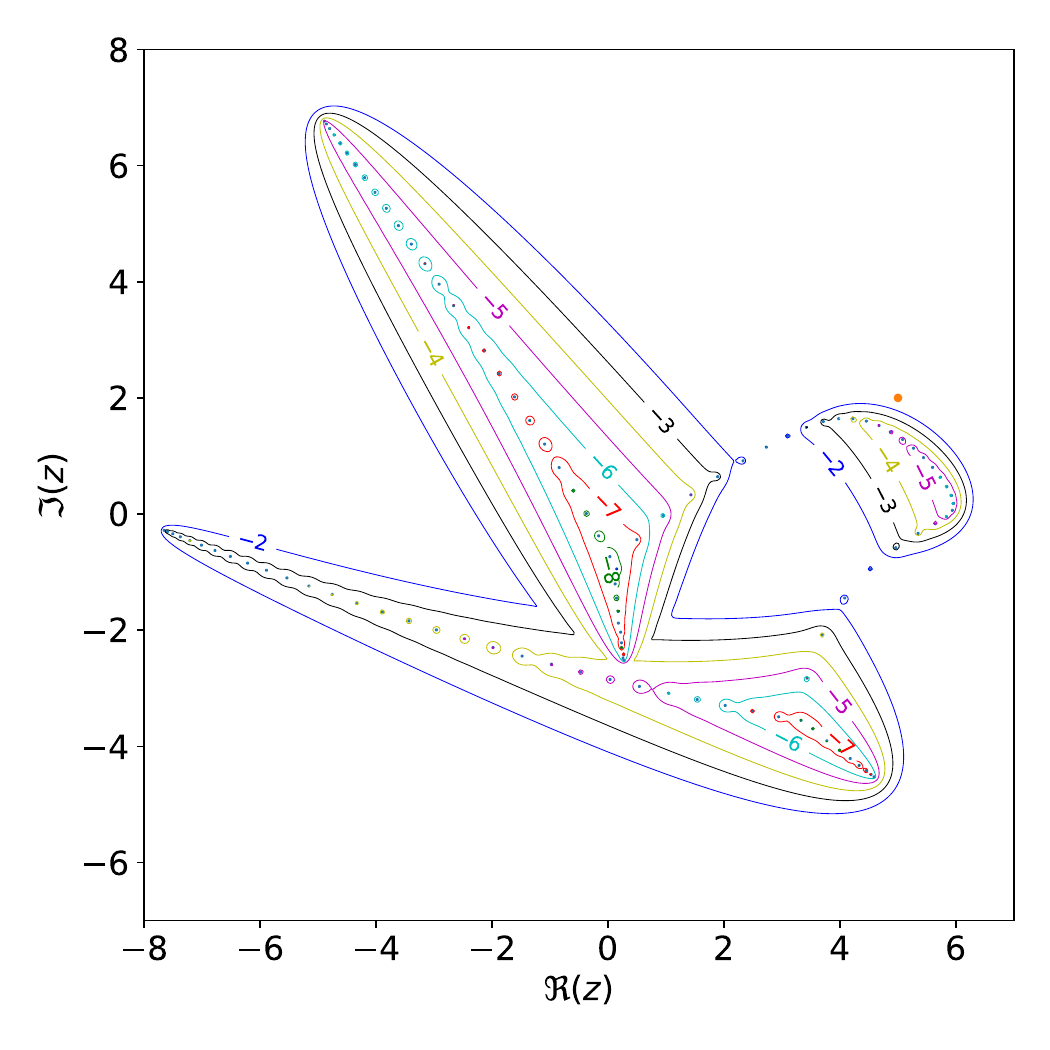} }}%
    \qquad
    \subfloat{{\includegraphics[width=0.45\linewidth]{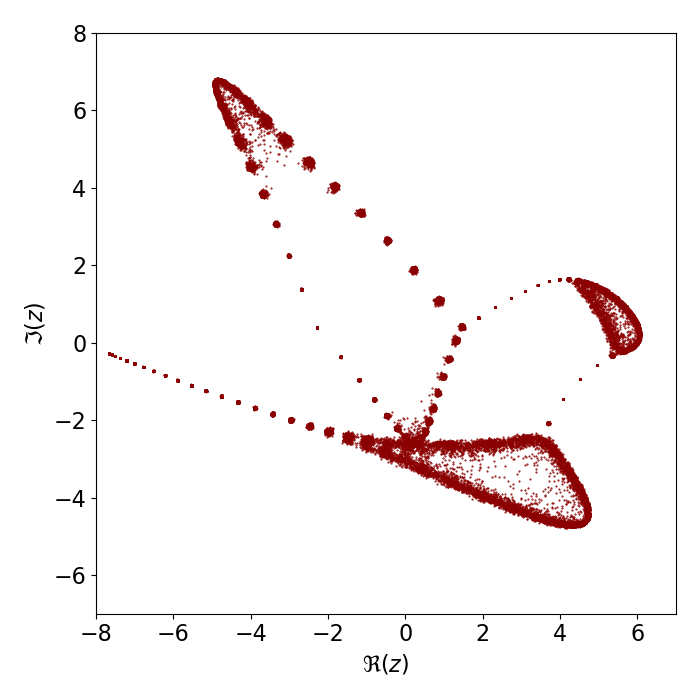} }}%
    \caption{Pseudospectrum of Toeplitz matrix for $\epsilon = 10^{-2}, 10^{-3}..$ (left) and the union of spectra of the same Toeplitz matrix, but with added noise of magnitude $10^{-2}$ in the matrix 2-norm. 500 hundred perturbed matrices were sampled. Sampling is an inefficient way to approximate pseudospectra.}%
    \label{fig:ps_sampling}%
\end{figure}

Let us look at a few simple examples. First, we will consider the following Toeplitz matrix:
\begin{align}
    \bm A = \begin{bmatrix}
        0 & 2i & -1 & 2 & 0 & 0 \\
        0 & 0  & 2i & -1& 0 & 0 \\
        -4& 0  & 0  & 2i& -1& 2 \\
       -2i& -4 & 0  & 0 & 2i& -1\\
         0& -2i& -4 & 0 & 0 & 2i\\
         0 & 0 & -2i& -4& 0 & 0 \\
    \end{bmatrix}.
\end{align}
We can see the pseudospectrum of the matrix in Fig. \ref{fig:ps_sampling}. The spectrum can be identified as the $\epsilon = 0$ pseudospectrum. The pseudospectrum was computed in two ways: Firstly by sampling the resolvent on a grid of complex numbers and computing the smallest singular value and via sampling perturbed matrices. While the sampling approach is very inefficient, it does lead to similar patterns as the SVD-based approach and one can see how the matrix responds differently to perturbations on different parts of the complex plane. In these and any other plots, $\Re(\cdot)$ and $\Im(\cdot)$ denote real and imaginary part of a complex quantity.

We had already introduced the concept of pseudomodes previously. We shall now gain some more understanding with a second example. Furthermore, we consider a sparse matrix that is only populated on its main diagonal and the first upper diagonal, similar to a matrix one would get by applying a one-sided finite difference scheme to a differential operator. 

\begin{align}
    x_{j} = \frac{2\pi}{N}, \hspace{0.2in} 1 \leq j \leq N \\
    A_{jj} = x_{j}, \hspace{0.2in} A_{j,j+1} = 0.5 x_{j}
\end{align}

\begin{figure}%
    \centering
    \subfloat{{\includegraphics[width=0.4\linewidth]{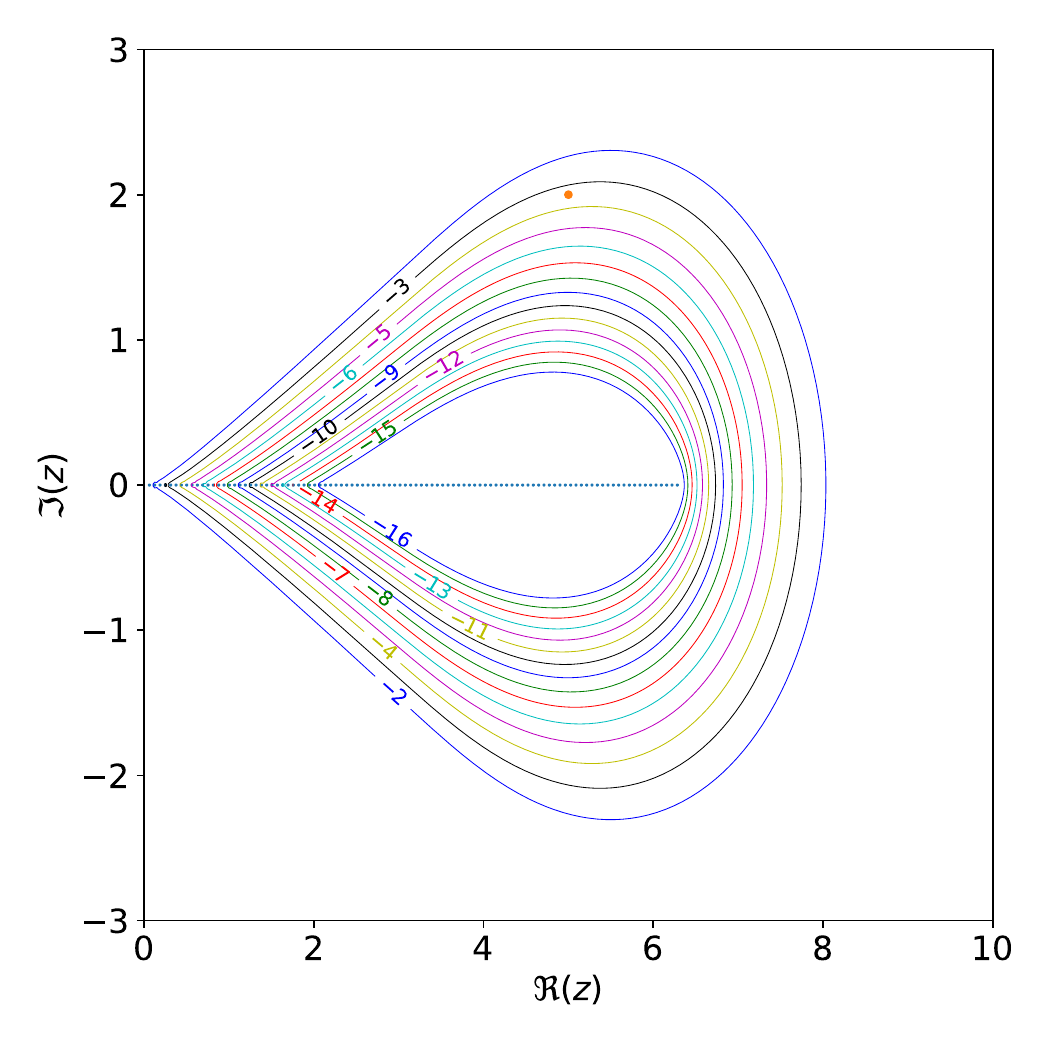} }}%
    \qquad
    \subfloat{{\includegraphics[width=0.4\linewidth]{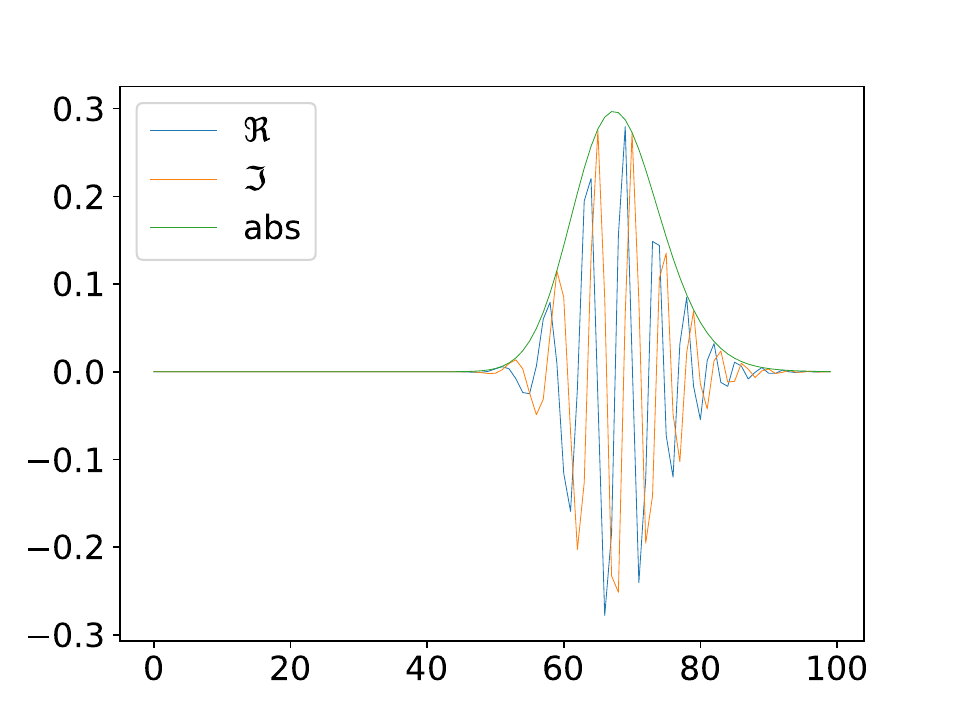} }}%
    \caption{Pseudo spectrum of the twisted toeplitz matrix for $\epsilon = 10^{-2}, 10^{-3}..$ (left) and a pseudomode of the same matrix.}%
    \label{fig:pseudomode}%
\end{figure}

The pseudospectrum and spectrum can be seen in Fig. The pseudospectrum and spectrum can be seen in Fig. \ref{fig:pseudomode}, along with a pseudomode of said matrix. We can see that the pseudomode is similar to a wave packet bounded by an exponential curve. Formally it fulfills the condition: $\vert\vert A \mathbf{v} - \lambda \mathbf{v}\vert\vert = \epsilon \vert\vert \mathbf{v}\vert \vert$ for a given pseudoeigenvalue $\lambda$. In fact, we can show that this pseudomode is proportional to a exponential wavepacket. This means it is very localized. For certain parts of the complex plane, such localized pseudomodes exist; for others, they are not necessarily localized. There exists a mathematical criterion, called the twist condition to make this distinction for strongly structured matrices, such as the twisted Toeplitz matrix at hand. This twist condition has been successfully used for simple fluid-based systems \cite{OBRIST_SCHMID_2010}, unfortunately the formalism does not extend in a straightforward way to our MHD system. A related result that deals with coupled systems and a modified mass matrix of a similar form as our system is \cite{Dawson_McKeon_2019}, which shows that while it is possible to obtain some bounds and geometric intuition using the twist condition, the framework is not mature and straightforward enough to apply here, and the scope of the work did not permit extensive studies on this topic. It shall be noted that we are dealing with a linear system that includes non-periodic boundary conditions, non-constant coefficients, arises from two coupled states, and contains an implicit inversion of a differential operator. These factors very much limit our ability to derive analytical results from the resulting system matrix. More generally, analytic characterizations of the ideal MHD spectrum for realistic geometries can be very complex~\cite{freidberg2014ideal}. 

These pseudomodes can be seen as generalizations of eigenvectors in the same way that pseudo-eigenvalues can be seen as generalizations of eigenvalues; they represent directions, that, when excited, lead to a strong system response of magnitude $\mathcal{O}(1/\epsilon)$. We shall see that some of these pseudomodes are able to trigger strong transient growth.

\subsection{\label{sec:energynorm}Energy and $L_{2}$ norm}

As stated in the previous section, the pseudospectrum and, therefore, by extension, the transient behavior of the system depends on the norm chosen. Since computing the discrete $L_2$ norm of a vector is straightforward and all previously mentioned results for pseudospectra rely on using the discrete $L_2$ norm, we shall formulate our framework in such a way that the Euclidean 2-norm is a good approximation of whatever physical norm we are interested in. We will do this by transforming our linear system into a coordinate system in which the discrete $L_2$ norm is equivalent to a converging approximation of a desired continuous norm. The simplest norm one could consider that has any physical interpretation is simply the integral of the state vector, in our case, the two stream functions $\phi, \psi$. Looking at only one of the stream functions for simplicity, we get the following desired expression (repeated indices imply summation):
\begin{align}
    \int_{-d}^{d} \vert \psi \vert ^{2} dy \approx w_{i}^2\vert \psi_{i} \vert^{2} = \vert\vert \bm W \psi \vert\vert_{2}^{2}, \hspace{0.1in} \bm W = \mathrm{diag}\{w^{\frac{1}{2}}\},
\end{align}
after choosing numerical quadrature weights, $w_i$. For an evenly spaced grid, this would simply amount to $w_i^2 = h$ where $h$ is the grid spacing, recovering a mid-point integration scheme. In the case of the Chebyshev expansion used in this work, when using the matrices that map from physical space to physical space, as defined in Sec.\ref{section:Chebyshev_matrix}, we will use the Gauss-Lobatto weights, defined as follows, with $x_j$ being the interpolation points and $N$ the total amount of grid points:
\begin{align}
    w_j^{2} &= \frac{\pi \sqrt{d^2-x_{j}^{2}}}{2(N+1)}.
\end{align}
The factor two stems from the fact that our state vector contains two variables, and would be changed to one for a system with one state variable. This induces a matrix norm, with the inverse of $\bm W$ being calculated analytically since $\bm W$ is a diagonal matrix:
\begin{align}
    \vert\vert \bm A \vert\vert = \vert\vert \bm W \bm A \bm W^{-1} \vert \vert_{2}.
\end{align}
 This means we can use this matrix norm to transform our system is such a way that the Euclidean norm now represents the integral of the state vector, accounting for the uneven spacing of the Chebyshev grid \cite{Trefethen_1999}.

Getting a system whose Euclidean norm converges to the continuous energy norm is more involved. Let us again look at a single component of the state vector $\phi$ and compute the $L_2$ norm, but directly using the coefficients of the pseudospectral approximation instead of function values at the physical grid points. Let us recall that the approximation via pseudospectral method is given by:
\begin{align}
    \phi &= c_iT_i.
\end{align}
That means we can write the desired integral of $\phi$ as follows,
\begin{align}
    \vert\vert \phi \vert\vert^2_2 &= \int_{-d}^{d} \vert \phi \vert^2 dy  = d\cdot\int_{-1}^{1}c^*_j c_k T_jT_k dy = \vert\vert \bm F \mathbf{c} \vert\vert^{2}_{2} \\ 
    &= c^*_j M_{jk} c_k = \mathbf{c}^H \bm M \mathbf{c}, \notag
\end{align}
where the matrix $\bm F$ fulfills the factorization $\bm F^H \bm F = \bm M$ with:
\begin{align}
    M_{ji}& = d \cdot \int_{-1}^{1} T_{i}(y) T_{j}(y) dy = \begin{cases}
    0 & \text{if } \textit{j + k odd, } \\
    \frac{d}{1-(j+k)^2} + \frac{d}{1-(j-k)^2} & \text{j + k even}
\end{cases}.
\end{align}
The factor $d$ being present is due to the dimension of the computational domain in the $y$-direction. This derivation shows how we can easily get integrals from the coefficients of a spectral series, due to the properties of the basis functions used. Concretely, we want to look at a norm corresponding to the total energy of the perturbed velocity and magnetic fields, i.e. the following expression:
\begin{align}
    E_{V} = \frac{1}{2} \int_{-d}^{d} (\vert \mathbf{u} \vert^2 + \vert \mathbf{b} \vert^{2}) dy \label{eq:energy_integral}.
\end{align}
Using the solenoidal constraint and the incompressibility constraint, along with Parsevals theorem, we can formulate Eq. \ref{eq:energy_integral} as a function of the stream function and vector potential:

\begin{align}
    E_{V} = \int_{k} \frac{1}{2k_x^2} \int_{-d}^{d} (\vert \partial_y \psi \vert^2 + k_x^2 \vert \psi \vert^2 + \vert \partial_y \phi \vert^2 + k_x^2 \vert \phi \vert^2)dydk_x
\end{align}

The energy density is therefore:

\begin{align}
    E = \frac{1}{2k_x^2} \int_{-d}^{d} (\vert \partial_{y} \psi \vert^2 + k_x^2 \vert \psi \vert^2 + \vert \partial_{y} \phi \vert^2 + k_x^2 \vert \phi \vert^2)dy
\end{align}

We will again look at a single component, $\phi$ to keep things simple. The energy norm of $\phi$ is defined to be the following expression:
\begin{align}
    \vert\vert \phi \vert\vert^2_H &= \int_{-d}^{d} \vert \partial_y \phi \vert^{2} + k_x^2\vert \phi \vert^2 dy  = \vert\vert \bm F \mathbf{c} \vert\vert^{2}_{2}.
\end{align}
Where $\bm F^H \bm F = \bm D^H \bm M \bm D + k^2 \bm M$ and the matrix $\bm D$ transforms the coefficients of a Chebyshev series into the coefficients of the derivative of the same series, with this matrix being easily obtained from the recurrence relations, see \cite{canuto2007spectral}:

\begin{align}
    D_{ij} = \begin{cases}
    j & \text{if } \textit{i = 0 and j odd, } \\
    2 j  & \text{if }  \textit{j $>$ i} \end{cases}
\end{align}

And we can therefore approximate an energy norm, using the same technique as with the single variable, but this time for the entire state vector. Because we are interested in restricting our dynamics to a subset of the spectrum, we will include this restriction process in our derivation. Let us assume we have computed the entire discrete spectrum of the differential operator, consisting of all eigenpairs: $(\mathbf{\tilde{v}}, \sigma)$ and a subset of the eigenpairs we want to use: $\mathcal{S}^{N} =  \{ \mathbf{\tilde{v}}_0, ...\mathbf{\tilde{v}}_N \}$. Any vector function $\mathbf{v} \in span\{\mathcal{S}^N\}$ can be expanded using these eigenvectors as basis functions, i.e., in modal coordinates:

\begin{align}
    \mathbf{v} = \kappa_i \mathbf{\tilde{v}}_i
\end{align}

Any two eigenvectors $\mathbf{\tilde{v}}_i$ have an inner product induced by the energy norm:
\begin{align}
    (\mathbf{\tilde{v}}_i, \mathbf{\tilde{v}}_j)_{E} = Q_{ij} = \frac{1}{2k_x^2} \int_{-d}^{d} \mathbf{\tilde{v}}_i \mathcal{Q} \mathbf{\tilde{v}}_jdy
\end{align}

The matrix $\mathcal{Q}$ can be derived by applying integration by parts to Eq.\ref{eq:energy_integral} and using the boundary conditions:
\begin{align}
    \int_{-d}^{d} (\vert \bm D \psi \vert^2 + k_x^2 \vert \psi \vert^2 + \vert \bm D \phi \vert^2 + k_x^2 \vert \phi \vert^2)dy &= \left[ \psi \cdot \bm D\psi + \phi \cdot \bm D\phi \right]_{-d}^{d} \\ \notag + \int_{-d}^{d} k_x^2 (\phi^{H}\phi + \psi^{H} \psi) - (\psi^H \bm D \psi + \phi^H \bm D \psi) dy \\ \notag
    \mathcal{Q} &= \begin{bmatrix}
          k_{x}^{2} - \bm D^2 & 0 \\
            0 & k_{x}^{2} - \bm D^2 \\
        \end{bmatrix} \notag
\end{align}
We now define a matrix $Q_{ij}$ whose entries correspond to the inner products of eigenvectors $\mathbf{v}_i, \mathbf{v}_j$, 
\begin{align}
    \label{sec:Q_matrix}
    Q_{ij} = \int_{-d}^d \tilde{\bm v}_i\mathcal{\bm Q}\tilde{\bm v}_jdy.
\end{align}
This matrix is hermitian and positive-definite, and therefore can be factored using, for example, a Cholesky decomposition, as $\bm Q = \bm F^H \bm F$. Taking this matrix, we can expand any linear combination of basis vectors using the series $\mathbf{v} = \kappa_i \mathbf{\tilde{v}}_i$ and get the energy norm of the vector, using the Euclidean 2-norm as follows again, also inducing a matrix norm:
\begin{align}
    (\mathbf{v}, \mathbf{v})_{E}&= \frac{1}{2k^2} \int_{-d}^{d} \mathbf{v}^H \mathcal{Q} \mathbf{v}dy = \boldsymbol \kappa^H \bm Q \boldsymbol \kappa = \boldsymbol \kappa^H \bm F^H \bm F \boldsymbol \kappa, \\ \notag
    \vert\vert\ \mathbf{v}\vert\vert_E &= \vert\vert \bm F \boldsymbol \kappa \vert\vert_2, \\ \notag
    \vert\vert \bm A \vert\vert_{E} &= \vert\vert \bm F \bm A \bm F^{-1} \vert\vert_{2}. \notag
\end{align}
It is noted that we have to perform two operations on the matrix $\bm Q$: We need to factor it as $\bm F^H \bm F=\bm Q$ and we need to obtain the inverse matrix $\bm F^{-1}$. Since the Cholesky decomposition fails, if a single eigenvalue of $\bm Q$ is negative, even if this is only due to numerical noise, an alternative approach using the SVD is considered, which is more robust. It uses the fact that the left and right singular vectors of a symmetric, definite matrix are identical: 
\begin{align}
    \bm Q &= \bm U \boldsymbol \Sigma \bm V^H \\ \notag
    \bm U &= \bm V \\ \notag
    \bm F &= \sqrt{ \boldsymbol \Sigma} \bm U^H \\ \notag
    \bm F^H \bm F &= \bm U\sqrt{\boldsymbol \Sigma} \sqrt{\boldsymbol \Sigma} \bm U^H = \bm U \boldsymbol \Sigma \bm V^H. \notag
\end{align}
The inverse matrix $\bm F^{-1}$ was obtained via pseudoinverse. For the pseudoinverse, a tolerance value can be chosen that decides which singular values are considered for the inversion. A very similar approach was used for hydrodynamic calculations in \cite{Schmidhenningson}. For the sake of reproducibility, we used a similar procedure. This requires us to reformulate the numerical system in terms of expansion coefficients, and unfortunately, we cannot re-use the system formulated in terms of physical variables, which was used with the $L_2$ norm. It is potentially possible to formulate the energy norm in physical space using, for example, an auxiliary state variable, although that would double the size of the system matrices. We closely followed the matrix-based approach given in detail in \cite{energy_norm_derivation}.

\subsection{\label{sec:transient_theory}Quantifying transient growth}

We have demonstrated the need to look at matrices with other tools than eigenvalue analysis, especially in the presence of uncertainties or non-normal systems, as described in the previous sections. For engineers, the most important aspect might be the possibility of non-modal effects, i.e., transient growth. Simply examining the pseudospectrum of a matrix can tell us whether there is a possibility of non-normal effects, but it does not directly tell us how strong it will be at which time or which structures in the initial conditions encourage it. Recall that while modal stability should be invariant with respect to initial conditions, non-modal stability can depend very heavily on the exact nature of the perturbations. Having derived the necessary formulas to include any physics we care about in the Euclidean norm, we can now make use of the bounds and formulas derived for pseudospectra, using the Euclidean norm.

First, we will show some results that give us a more quantitative picture of possible transient effects as motivation. We can show that the norm of the matrix exponential gives us a precise upper bound on the largest possible transient growth of all possible initial conditions. This is to be understood in the sense of an optimization problem; the matrix exponential represents the maximum amount of growth possible at a time $t$ after optimizing over the set of all possible initial conditions.

\begin{align}
    \sup_{\mathbf{v}(0)} \frac{\vert\vert\mathbf{v}(t)\vert\vert^{2}}{\vert\vert\mathbf{v}(0)\vert\vert^{2}} =  \sup_{\mathbf{v}(0)}\frac{\vert\vert \mathrm{exp}(t\cdot M^{-1}L)\mathbf{v}(0)\vert\vert^{2}}{\vert\vert\mathbf{v}(0)\vert\vert^{2}} = \vert\vert \mathrm{exp}(t\cdot M^{-1}L)\vert\vert^{2}
\end{align}

\begin{figure}%
    \centering
    {\includegraphics[width=0.6\linewidth]{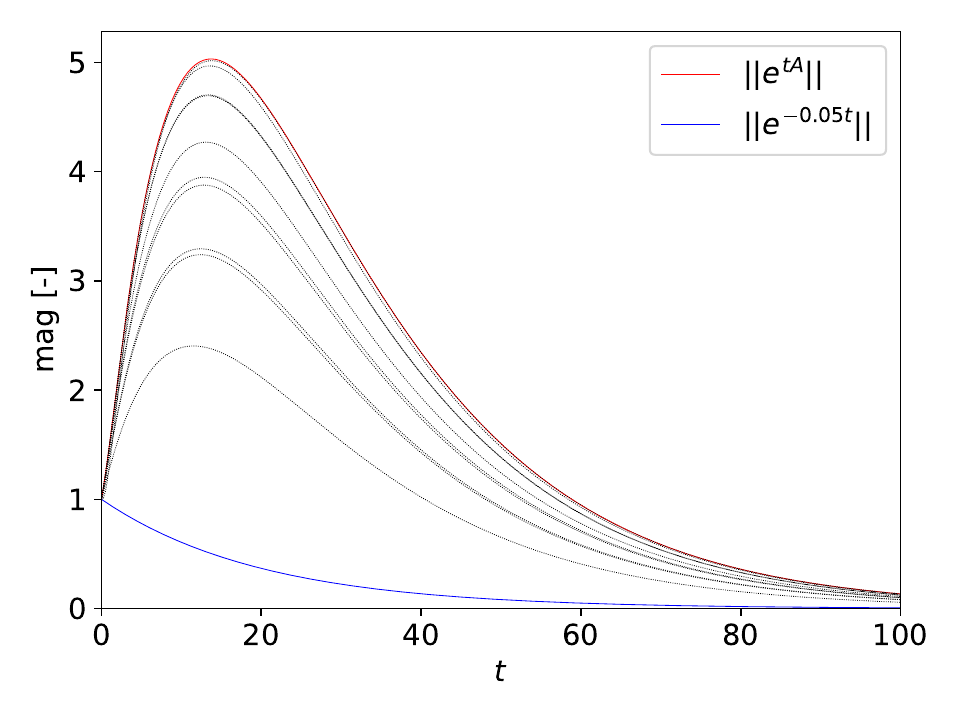} }%
    \caption{Solutions to the toy problem. The upper bound given by the exponential is never breached; 10 trajectories with random initial conditions were sampled and solved. The blue curve corresponds to the slowest-decaying eigenvalue. All sampled initial conditions exhibit transient growth.}%
    \label{fig:toy_bounds}%
\end{figure}

The above formula is demonstrated with the toy example from earlier. Sampling random initial conditions (IC) for the toy problem \ref{eq:toy_problem} and evolving them over time shows that all sampled ICs exhibit transient growth before decaying, as can be seen in Fig. \ref{fig:toy_bounds}. The upper bound given by the matrix exponential norm is never breached.

We introduce two numbers we call the spectral abscissa $\alpha(A) = \sup_{z\in \sigma(A)} \Re(z)$ and the $\epsilon$-pseudospectral abscissa $\alpha_{\epsilon}(A) = \sup_{z \in \sigma_{\epsilon}} \Re(z)$. The spectral abscissa is the eigenvalue with the largest real part $\Re(z)$, i.e., the fastest growing eigenvalue, the $\epsilon$ - pseudospectral abscissa is the part of the $\epsilon$ - pseudospectrum that protrudes the furthest into the unstable part of the complex plane and therefor the non-modal analogue of the spectral abscissa. We can derive two simple bounds on the magnitude of $\vert\vert\mathrm{exp}(t\cdot M^{-1}L)\vert\vert^{2}$:

\begin{align}
    \vert\vert\mathrm{exp}(t\cdot M^{-1}L)\vert\vert^{2} &\geq \mathrm{exp}(t\alpha(A)) &\forall t \geq 0 \\
    \sup_{t \geq 0}(\vert\vert\mathrm{exp}(t\cdot M^{-1}L)\vert\vert^{2}) &\geq \frac{\alpha_{\epsilon}(A)}{\epsilon} & \forall \epsilon > 0 \label{eq:tg_bound_lower}
\end{align}

A useful bound in practice is Eq. \ref{eq:tg_bound_lower}, which tells us that transient growth will occur if the spectra of perturbed matrices protrude into the right-hand side of the complex plane and gives us a lower bound on the possible magnitude. They should be understood as a lower bound that \textit{will} be breached by some initial condition at some time. It does not tell us anything about when this will happen or what kind of initial condition leads to that growth, and in general, it is to be used as an indicator rather than a quantitative instrument.

The question remains: which initial conditions will trigger the highest possible initial growth at time $t$. We are considering the linear, generalized eigenvalue problem $\sigma \mathcal{M} \mathbf{v} = \mathcal{L} \mathbf{v}$. We can transform it into standard form if $\mathcal{M}^{-1}$ exists: $\sigma \mathbf{v} = \mathcal{M}^{-1} \mathcal{L} \mathbf{v} = \bm A \mathbf{v}$. 

The optimal initial conditions for transient growth at $t=t_0$ are derived as follows: The upper bound for our growth is $\vert\vert \mathrm{exp}(t_0 A)\vert\vert$, so initial conditions with the desired properties will lead to a solution with that magnitude at $t_0$. So, we can impose $\mathbf{v} = \vert\vert \mathrm{exp}(t_0 \bm A)\vert\vert \mathbf{v}_{0}$. The singular value decomposition of a matrix is the following: $\bm A \bm V= \bm U \boldsymbol \Sigma$. If we look only at the leading singular vectors, denoted $\mathbf{v}_{1}$ and $\mathbf{u}_{1}$, we get the expression $\bm A\mathbf{v}_{1} = \mu \mathbf{u}_{1}$, where $\mu$ is the largest singular value. Setting $\bm A = \mathrm{exp}(t \bm A)$, this corresponds exactly to the condition we have for our maximum initial growth condition and solution at $t_{0}$. Therefore, $\mathbf{v}_{1}$ is the initial condition leading to optimal growth, by producing the state $\mathbf{u}_1$ with a norm amplified by the factor $\mu$ \cite{Schmidhenningson}. All the above calculations assume that the matrix norm is meaningful, i.e., that the matrix has been transformed such that the norm represents, for example, energy.

\subsection{\label{sec:PS_ALGO}Quick ways to compute pseudospectra}

Since pseudospectra are a tool as important as eigenvalues, we should make similar efforts to efficiently obtain them, as has been done with various eigenvalue solvers. It turns out that in many cases, pseudospectra can be obtained with a similar effort as eigenvalues.

Computing a pseudospectrum of a matrix $\bm A$ is simple in principle: Choose a grid of values in $\mathbb{C}$ and compute the SVD of the matrix resolvent at each point, find the minimum singular value, and plot the result. This procedure is not optimal. We only need the smallest singular value, so we should not compute the full SVD. Computing the whole SVD costs approximately $\mathcal{O}(N^{3})$, where $N$ is the dimension of the matrix at hand. Given a grid with $\nu$ points in each direction, we have a total cost of $\mathcal{O} (\nu^{2}N^{3})$. Performing a Schur factorization costs around $\mathcal{O}(N^{3})$, but in turn allows us to get the minimum singular values in $\mathcal{O}(N^{2})$ operations. After performing the Schur factorization, all that is left to do is compute the minimum singular values at locations of interest.

This value $\sigma_{min}$ can be computed in $\mathcal{O}(N^{2})$ operations. Since $\sigma_{min} = \sqrt{\lambda_{min}((z-T)^{H}(z-T))}$ in the case of an upper triangular matrix, we can do this via finding the minimum eigenvalue via iteration. A common choice is inverse Lanczos iteration, which has good convergence properties. This procedure was presented in \cite{Trefethen_1999}, along with a MATLAB implementation. The corresponding pseudocode is presented in \ref{alg:PS_fast}. The parameter $tol$ is used as a convergence criterion for the Lanczos iteration and $\nu$ is a variable that stores the dimension of the grid of sampled values of $z$, assuming a square, evenly spaced grid. While the algorithm is fast, it is certainly not as robust as computing the SVD, which is possible for any value of $z$ at which the matrix resolvent is well defined. When a large number of iterations is needed to achieve convergence of the inverse Lanczos iteration, it is possible to quickly see diminishing returns in computation time. The result of algorithm \ref{alg:PS_fast} is the norm of the matrix resolvent, evaluated at an evenly spaced grid in the complex plane. All figures of pseudospectra in this thesis were produced by plotting approximate contour plots with this discrete grid.

\begin{algorithm}
\caption{Pseudospectrum with Lanczos iteration}\label{alg:PS_fast}
\begin{algorithmic}
\State $T, Z = \texttt{SCHUR}(\bm A) $
\State $tol = 1e^{-3}$

\For{$i$ in $0, ..., \nu$}
\For{$j$ in $0, ..., \nu$}
    \State $z \gets z_{new} $
    \State $T_{1} \gets (zI-T)$
    \State $T_{2} \gets T_{1}^{H}$
    \State $\sigma_{old} \gets 0$
    \State $\mathbf{q}_{old} \gets \texttt{zeros}(N)$
    \State $\beta \gets 0$
    \State $\mathbf{q} \gets \texttt{random}(N), s.t. \vert\vert q\vert\vert = 1$
    \While{$k \leq maxiter$}
        \State $\mathbf{v} \gets \texttt{solve}(T_{1}, \texttt{solve}(T_{2}, \mathbf{q})) - \beta \cdot \mathbf{q}_{old}$
        \State $\alpha \gets \mathbf{q}^{*} \cdot \mathbf{v}$
        \State $\mathbf{v} \gets \mathbf{v} - \alpha \cdot \mathbf{q}$
        \State $\beta \gets \vert\vert \mathbf{q} \vert \vert$
        \State $\mathbf{q}_{old} \gets \mathbf{q}$
        \State $\mathbf{q} \gets \mathbf{v} / \beta$
        \State $H(k + 1, k) \gets \beta$
        \State $H(k, k + 1) \gets \beta$
        \State $H(k, k) \gets \alpha$
        \State $\sigma \gets \sigma_{max}(H(0:k + 1, 0:k + 1)))$
        \If{$\vert \sigma_{old} / (\sigma -1)\vert < tol$}
            \State $\mathbf{break}$
       
        \EndIf
        \State $ \sigma_{old} \gets \sigma$    
    \EndWhile
    \State $R(i, j) = 1/ \sqrt{\sigma}$     
    \EndFor
\EndFor
\end{algorithmic}
\end{algorithm}

Much of the time, there is a part of the spectrum we are particularly interested in. In the case of stability analysis, we typically look at parts close to the imaginary axis. Also, spurious eigenvalues introduced by discretization typically have very large eigenvalues and are not of interest to us. One approach to further increase computation speed is to perform an orthogonal projection of the system $A$ onto a subset of its eigenvectors, $\mathbb{S}^{n} = \{\mathbf{v}_{0}, ..., \mathbf{v}_{n} \}$, where $\bm A \bm V = \bm V \bm D$ and $n << N$. We can then factor $\bm V=\bm Q \bm R$ using a QR-decomposition. Then we can project our full dynamics matrix $\bm A$ onto $\mathbb{S}^{n}$ via $\bm T=\bm R \bm D \bm R^{-1}$ \cite{Trefethen_1999}. In practice, the inversion of $\bm R$ is a sensitive computation for ill-conditioned $R$. If we are working in coefficient space, this restriction process was already covered in Sec. \ref{sec:energynorm} and is even simpler.

\subsection{\label{sec:resolvent}Receptivity and Pseudoresonance: Resolvent analysis}

Apart from transient growth, non-normal operators allow for pseudoresonance, i.e., nonmodal effects that get excited by harmonic forcing. We were able to write the nonlinear system Eq. \ref{eq:reduced_MHD_2d} in the following form, including the nonlinearities as harmonic forcing:

\begin{align}
     \begin{bmatrix}
        \psi \\
        \phi 
        \end{bmatrix} = \mathcal{H}(\omega) \cdot \mathbf{\hat{G}} , \quad \mathbf{\hat{G}} = \mathcal{M}^{-1} \cdot \mathbf{\hat{F}}, \\  \notag
        \mathcal{H}(\omega) = (i\omega\cdot\mathbf{1} - \mathcal{M}^{-1}\mathcal{L})^{-1}.
\end{align}

For $k_x$ and $\omega$ such that $\mathcal{H}$ is not singular, we can compute the resolvent operator $\mathcal{H}(\omega, k_{x})$. Assuming the system is spectrally stable, the resolvent operator maps harmonic forcing input modes to the corresponding asymptotic response modes \cite{Schmidhenningson}. This operator can be decomposed using the SVD to obtain the spatial shapes and corresponding amplification factors, similarly to the derivation of optimal initial conditions:
\begin{align}
\label{eq:resolvent_decomp}
    \mathcal{H}(\omega, k_{x}) &= \Psi \Sigma \Phi,\\ \notag
    R(\omega) &= \underset{\mathbf{q}_{in}}{\textrm{max}} \frac{\vert\vert \mathbf{q}_{out}(t) \vert\vert^{2}}{\vert\vert \mathbf{q}_{in} \vert\vert^{2}} = \underset{\mathbf{q}_{in}}{\textrm{max}} \frac{\vert\vert (i\omega \bm I - \mathcal{M}^{-1}\mathcal{L})^{-1} \mathbf{q}_{in} \mathrm{exp}(i\omega t) \vert\vert^{2}}{\vert\vert \mathbf{q}_{in} \vert\vert^{2}}, \\
    &= \vert\vert (i\omega \bm I - \mathcal{M}^{-1}\mathcal{L})^{-1} \vert\vert^{2}, \notag
\end{align}

where $R(\omega)$ is the function we maximize by choosing $\phi_1$ as input shape of the forcing with frequency $\omega$. For a normal system, the resolvent norm decays as $\frac{1}{dist(z, \sigma)}$, where $dist(z, \sigma)$ is the distance between the complex number $z$ and the eigenvalue closest to it. Intuitively, this means that resonance is only possible when an eigenfrequency of the system is excited. In control systems, one typically uses a bode plot to visualize the system response to harmonic inputs. If the system is non-normal, it is possible to observe pseudoresonances, i.e., strong amplification of harmonic forcing at frequencies that are far away from the nearest eigenfrequency. This concept is particularly relevant, as many signals can be approximated by a superposition of real frequencies \cite{trefethen_stabiltiy}.

The decomposition Eq. \eqref{eq:resolvent_decomp} gives us a hierarchical ordering of input $ \Phi = \{ \phi_{1}, .., \phi_{n} \}$ and output modes $ \Psi = \{ \psi_{1}, .., \psi_{n} \}$, according to their amplification factor, the corresponding singular value. Since the SVD orders modes according to $l_{2}$ - norm, the system matrix should, again, be transformed in a manner that makes this norm a physically relevant quantity, such as the RMS of the state variables or an energy measure. This scaling is typically absorbed into a weighting matrix $W$, as described in \ref{sec:energynorm}. Using the resolvent operator, we can rewrite Eq. \ref{eq:reduced_MHD_2d} as follows:
\begin{align}
    \mathbf{x}_{\omega, k_x} = \begin{bmatrix}
    \phi \\ \psi
\end{bmatrix}_{\omega, k_x} = \mathcal{H}(\omega, k_x) \mathbf{G}_{\omega, k_x}. \label{eq:resolvent_system}
\end{align}

Since the elements of $\Phi$ and $\Psi$ form an orthonormal basis, we can write Eq. \ref{eq:resolvent_system} in that basis, by constructing both $\mathbf{G}_{\omega, k_x}$ and $\mathbf{x}_{\omega, k_x}$ from the resolvent basis, as follows:

\begin{align}
    \mathbf{G}_{k_x, \omega} = \sum_{m}\phi_{k_x, \omega, m} \xi_{k_x, \omega, m}, \\
    \mathbf{x}_{\omega, k_x} = \sum_{m}\psi_{k_x, \omega, m} \sigma_{k_x, \omega, m} \xi_{k_x, \omega, m}. \notag
\end{align}

Thus, the perturbation vector can be written as a linear combination of singular response modes, weighted by an unknown set of forcing coefficients. This decomposition has been used with great success for reduced-order modeling, control and in explaining the linear origins of wall-bounded turbulence \cite{McKEON_SHARMA_2010}\cite{Gomez_Blackburn_Rudman_Sharma_McKeon_2016}\cite{Schmidt_Towne_Rigas_Colonius_Bres_2018} \cite{Bae_Dawson_McKeon_2020}.

\subsection{\label{sec:numerics}Numerical Algorithms}

We will now focus on methods to convert continuous differential operators into matrix equations. Each has pros and contras associated with it. We have chosen two that are well-suited to our application. The first is an approximation using Chebyshev polynomials; the second is a compact-finite difference scheme. In this thesis, we have two tasks to fulfill: discretize an eigenvalue problem and solve a full nonlinear transient PDE. The first can be achieved using matrix representations, with orthogonal polynomials being chosen due to their extensive use in the hydrodynamic stability literature and good convergence properties. The latter is much better done without using matrices, and the open-source spectral solver Dedalus \cite{dedalus} was chosen. It is open source and gives the user a large amount of control and access to the solution procedure. We shall not go in depth on the solution procedure that Dedalus uses, as it is very well described in the papers introducing and benchmarking the solver. We shall only briefly outline why, unfortunately, Dedalus is not the best option to calculate pseudospectra and related quantities.

\subsubsection{\label{section:Chebyshev_matrix}Chebyshev differentiation matrices}

Suppose we have a function, evaluated at the grid points given by the following distribution:

\begin{align}
    x_{j} = cos(j\pi / N), \hspace{0.2in} j = 0, 1, ..., N \label{eq:cheb_points}.
\end{align}

we will call this set of function evaluations $v_{j}$. We now want to do two things: Approximate $v_{j}$ by a polynomial and obtain the derivative of that polynomial at the points $x_{j}$. There exists a unique polynomial of degree $N$ that satisfies $p(x_{j})=v_{j}$. We can find a basis $\{l_{0}, ...l_{N}\}$ to express this interpolant via Lagrange interpolation as follows:

\begin{align}
   l_{j}(x) = \prod^{N}_{i=0, i\neq j} \frac{x-x_{i}}{x_{j} - x_{i}}.
\end{align}

The full interpolating polynomial is obtained by a weighted sum of these basis elements. Each basis function is 1 at $x = x_{j}$, the interpolation point, and 0 at all other $x$. The derivative of each basis function is easily obtained after transforming the equation using the logarithm:
\begin{align}
    log(l_{j}(x)) &= \sum^{N}_{i=0, i\neq j} \log\left(\frac{x-x_{i}}{x_{j} - x_{i}}\right),\\ \notag
    log(l_{j}(x))^{\prime} &= \frac{l_{j}^{\prime}}{l_{j}} = \sum^{N}_{i=0, i\neq j} \frac{1}{x_{j} - x_{i}},\\ \notag
    l_{j}(x)^{\prime} &= l_{j}(x) \sum^{N}_{i=0, i\neq j} \frac{1}{x_{j} - x_{i}}. \notag
\end{align}
Once we have the derivative of an arbitrary basis element, we can simply compute $w_{j} = p^{'}(x_{j})$, where $p(x)$ is the approximation built using the basis elements $l_j$. This operation can be represented by a matrix, after settling on using the grid point distribution given by Eq. \ref{eq:cheb_points}:
\begin{align}
    w &= \bm D_{N}v \\
    (D_{N})_{00} &= \frac{2N^{2}+1}{6}, \quad \notag (D_{N})_{00} = -\frac{2N^{2}+1}{6} ,\\ \notag
(D_{N})_{jj} &= \frac{-x_{j}}{2(1-x_{j}^{2})}, j=0, 1, ..., N-1, \\ \notag
(D_{N})_{ij} &= i\neq j  i,j = 0, ...., N-1.
\end{align}
For second-order equations, we might want to incorporate boundary conditions of the form $p(\pm 1) = 0$. In this case, we can simply ignore $w_{0, N}$ and set $v_{0, N} = 0$. This is equivalent to deleting the first and last rows and columns of $\bm D^{2} = \bm D \cdot \bm D$.

The procedure for obtaining a fourth-order differentiation matrix that incorporates the boundary conditions $p^{\prime}(\pm 1) = p(\pm 1) = 0$ is slightly more involved. We will focus on finding an interpolant function $p$ that satisfies these conditions. Set $p(x) = (1-x^{2})q(x)$. We can differentiate this expression:

\begin{align}
    p^{\prime} &= (1-x^{2})q^{\prime} - 2xq, \\ \notag
    p^{\prime\prime} &= (1-x^{2})q^{\prime\prime} - 4xq^{\prime} - 2q, \\ \notag
    p^{\prime\prime\prime} &= (1-x^{2})q^{\prime\prime\prime} - 6xq{\prime\prime} - 6q^{\prime}, \\ \notag
    p^{\prime\prime\prime\prime} &= (1-x^{2})q^{\prime\prime\prime\prime} - 8xq^{\prime\prime\prime} - 12q^{\prime\prime}, \\ \notag
\end{align} 

and verify that our boundary conditions are met. Using the fact that $q=\frac{v}{1-x^{2}}$ we can again construct a differentiation matrix \cite{tref_spectralmethods}.

\subsubsection{\label{subsec:Chebyshev_coeff}Chebyshev pseudospectral methods}

We can also directly use Chebyshev polynomials as basis functions, use the grid points given by Eq. \ref{eq:cheb_points} and write a series approximation of a function as a weighted sum of polynomials $T_{k} : a \rightarrow b, a\in[-1, 1], b\in[-1, 1]$. The expansion of a function with respect to Chebyshev polynomials is as follows:

\begin{align}
    u(x) &\simeq \sum^{\infty}_{k=0} a_{k}T_{k}(x), \hspace{0.1in} 
    u(x)_{N} \simeq \sum^{N}_{k=0} a_{k}T_{k}(x), \\ \notag
    T_{k}(x) &= \cos(k\theta), \hspace{0.3in} \theta = \arccos(x). \label{eq:chebpoints} \notag
\end{align}

This series of Chebyshev polynomials will converge, as long as there are no singularities in $u(x)$ in the interval $x\in [-1, 1]$. Typically, this will lead to a fairly good approximation for a low value of $N$. 

The pseudospectral approach requires that the residual be zero at a set of collocation points by choosing the coefficients of the expansion in such a manner that the differential equation is exactly satisfied at the grid points. Typically, these points are chosen to be the ones specified in Eq. \ref{eq:chebpoints}. This yields a system of equations we can solve for the unknown coefficients $a_{i}$, that can be written in matrix form:

\begin{align}
    \bm H  \mathbf{a} &= u(x), \\
    H_{ij} &= T_{j}(x_{i}). \notag
\end{align}

Generalizing this concept in order to approximate solutions of a linear differential operator $Lu = f$ is simple: construct $\bm Lu_{N}(x_{i}) = f(x_{i})$ and again solve a linear system for the coefficient vector. In contrast to simple function interpolation, when applying the Chebyshev expansion to a differential operator, boundary conditions have to be considered. For homogenous Dirichlet boundary conditions, we would have the  set of constraints at the boundary nodes:

\begin{align}
    \sum^{N}_{k=0} a_{k}T_{k}(x_{0}) &= 0,  \quad
    \sum^{N}_{k=0} a_{k}T_{k}(x_{N}) = 0
\end{align}

This need for an extra set of equations could be circumvented by modifying the set of basis functions to only include ones that identically fulfill the boundary conditions. Especially for simple, homogenous boundary conditions, this is simple, and we implicitly used this approach previously when we derived the fourth-order differentiation matrix.

Apart from thinking about implementing boundary conditions, we will also need the derivatives of our basis functions. They are given by the recurrence relation:

\begin{align}
    T_{0}^{k}(y_j) &= 0, \hspace{0.1in} T_{1}^{k}(y_{j}) = T_{0}^{(k-1)}(y_{j}), \\ \notag
    T_{2}^{k}(y_j) &= 4T_{1}^{(k-1)}(y_{j}),\\ \notag
    T_{n}^{k}(y_j) &= 2nT_{n-1}^{(k-1)} (y_j) + \frac{n}{n-1} T_{n-1}^{k}(y_j) \hspace{0.1in} n=3, 4, ...\\ \notag
\end{align}

Mapping from a domain with boundaries $-d, d$ to one with $-1, 1$ is done by using the coordinate transform $\hat{y} = y \cdot d$. If this technique is used, one must include this coordinate transform when computing derivatives: $T_{n}^{k}(\hat{y}_j) = \frac{1}{d^{k}} T_{n}^{k}(y_j)$. Assuming we have a vector $\mathbf{a}$ of coefficients corresponding to an expansion in Chebyshev polynomials, this recurrence can be written in matrix form:

\begin{align}
    \mathbf{a^{(1)}} &= D \mathbf{a}, \\
    D_{ij} &= \begin{cases}
    j & \text{if } \textit{i = 0 and j odd, } \\
    2 j  & \text{if }  \textit{j $>$ i}. \end{cases}
\end{align}

Having obtained the coefficients of the series expansion, we can obtain our solution in `physical' space, by evaluating the resulting polynomial on a grid of x-values. This can be done via multiplication with an evaluation matrix:

\begin{align}
    \mathbf{u}(x) &= \bm E\mathbf{a}, \\
    E_{ij} &= T_{j}(x_{i}).
\end{align}

We notice that we could skip the intermediate step of solving for coefficients, and directly obtain a matrix that differentiates an input, by composing  $\bm D = \bm E \bm H^{-1}$. This matrix was introduced in the previous section. For the eigenvalue problem at hand, we obtain a generalized eigenvalue problem consisting of two matrices, with either four (inviscid) or six (viscous) rows being used to enforce boundary conditions. When solving the discrete eigenvalue problem, the modes associated with the parts of the matrix used for boundary conditions are not of interest. Multiplying the corresponding rows with a large complex number allows us to assign them very large eigenvalues to make them easy to filter out by simply not including them in $\mathcal{S}^{N}$.

\subsubsection{\label{section:compact}Compact finite difference matrices}

A second approach to obtaining high-order derivative approximations is known as compact finite-difference methods. While it is easy to construct finite-difference schemes of arbitrarily high order via Taylor-approximation, they require larger and larger stencils, and this typically leads to problems at the boundaries of the computational domain. While Chebyshev methods can deal with boundary conditions by using the known behavior of the interpolant function at the boundaries, this is not typically possible for a general finite-difference scheme. Compact finite difference schemes \cite{WANG20161843} \cite{LELE199216} strike a good balance: high orders are achievable, with Dirichlet boundary conditions being very easy to implement. A common drawback of compact schemes is the need to invert a matrix to obtain the dense derivative matrix, but in our application, this is not of great concern. Moreover, the matrices are of low condition number, making the inversions feasible. The resulting matrix is also fairly well conditioned, making operations involving inversions of derivative matrices possible and numerically safe. This stands in contrast with Chebyshev matrices, which quickly become very challenging to operate on.

The grid on which we use compact finite difference schemes is evenly spaced:
\begin{align}
    x_{i} = a + ih \hspace{0.3in} 0 < i < M \\
    h = \frac{b-a}{M}   \hspace{0.3in} a \leq x \leq b
\end{align}

The general form for a first-order derivative compact scheme is as follows:

\begin{align}
    \beta u^{\prime}_{i-2} + \alpha u^{\prime}_{i-1} + u^{\prime}_{i} + \alpha u^{\prime}_{i+1} + \beta u^{\prime}_{i+2} = 
    c \frac{u_{i+3} - u_{i-3}}{6h} + b \frac{u_{i+2} - u_{i-2}}{4h} + a \frac{u_{i+1} - u_{i-1}}{2h}
\end{align}

The constants can now be chosen to obtain a scheme of second, fourth or sixth order. A fourth order scheme is obtained of the following form:
\begin{align}
    \frac{1}{4} u^{\prime}_{i-1} + u^{\prime}_{i-1} + \frac{1}{4} u^{\prime}_{i+1} = \frac{3}{4h}(u_{i+1} - u_{i-1})
\end{align}

The general form for a second-order derivative compact scheme is as follows:

\begin{align}
    \beta u^{\prime\prime}_{i-2} + \alpha u^{\prime\prime}_{i-1} + u^{\prime\prime}_{i} + \alpha u^{\prime\prime}_{i+1} + \beta u^{\prime\prime}_{i+2} = 
    c \frac{u_{i+3} - 2u_{i} + u_{i-3}}{9h^{2}} + b \frac{u_{i+2} - 2u_{i} + u_{i-2}}{4h^2} + a \frac{u_{i+1} - 2u_{i} + u_{i-1}}{h^2}
\end{align}

The constants can now be chosen to obtain a scheme of second, fourth or sixth order. A fourth-order scheme is obtained in the following form:
\begin{align}
    \frac{1}{10} u^{\prime\prime}_{i-1} + u^{\prime\prime}_{i-1} + \frac{1}{10} u^{\prime\prime}_{i+1} = \frac{6}{5h^2}(u_{i+1} - 2u_{i} + u_{i-1})
\end{align}

We can now write the approximate second derivative using the following relation:

\begin{align}
    \bm A \mathbf{U}^{\prime\prime} = \frac{12}{h^2}\bm B \mathbf{U} \\
    \mathbf{U}^{\prime\prime} = \bm A^{-1} \frac{12}{h^2} \bm B \mathbf{U}
\end{align}

Boundary conditions can either be incorporated into the first and last row of matrices $\bm A$ and $\bm B$, or we can restrict our computations to the interior nodes of the domain and separately handle boundary conditions for simple cases of zero-valued Dirichlet boundary conditions, similarly to the strategy with the Chebyshev derivative matrices.

\begin{figure}[h]
    \centering
    \includegraphics[width=0.5\textwidth]{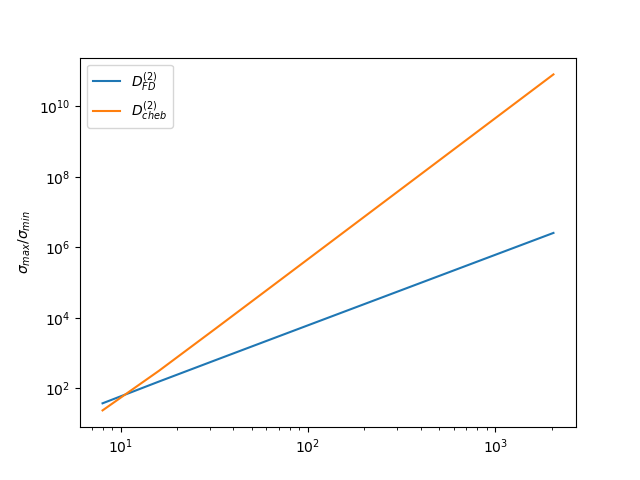}
    \caption{Condition number of second derivative matrix for a matrix derived from Chebyshev polynomials, along with a matrix resulting from a compact finite difference scheme.}
    \label{fig:cond}
\end{figure}

\subsubsection{\label{subsec:Dedalus}Discretizing DAE with spectral methods: The Chebyshev-$\tau$ method}

Let us suppose for now that we have a linear differential equation of the following form:

\begin{align}
    Lu &\equiv \partial_{xx}u - \kappa u = 0 , \hspace{0.2in} x \in (-1, 1) \\
    u(-1) &= u(1) = 0
\end{align}

We want to find the solution as a series of Chebyshev polynomials $T_{k}$:
\begin{align}
    u = \sum^{N+2}_{k=0} a_{k}T_{k}(x)
\end{align}

The Chebyshev polynomials $T_{k}$ are orthogonal with respect to the inner product

\begin{align}
    \langle f, g \rangle = \int^{1}_{-1} \frac{fg}{\sqrt{1-x^{2}}}dx
\end{align}

which is a weighted version of the standard inner product associated with the $L^{2}$ space of functions on $(-1, 1)$. If we set $\langle Lu, T_{i}\rangle = 0$ for $i=0..N$, we get N+1 equations. We have not yet considered the boundary conditions. We slightly modify our original problem to include the following:

\begin{align}
    Lu = \tau_{1}T_{N+1} + \tau_{2}T_{N+2}
\end{align}

Which gives us two additional equations, namely $\langle Lu, T_{N+j}= \tau_{j} \vert\vert T_{N+j} \vert\vert^{2}$ for $j=1, 2$. Only the last two polynomials of our expansion care about this modification, due to the orthogonality of Chebyshev polynomials. Chebyshev polynomials have to following values at the boundaries: $T_{n}(\pm1)=(\pm1)^{n}$, this gives us, together with the boundary conditions, our last two equations:

\begin{align}
    u(-1) &= \sum^{N+2}_{n=0} (-1)^{n}u_{n}, \hspace{0.3in} u(1) = \sum^{N+2}_{n=0} (1)^{n}u_{n}
\end{align}

In total, using this method, we have a system of $N+3$ equations for just as many $u_{k}$ coefficients. The terms $\tau_i$ can be used to accommodate a wide variety of boundary conditions, unfortunately this inevitably lead to rows filled with zeros that produce singular matrices for the generalized eigenvalue problem. This scenario, where the system dynamics are given by $(\bm A, \bm B)$ such that $\partial_{t} \bm B  = \bm A\mathbf{x}$, is known as a matrix pencil. If $\bm B$ is singular, such as the system matrices generated by Dedalus, computing pseudospectra and other quantities that require $\bm B^{-1} \bm A$ becomes significantly more involved than simply inverting the non-singular matrix $\bm B$, obtained from \ref{section:Chebyshev_matrix} or \ref{subsec:Chebyshev_coeff}, which can already be an unwanted source of round-off error. The methods used to handle matrix pencils exist and are outlined in \cite{TrefethenEmbree+2005} and \cite{Dorsselaer_2003}. 

After a brief period of experiments, it was concluded that this method was not ideal for the linear calculations needed for this work. The main hurdles were very large system matrices, due to the first-order reformulation Dedalus uses, combined with the aforementioned singular nature of the matrix, and the numerical difficulty of computing the inner products of very high-order Chebyshev polynomials. All in all, this led us to doubt the potential of Dedalus for the linear calculations, and due to the relative simplicity of the problem, we chose to use our own implementation, using the procedures outlined in Sec. \ref{subsec:Chebyshev_coeff} and \ref{section:Chebyshev_matrix}.

\section{Results}

\subsection{Linearized System}

\subsubsection{Convergence of discretized norms}

\begin{figure}[htbp]
  \centering
  \begin{minipage}[b]{0.49\textwidth}
    \centering
    \includegraphics[width=\textwidth]{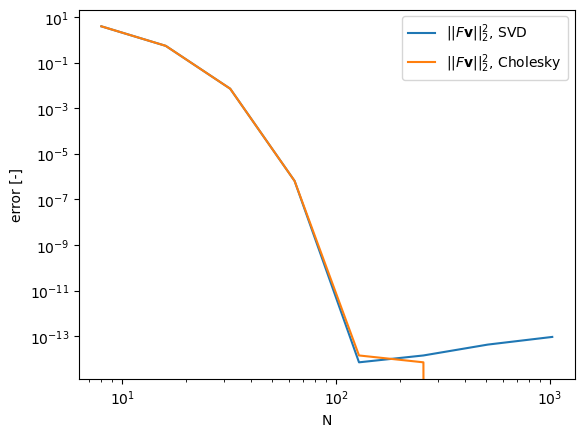}
    \caption{Convergence of $L_2$ quadrature, using Chebyshev expansion coefficients. The errors were computed using matrices obtained by Cholesky factorization and using an SVD.}
    \label{fig:L2_quadrature_cheb}
  \end{minipage}
  \hfill
  \begin{minipage}[b]{0.49\textwidth}
    \centering
    \includegraphics[width=\textwidth]{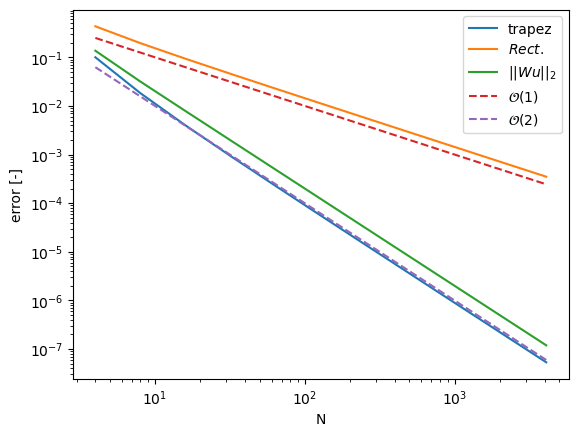}
    \caption{Convergence quadrature using the system formulated in physical space, i.e. Gauss-Chebyshev quadrature. For comparison, the curve labeled `Rect.' corresponds to taking the Euclidean 2 norm as a proxy for the continuous norm, with the expected first-order convergence being observed.}
    \label{fig:L2_quadrature_w}
  \end{minipage}
\end{figure}

We shall first verify that our discrete approximations of the energy and $L_2$ norm indeed converge to the correct values. The coefficient-based computation converges spectrally, as the integration is exact and the error therefore exactly proportional to the interpolation error of the integrand. Figs. \ref{fig:L2_quadrature_cheb} illustrate that the SVD-based factorization seems to introduce a small amount of noise at higher resolutions, while the Cholesky factorization leads to an integration error \textit{identically} zero and is therefore not visible on a logarithmic plot.

The Chebyshev-Gauss quadrature used for the $L_2$ norm is of second order and an order of magnitude more accurate than using a naive integration on an evenly spaced grid, as can be seen in Fig. \ref{fig:L2_quadrature_w}. Higher accuracy for this formulation of the linear system could be achieved using Clenshaw-Curtis quadrature, as mentioned in \cite{TrefethenEmbree+2005}.

\begin{figure}[h]
    \centering
    \includegraphics[width=0.5\textwidth]{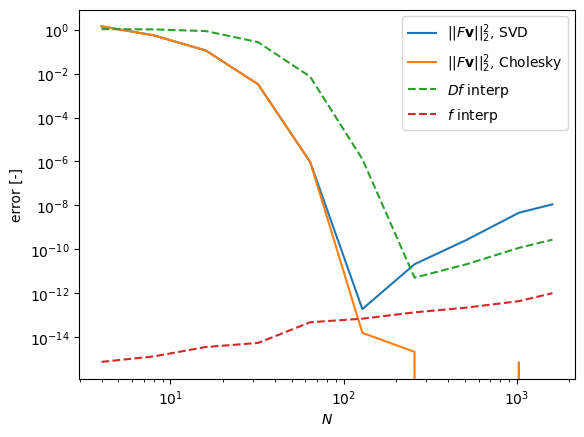}
    \caption{Convergence behaviour of a coefficient based approximation of the energy norm.}
    \label{fig:energy_quadrature}
\end{figure}

Once we look at the energy norm, the influence of round-off errors becomes more severe. While the Cholesky based weighting goes to identically zero error again, we get a significant error in our approximation when using the SVD approach, as seen in Fig. \ref{fig:energy_quadrature}. We can also see that the interpolation error of the derivative of the integrand is not monotonically going to zero, with this again being somewhat expected due to the ill-conditioned nature of Chebyshev polynomials and especially their derivatives, with any error introduced while computing expansion coefficients being amplified by the derivative. Since our norm approximation error should be proportional to the approximation error of the derivative, it is surprising that the norm error, using the Cholesky factorization, converges to be identically zero. A natural question to ask is why we would use an SVD-based factorization if Cholesky seemingly produces better results. The answer is very pragmatic: The Cholesky algorithm will fail as soon as the matrix to be factored is not numerically semi-positive definite (SPD), i.e., as soon as it has a negative eigenvalue. The matrices we want to factor are often very close to not being SPD for larger values of $N$, i.e., some of their eigenvalues are sufficiently small to be wrongfully computed as $\sigma_{wrong} = -\mathcal{O}(\epsilon_{machine})$. This will make a Cholesky factorization impossible using most standard computation routines. This type of ill-conditioning prevented us from using a Cholesky factorization when factoring $\mathbf{\tilde{v}}^H Q \mathbf{\tilde{v}}$

\subsubsection{Spectrum and Pseudospectrum of linearized MHD operator}

The spectrum and pseudospectrum of the operator described by Eq. \ref{eq:reduced_MHD_2d} was computed using a matrix based discretization at fixed wavenumbers $k_{x}$ with the Chebyshev differentiation matrices described in Sec. \ref{section:Chebyshev_matrix}, and with a coefficient based approach, described in Sec. \ref{subsec:Chebyshev_coeff}. The computations were carried out in the range of $k_{x} \in [0.2, 1.5]$ as this is a range of wavenumbers that should cover all possible regimes of stability. The Lundquist number was varied around a range of $S \in [10^{3}, 10^{5}]$, with higher values being increasingly inaccessible due to numerical limitations; with increasing $S$, the resistive layer shrinks in size, requiring higher wavenumbers to be resolved by the discretization scheme, in turn leading to worse-conditioned matrices.

\begin{figure}[htbp]
  \centering
  \begin{minipage}[b]{0.49\textwidth}
    \centering
    \includegraphics[width=\textwidth]{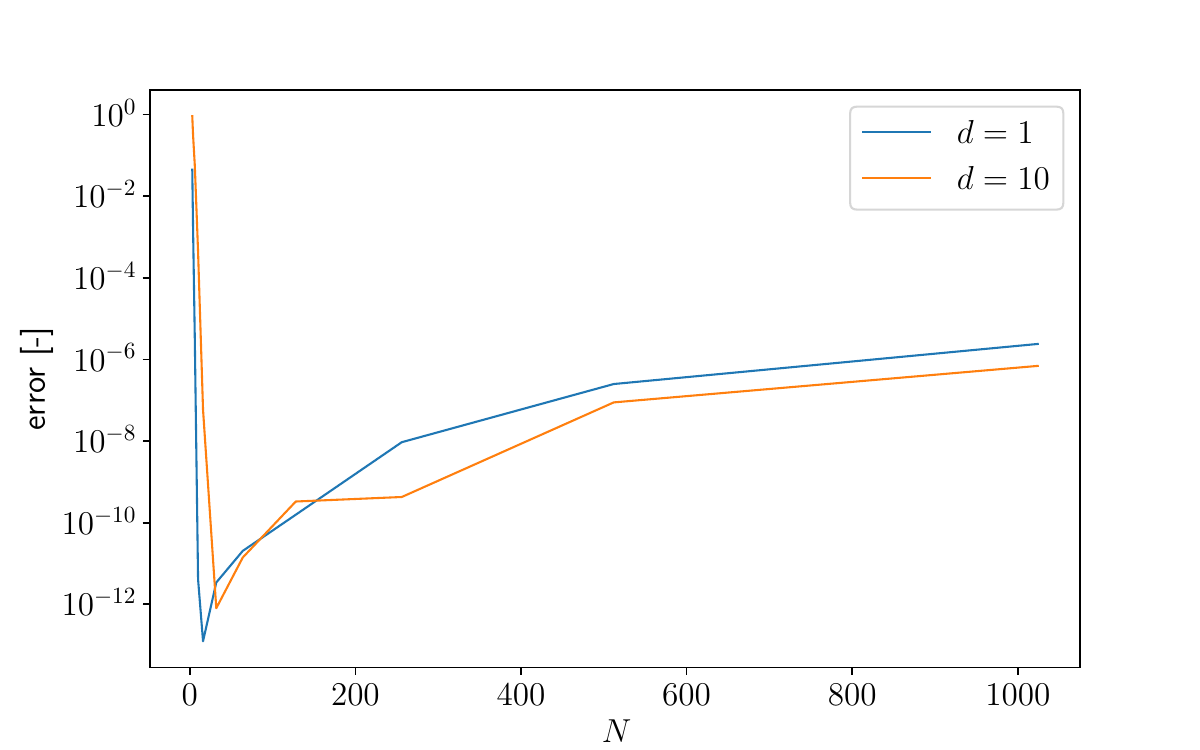}
    \caption{Convergence of fourth derivative spectral operator at different resolutions, for a very simple toy problem. A matrix inversion was used to solve the discretized BVP.}
    \label{fig:d4_bench}
  \end{minipage}
  \hfill
  \begin{minipage}[b]{0.49\textwidth}
    \centering
    \includegraphics[width=\textwidth]{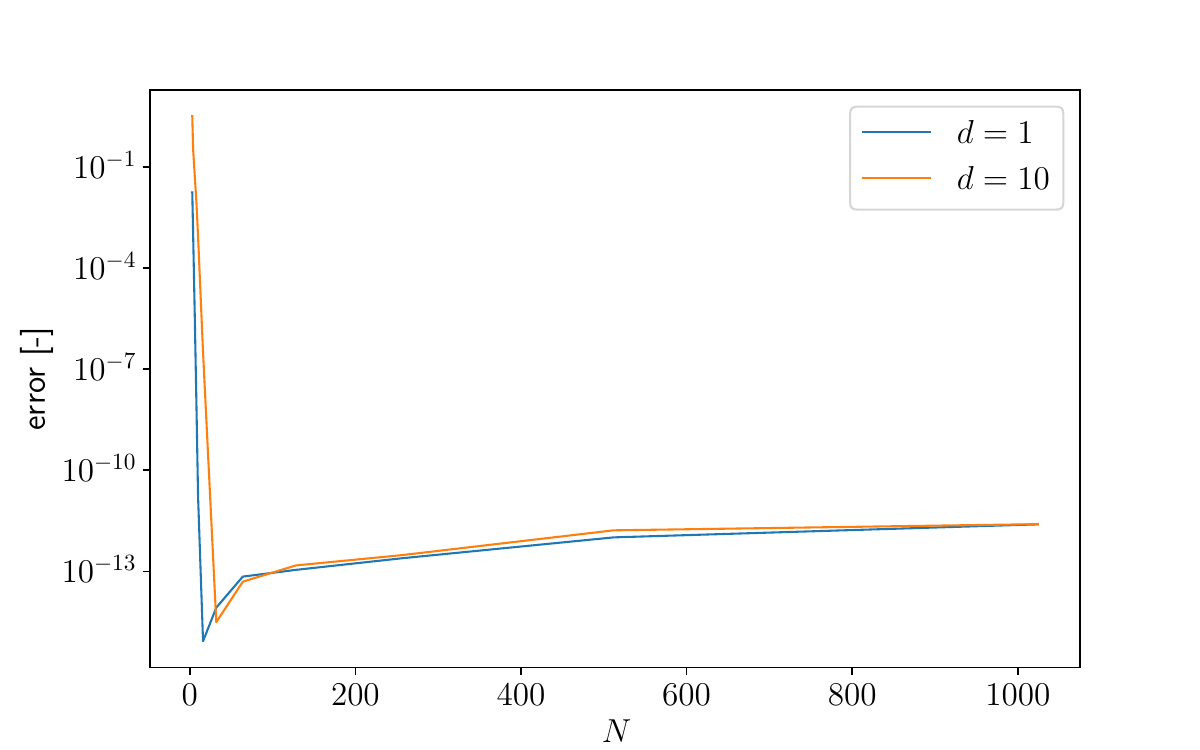}
    \caption{Convergence of second derivative spectral operator at different resolutions, for a very simple toy problem. A direct matrix inversion was used to solve the BVP.}
    \label{fig:d2_bench}
  \end{minipage}
\end{figure}

On the numerical stability and conditioning of Chebyshev matrices, the following arguments must be considered:. Fig. \ref{fig:d2_bench} shows the behavior of the second-derivative operator when used to solve a simple, second-order boundary value problem, which is solved by solving the discrete equation $D_{2}\mathbf{u} = \mathbf{f}$. Even at very high grid point numbers, the operator inversion is successful, although the condition number is in a range where operations such as inversions are considered numerically unsafe. The same cannot be said for the fourth-derivative operator, which is demonstrated for a similar problem in Fig. \ref{fig:d4_bench}. It is known that the condition numbers of the fourth-derivative matrix scale very badly with higher resolutions, with values larger than $10^{16}$ being obtained very quickly \cite{boyd2013chebyshev}. Inverting such a matrix is likely impossible numerically, especially when using standard data types, and even solving a linear system involving such a matrix can be nearly impossible \cite{trefethen1997numerical}. Unfortunately, for our application, we must obtain explicit inverse matrices. This limits us either to the inviscid (relatively well-conditioned due to lower order) case or parameter regimes that do not require high resolutions (low $S$ and $k_{x}$). The inviscid version of Eq. \ref{eq:reduced_MHD_2d} includes only second-order derivatives.

\begin{figure}[htbp]
  \centering
  \begin{minipage}[b]{0.49\textwidth}
    \centering
    \includegraphics[width=\textwidth]{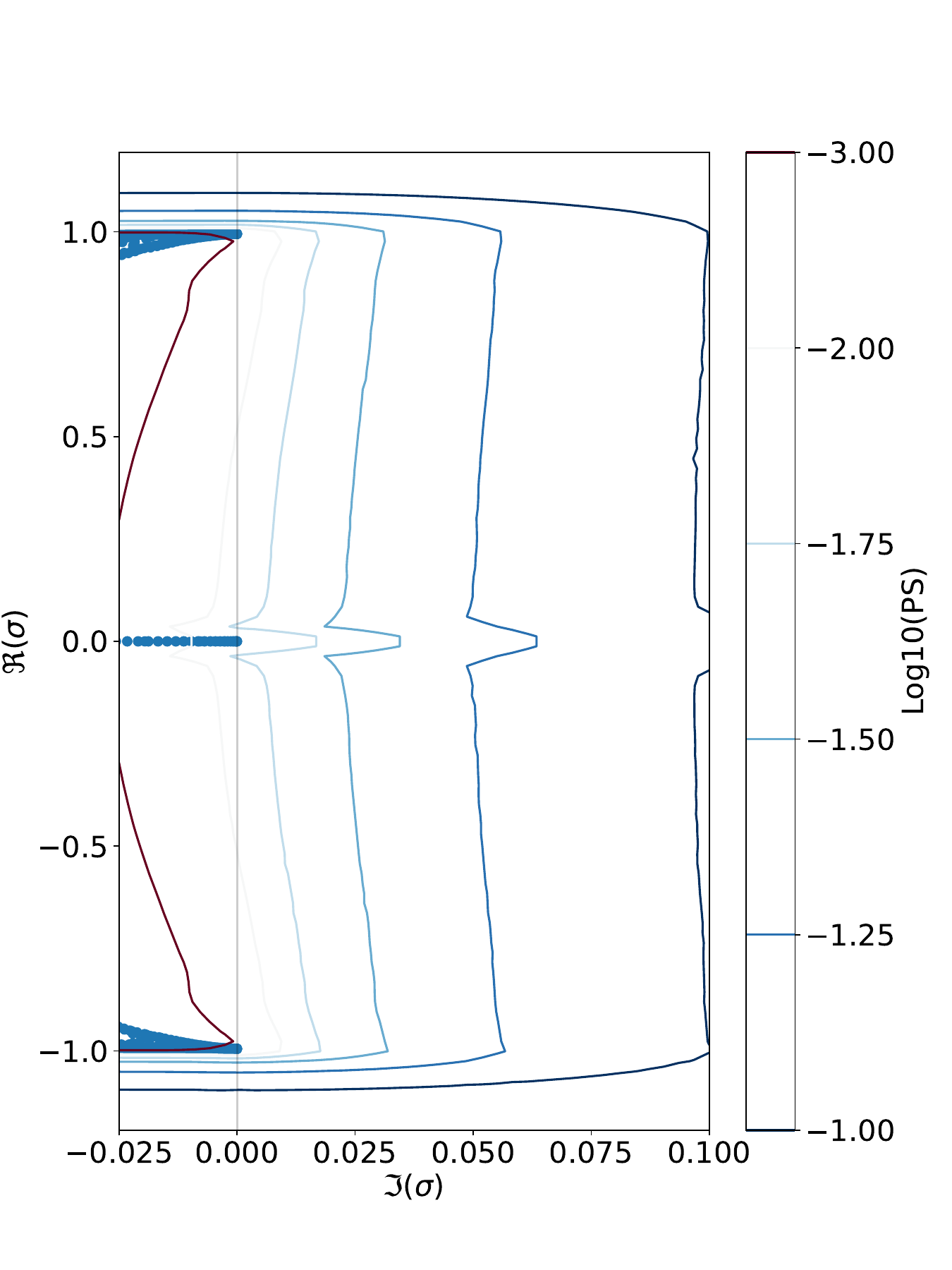}
    \caption{Section of the pseudospectrum, around the area of marginal stability. A tolence of $\epsilon = 10^{-6}$ was used for the pseudoinverse.}
    \label{fig:ps_zoom_1}
  \end{minipage}
  \hfill
  \begin{minipage}[b]{0.49\textwidth}
    \centering
    \includegraphics[width=\textwidth]{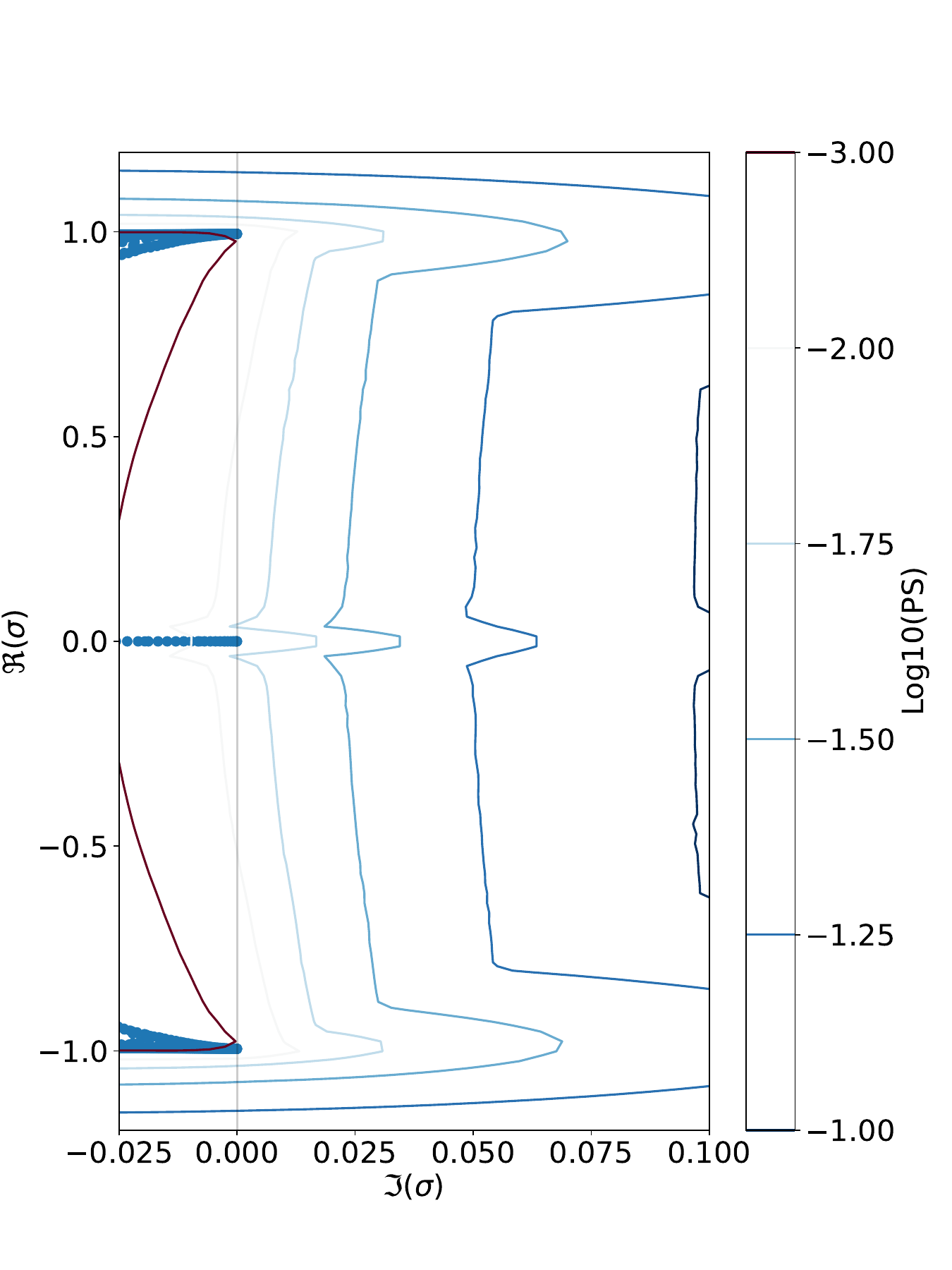}
    \caption{Section of the pseudospectrum, around the area of marginal stability. No truncation was used for the pseudoinverse.}
    \label{fig:ps_zoom_+}
  \end{minipage}
\end{figure}

We shall also consider an appropriate choice for the tolerance of the pseudoinverse, when factoring the inner product matrix for the coefficient-based approach. For certain parameter regimes, most notably for high values of $k_x \approx 1$ and $S>10^4$ the pseudospectra showed sensitivity to the choice of this parameter. An example can be seen in Figs. \ref{fig:ps_zoom_1} \ref{fig:ps_zoom_+}. The $\epsilon$-pseudospectra close to $\Re(\sigma) \pm k_x$ are sensitive with respect to the choice of the pseudoinverse cutoff. In Sec. \ref{sec:numerical_instability}, we will examine how this sensitivity manifests itself during the computation of pseudomodes and upper bounds on transient growth.

\begin{figure}[htbp]
    \centering
    \begin{minipage}{0.5\textwidth}
        \centering
        \includegraphics[width=\linewidth]{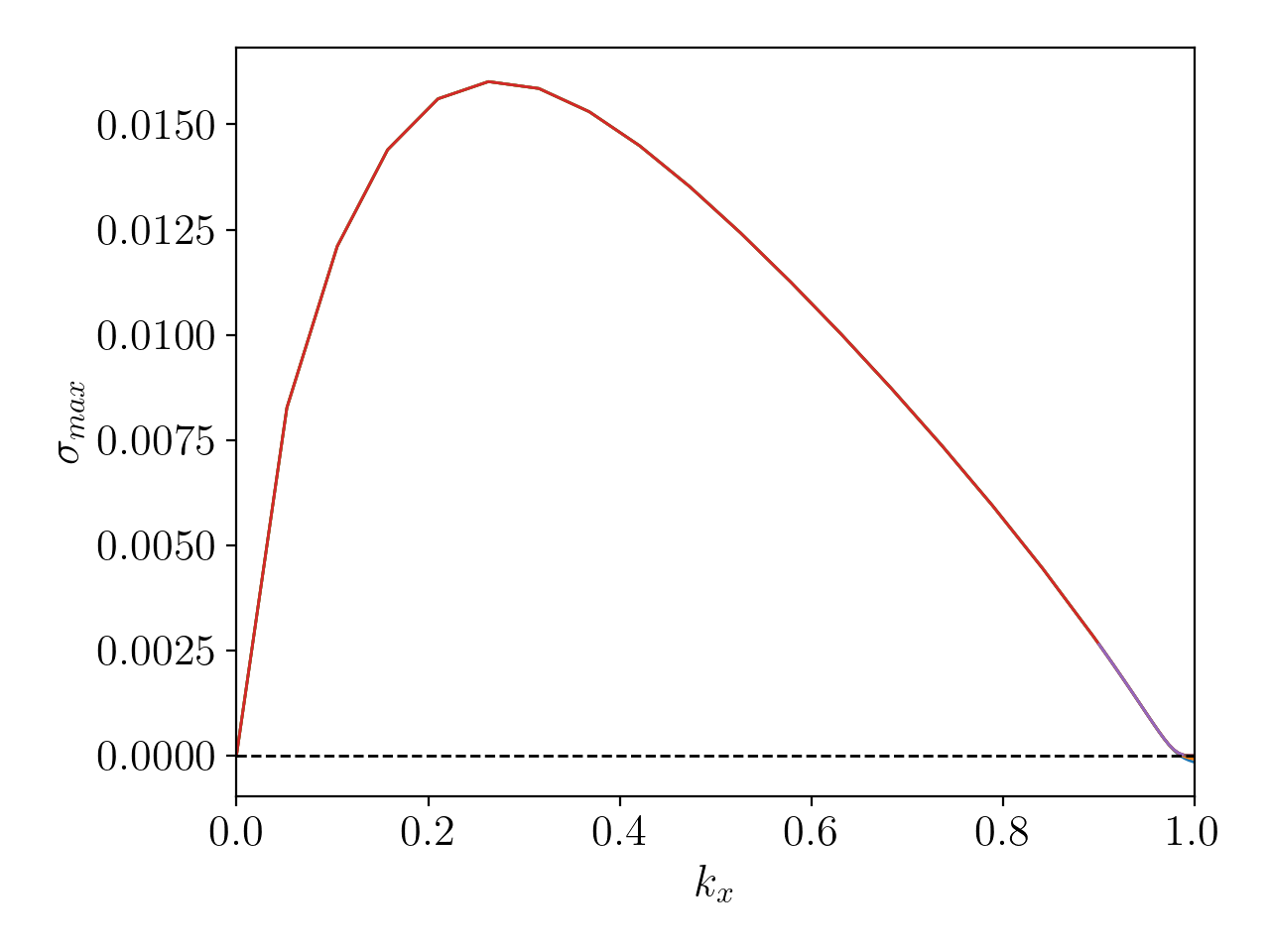}
        
    \end{minipage}\hfill
    \begin{minipage}{0.5\textwidth}
        \centering
        \includegraphics[width=\linewidth]{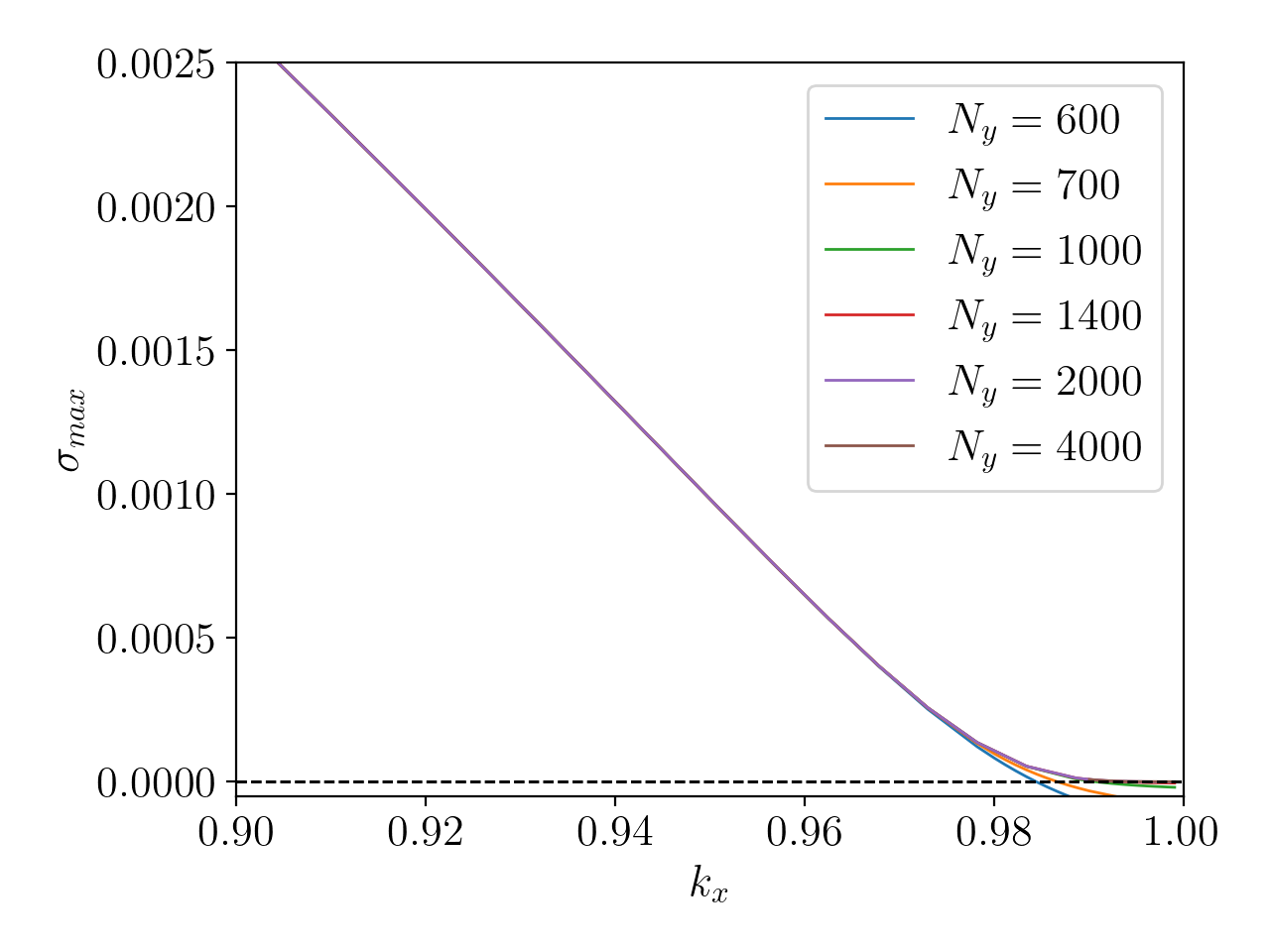}

    \end{minipage}

    \begin{minipage}{0.5\textwidth}
        \centering
        \includegraphics[width=\linewidth]{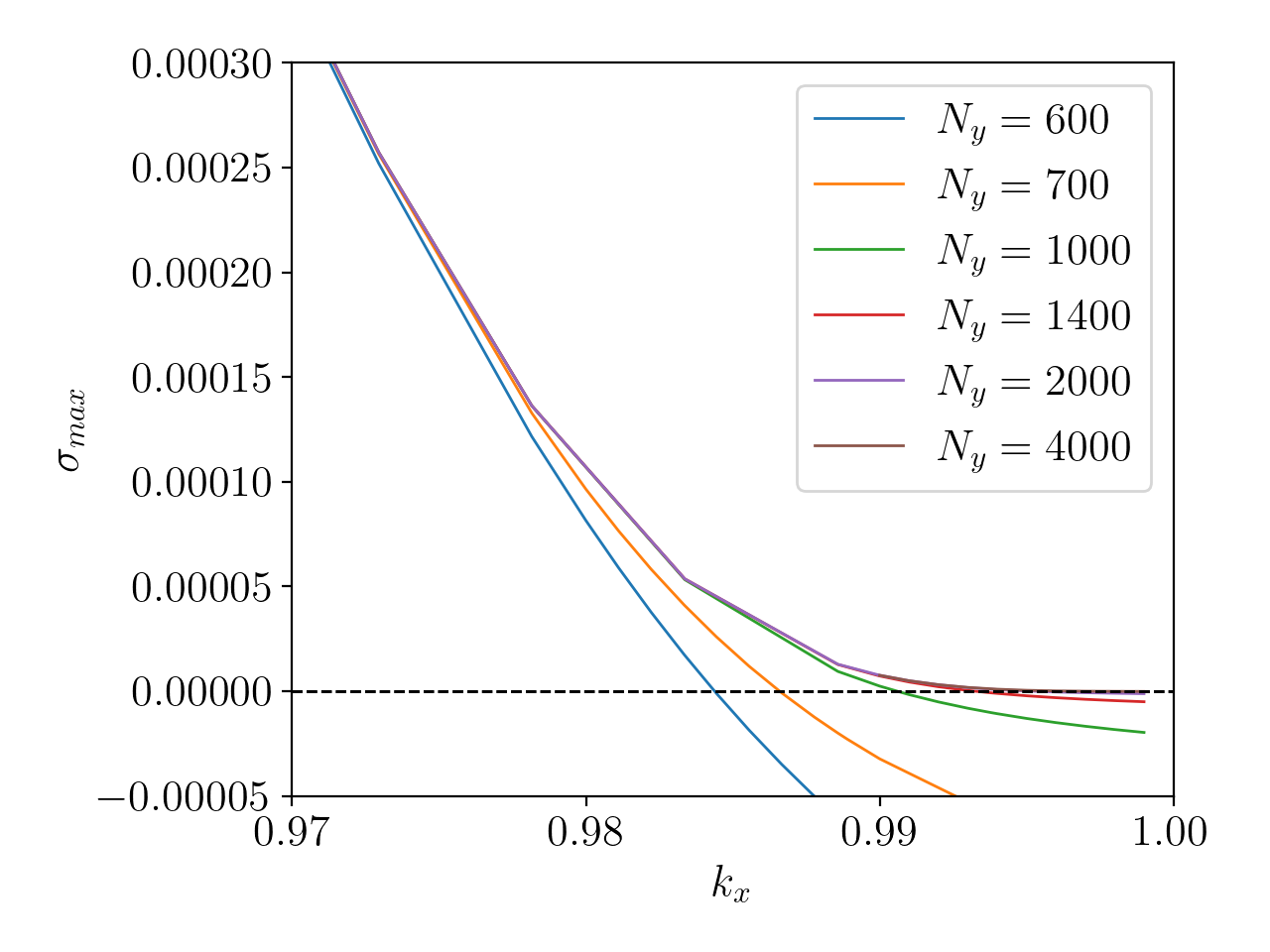}
        
    \end{minipage}\hfill
    \begin{minipage}{0.5\textwidth}
        \centering
        \includegraphics[width=\linewidth]{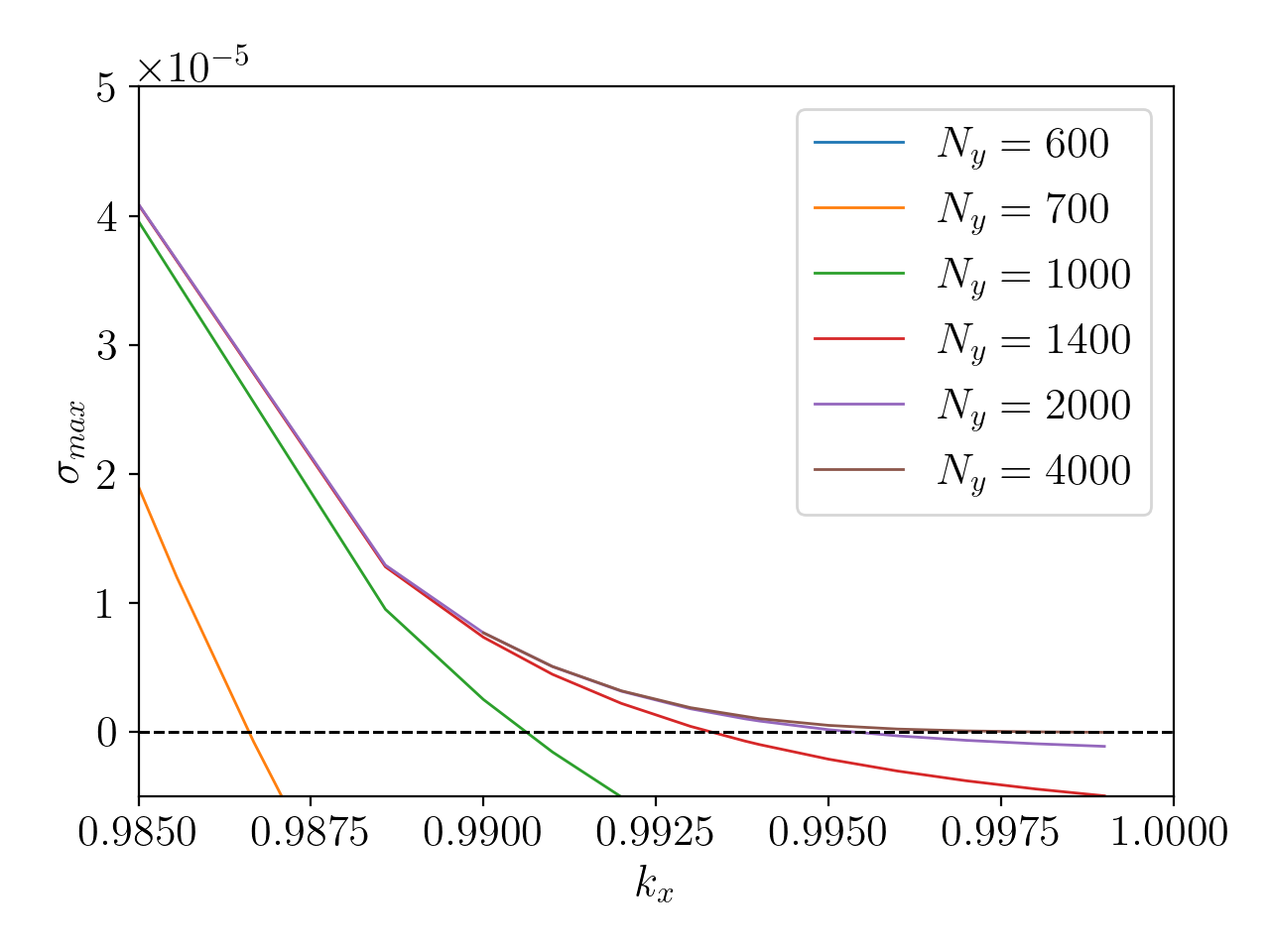}

    \end{minipage}
    \caption{Dispersion relation for linearized system at $S=10^3$. As $k_x \rightarrow 1$, higher resolutions are needed to obtain nonzero growth rates. Underresolving the system leads to marginal stability ($\sigma = 0$) at a slightly lower $k_x$ than would be predicted by the theory, although convergence to the theoretical value $k_x=1$ is observed.}
    \label{fig:dispersion}
\end{figure}

Solving the generalized eigenvalue problem, we obtain the shape of the asymptotically most unstable mode, often referred to as the tearing mode, along with its eigenvalue, the modal growth rate of the system. For the modal growth rate, extensive research has been conducted \cite{battacharjee1995} \cite{MacTaggart2020}, with dispersion relations being computed and verified numerically. Fig.\ref{fig:dispersion} shows a recreation of such a dispersion relation at $S=10^3$, with similar parameters used as in \cite{MacTaggart2020}. We can see that the tearing mode vanishes as $k_{x} \rightarrow 0$, since this physically corresponds to approaching the limit state of an infinitely long perturbation. As $k_{x} \rightarrow 1$, the growth rate vanishes as well, with the perturbation wavelength being too short to trigger the instability. The limit corresponding to $k_x=0$ is in practice impossible to verify with a numerical solver, as it corresponds to an infinitely large computational domain. Computing the fastest growth rate and the shape of the tearing mode becomes increasingly difficult at the limit of $k_x \rightarrow 1$, with the tearing mode becoming more and more localized, requiring higher grid resolutions. 

\begin{figure}[htbp]
  \centering
  \begin{minipage}[b]{0.49\textwidth}
    \centering
    \includegraphics[width=\textwidth]{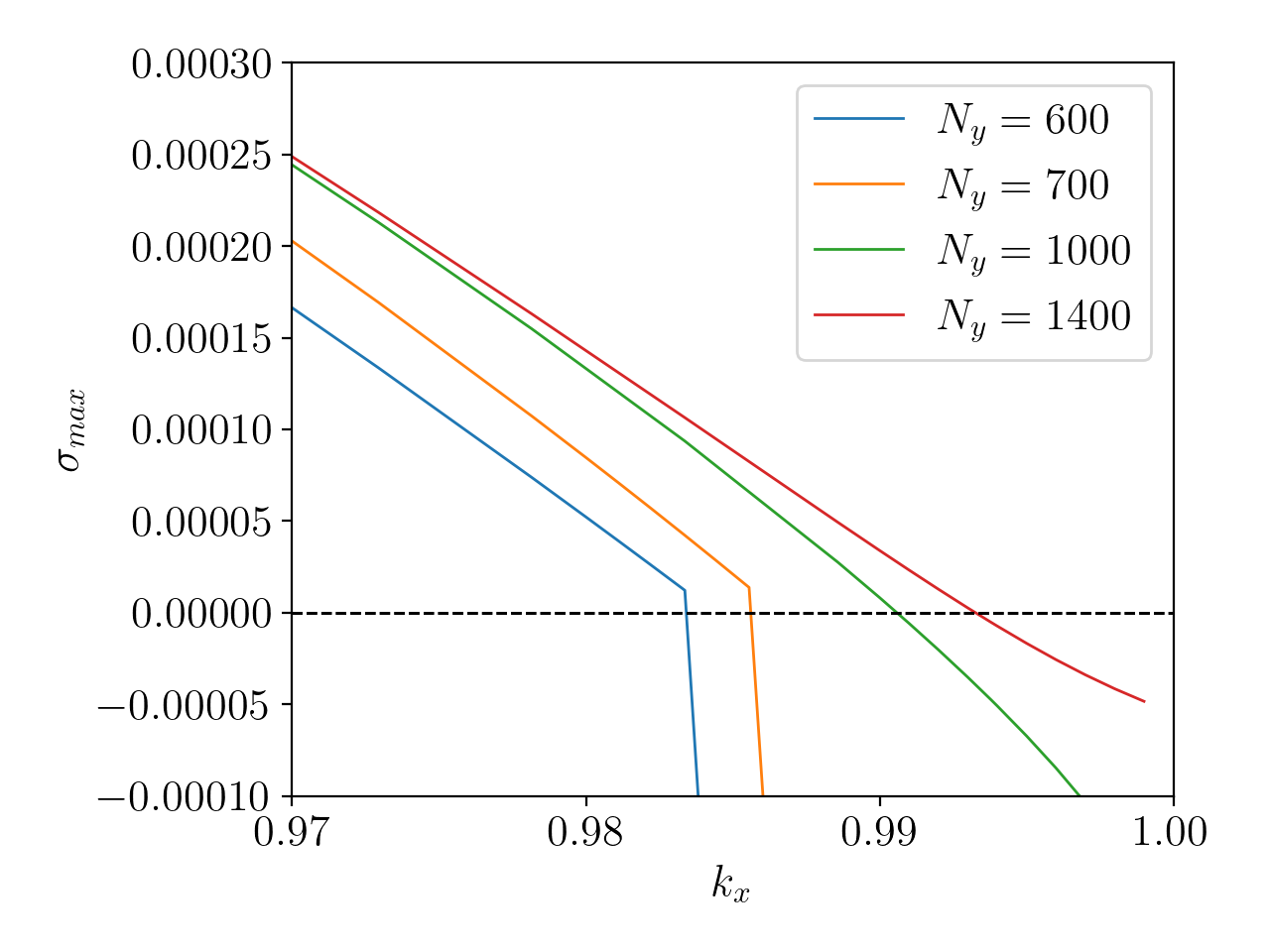}
    \label{fig:disp_s1e4}
  \end{minipage}
  \hfill
  \begin{minipage}[b]{0.49\textwidth}
    \centering
    \includegraphics[width=\textwidth]{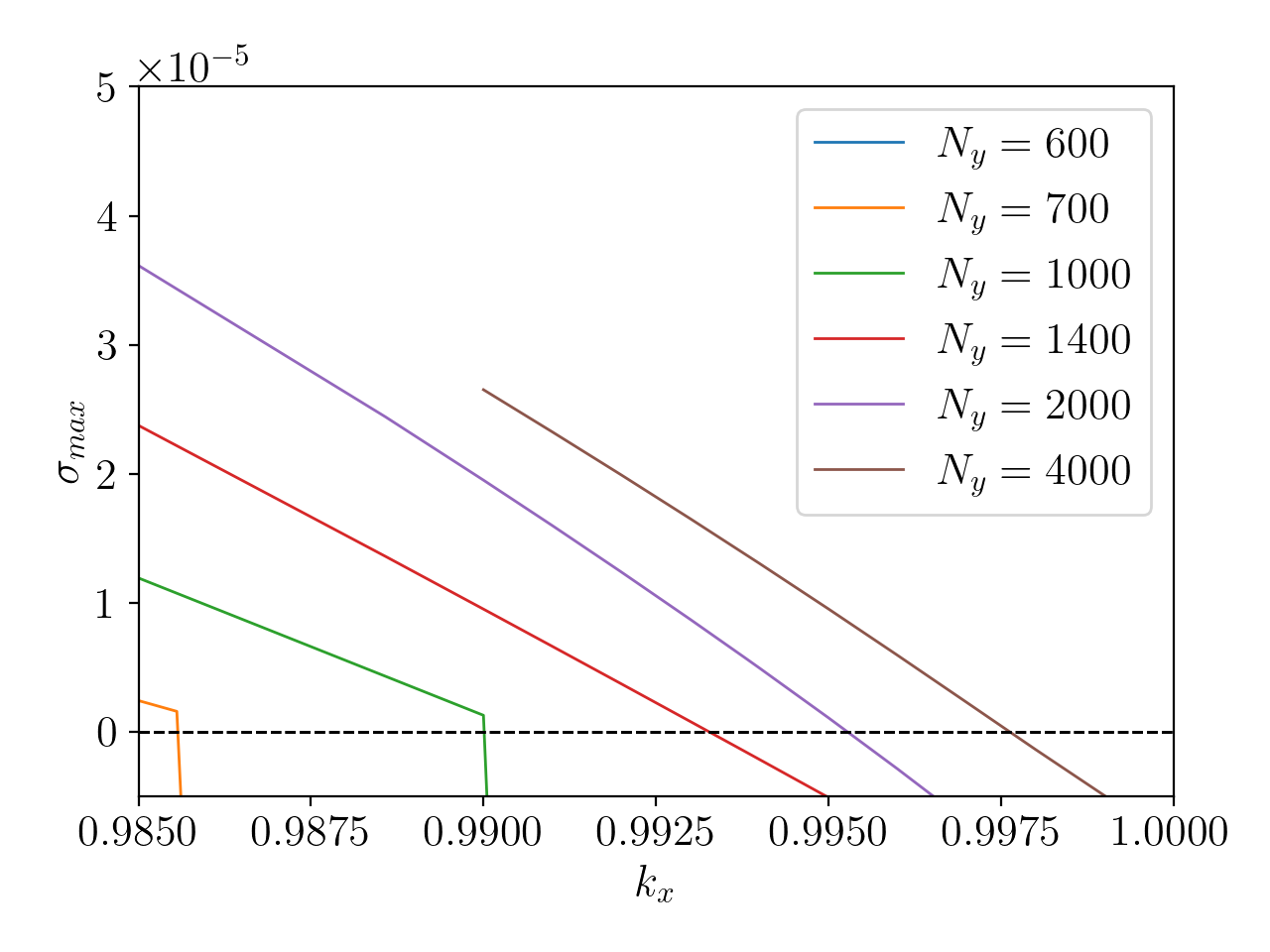}
    \label{fig:disp_s1e5}
  \end{minipage}
  \caption{Dispersion relations for $S=10^4$ (left) and $S=10^5$ (right). Only the most interesting part of the dispersion relation, around marginal stability, was included. A steep drop-off for the dispersion curve indicates that the eigenfunction corresponding to the tearing mode was not in the set of the ten least damped eigenfunctions. Otherwise, the growth rates smaller than zero indicate the decay rate associated with the eigenfunction corresponding to the tearing mode.}
  \label{fig:dispersion_2}
\end{figure}

From the dispersion relations in Figs. \ref{fig:dispersion} \ref{fig:dispersion_2} we can see that the exact wavenumber $k_x$ corresponding to marginal stability varies slightly with the resolution of the discretized system. At marginal stability, the tearing mode gets replaced by other eigenfunctions with eigenmodes very close to zero. We performed a search over the first ten eigenfunctions and used the decay rate associated with the tearing mode for stable wavenumbers. For some resolutions, the tearing mode was not found in those ten eigenfunctions, and therefore no smooth dispersion relation could be found.

\begin{figure}[h]
    \centering
    \includegraphics[width=0.99\textwidth]{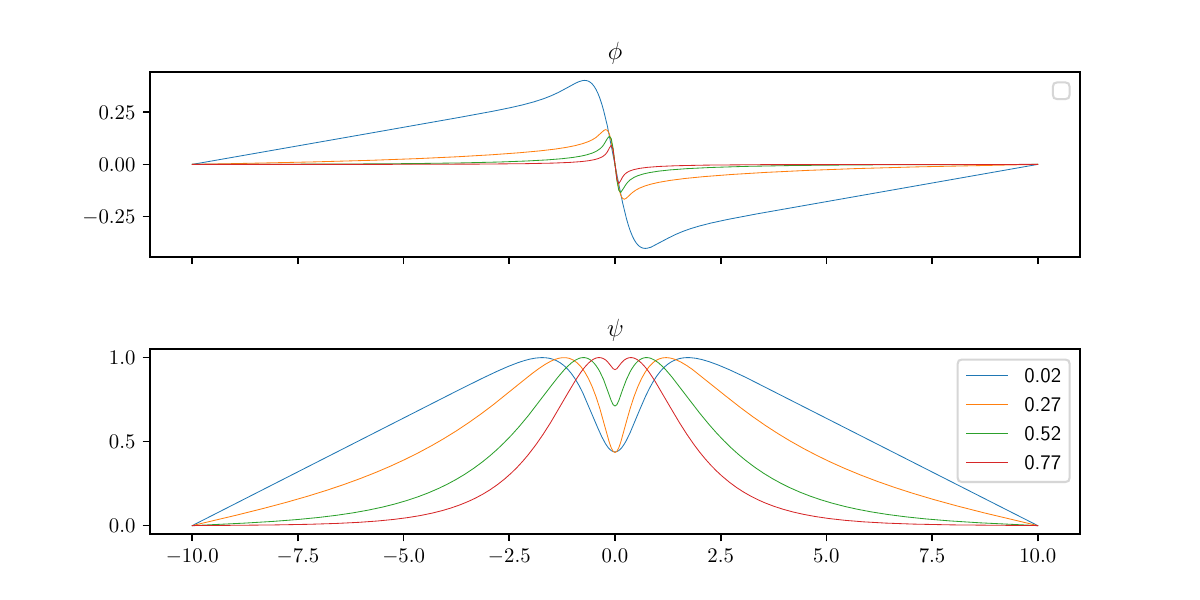}
    \caption{Most unstable eigenmodes of the stream function $\phi$ describing the velocity field and the vector potential $\psi$ describing the  magnetic field. The periodic wavenumber $k_{x}$ was varied to demonstrate the narrowing of the tearing mode with increasing perturbation wavenumber. The plot was produced using a Lundquist number of $S=10^3 $. These are the spatial structures that are responsible for the modal growth of the tearing instability.}
    \label{fig:eigenvectors}
\end{figure}

The eigenvectors of the linear MHD system are plotted in Fig. \ref{fig:eigenvectors}, for increasing periodic wavenumbers $k_{x}$. It shall be noted that both components of the vectors were normalized with respect to the maximum value of the magnetic vectorpotential $\psi$. With increasing perturbation wavenumber, the disipative layer becomes thinner. This effect is especially pronounced in the velocity field, with the area of large velocities being very small very quickly. The produced eigenvectors show similar structure to results in the literature, such as \cite{battacharjee1995}, \cite{furth_1963}, \cite{Zanna_2016} and \cite{secondary_instability_dahlburg}.

As the Lundquist number $S$ increases, we expect the thickness of the resistive layer to decrease along with the growth rate.

\begin{figure}[h]
    \centering
    \includegraphics[width=0.55\textwidth]{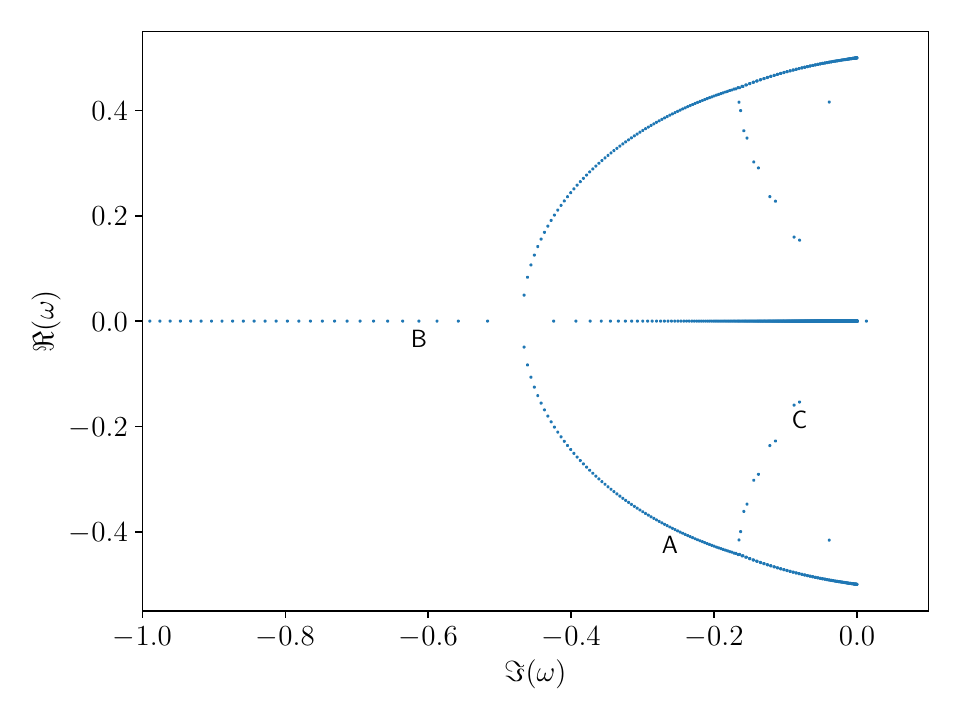}
    \caption{Spectrum of the linearized MHD operator. The labels correspond to different branches of the spectrum. The only unstable eigenvalue corresponds to the tearing mode, which is purely real.}
    \label{fig:sepctrum_labeled}
\end{figure}

An example of the spectrum, for an unstable $k_x$, can be seen in \ref{fig:sepctrum_labeled}. The purely real eigenvalue in the unstable part of the complex plane is the tearing mode. The spectrum has two distinct branches with eigenvalues with a large imaginary part that are connected to the central, real branch by two sparsely populated branches. The structure stays similar at all Lundquist numbers used in this thesis, with the main difference being the connecting branches approaching the $\Re(\sigma) = 0$. Varying $k_{x}$ manifests itself by changing the intersection points of branches with the purely real line. This crossing point will also be interesting when we compute pseudo modes. The connection point of branches B and A is situated at $\Im(\sigma) = \Re(\sigma) = k_x$.

\begin{figure}[htbp]
  \centering
  \begin{minipage}[b]{0.49\textwidth}
    \centering
    \includegraphics[width=\textwidth]{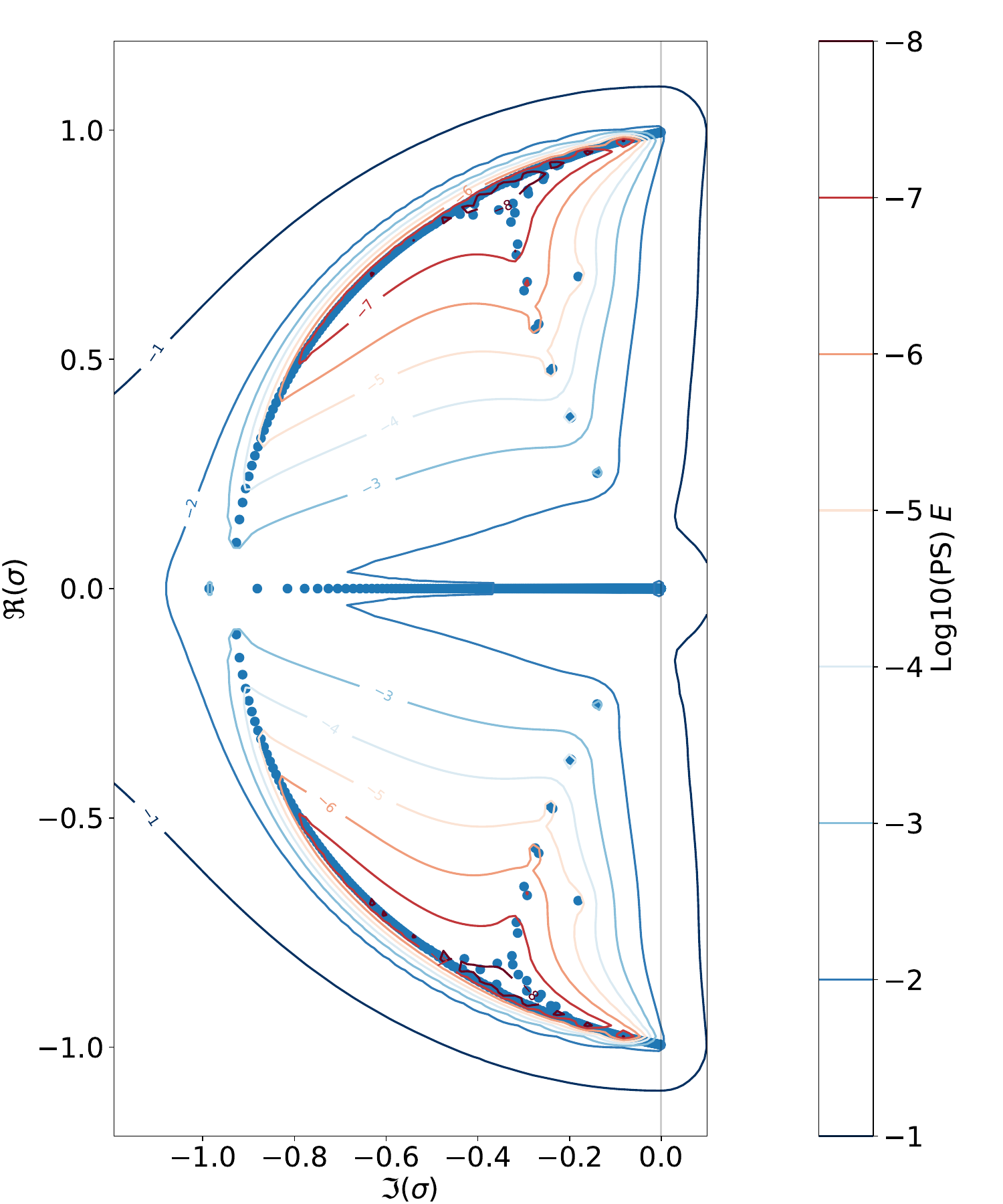}
  \end{minipage}
  \hfill
  \begin{minipage}[b]{0.49\textwidth}
    \centering
    \includegraphics[width=\textwidth]{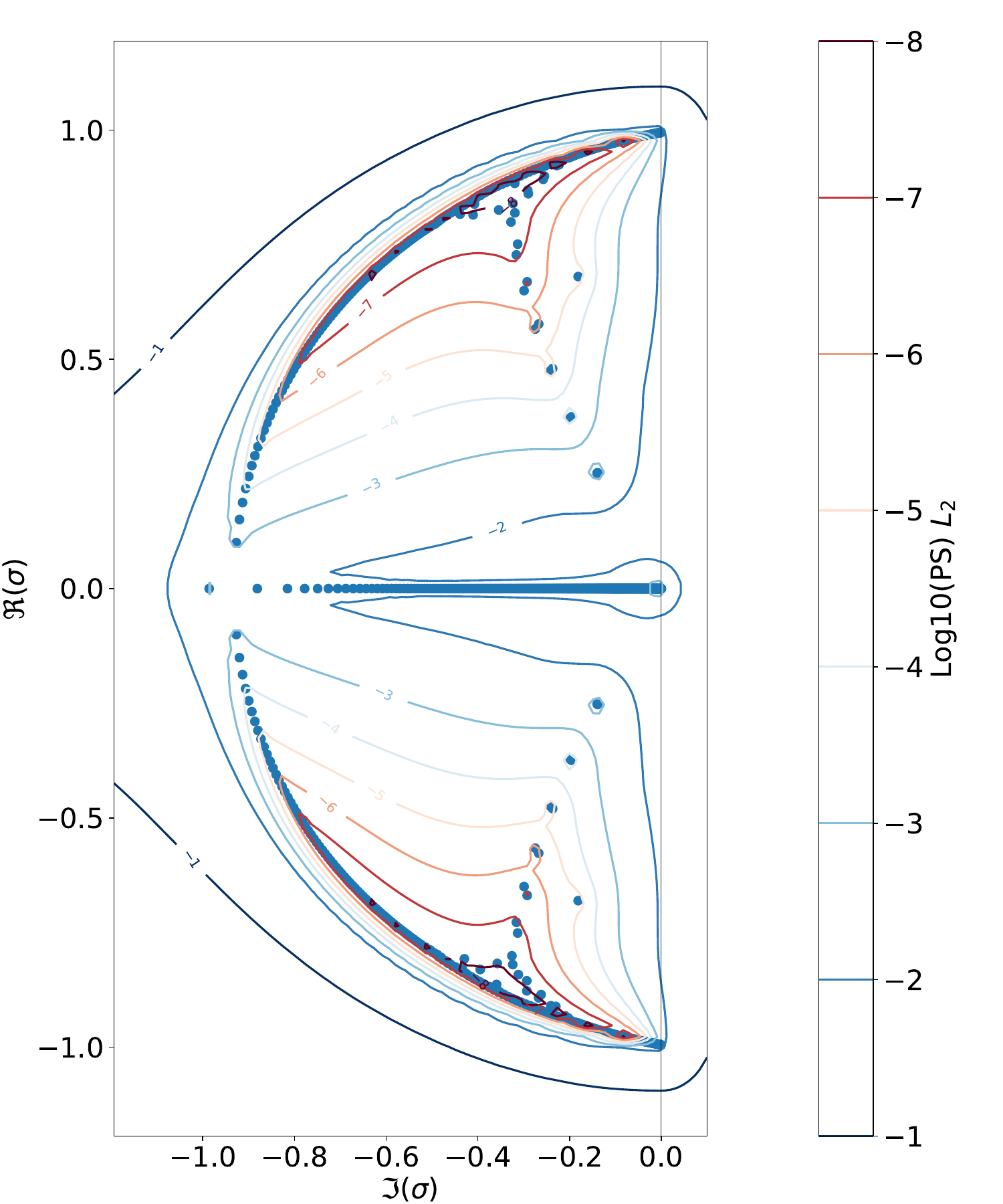}
  \end{minipage}
  \caption{Pseudospectra at $S=10^3$ and $k_x = 0.995$, which means there is no unstable tearing mode, as this wavenumber is slightly past the numerical bound for marginal stability. A grid resolution of $N=1024$ was used for both computations.}
  \label{fig:PS_s1e3_1024}
\end{figure}

\begin{figure}[htbp]
  \centering
  \begin{minipage}[b]{0.49\textwidth}
    \centering
    \includegraphics[width=\textwidth]{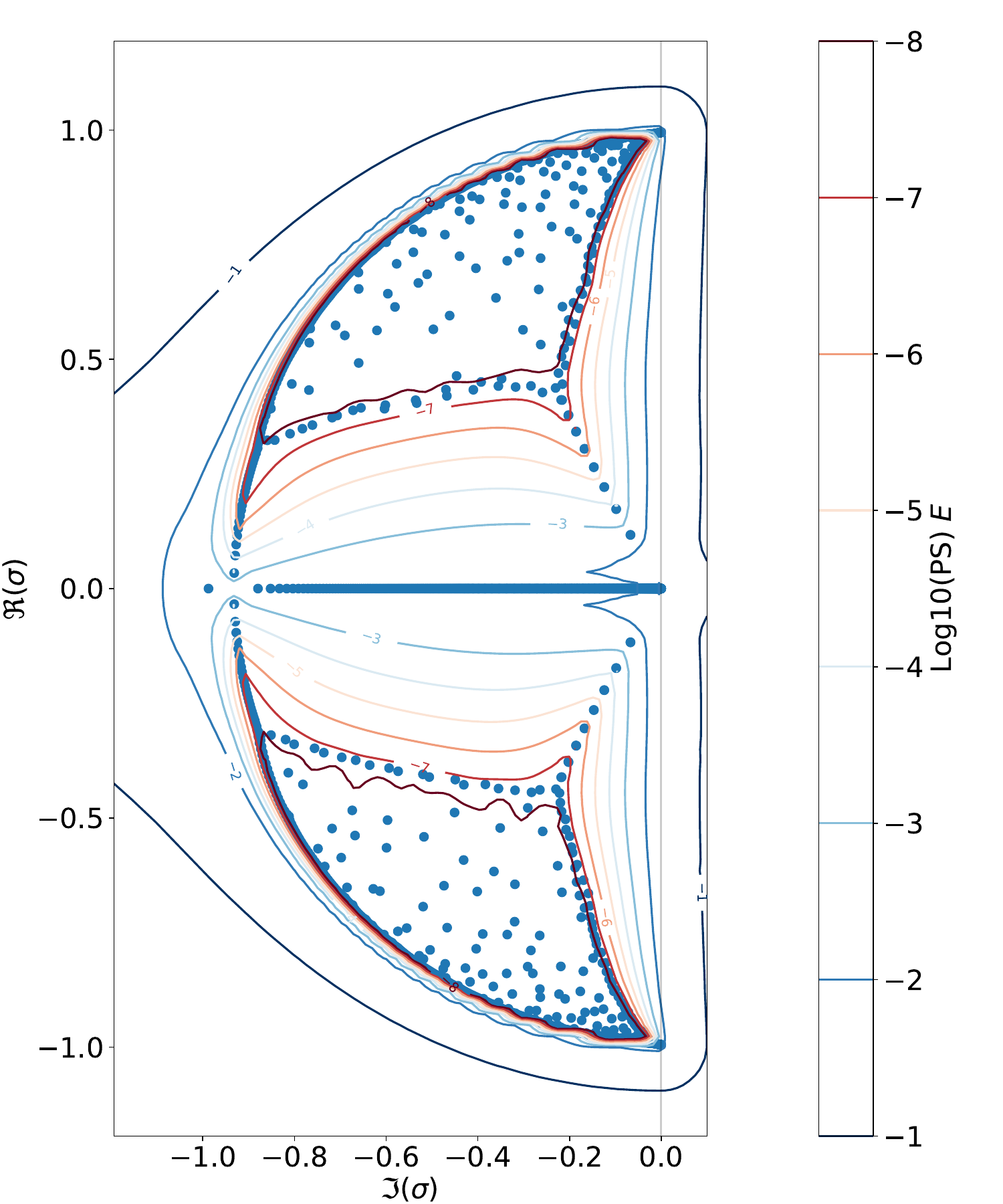}
  \end{minipage}
  \hfill
  \begin{minipage}[b]{0.49\textwidth}
    \centering
    \includegraphics[width=\textwidth]{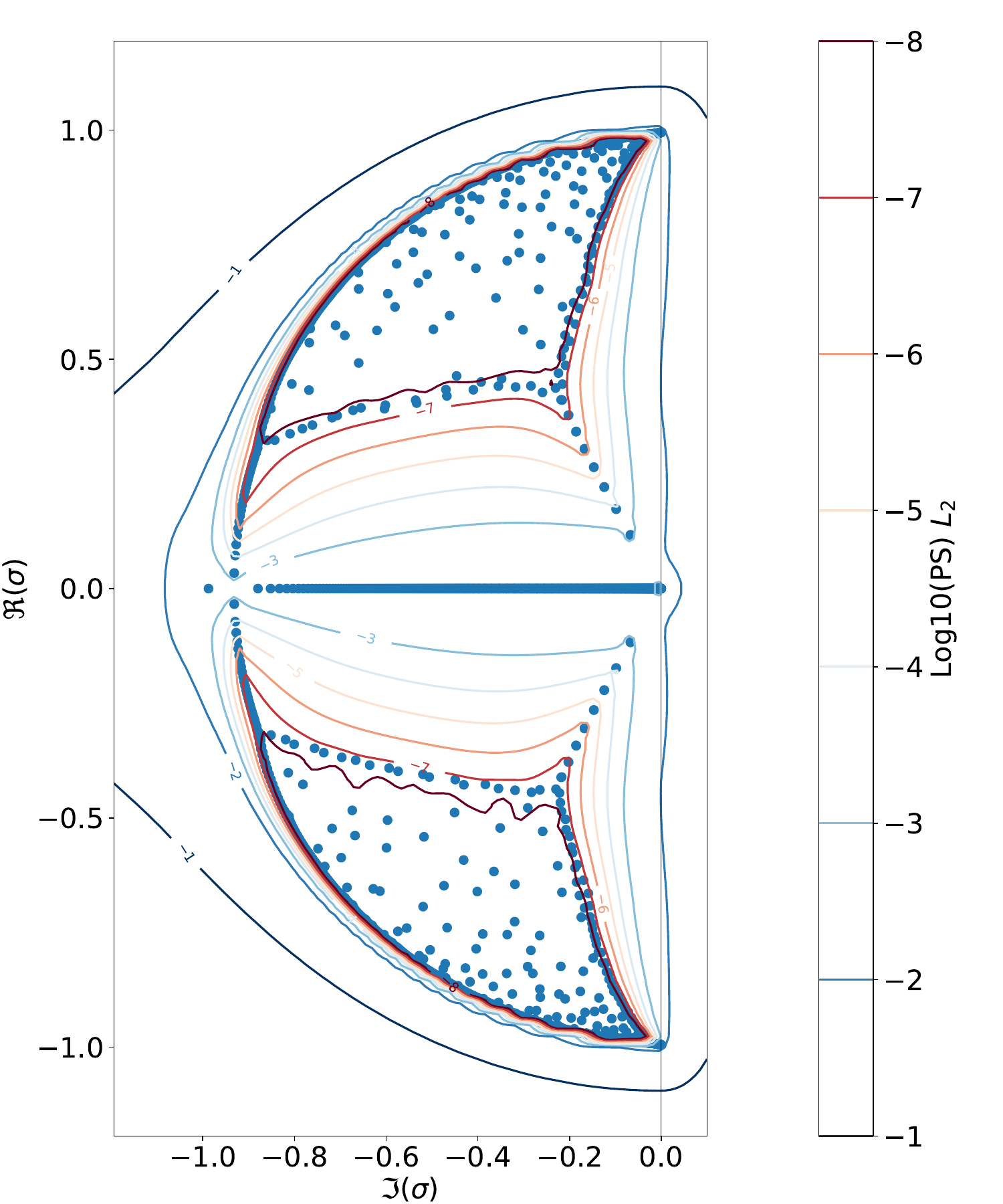}
  \end{minipage}
  \caption{Pseudospectra at $S=10^4$ and $k_x = 0.995$, which means there is no unstable tearing mode, as this wavenumber is slightly past the numerical bound for marginal stability. A grid resolution of $N=1600$ was used for both computations.}
  \label{fig:PS_s1e4_1600}
\end{figure}

\begin{figure}[htbp]
  \centering
  \begin{minipage}[b]{0.49\textwidth}
    \centering
    \includegraphics[width=\textwidth]{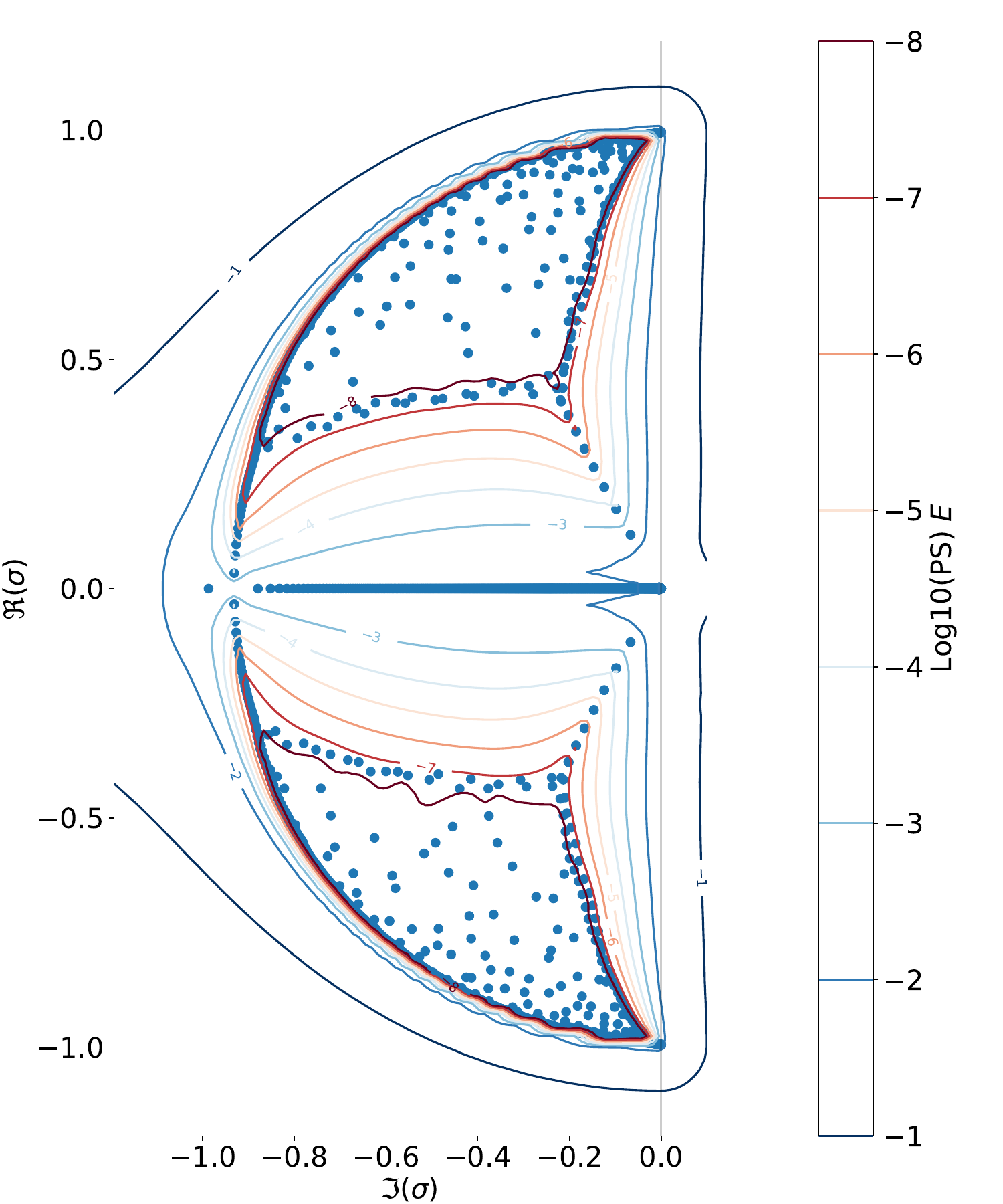}
  \end{minipage}
  \hfill
  \begin{minipage}[b]{0.49\textwidth}
    \centering
    \includegraphics[width=\textwidth]{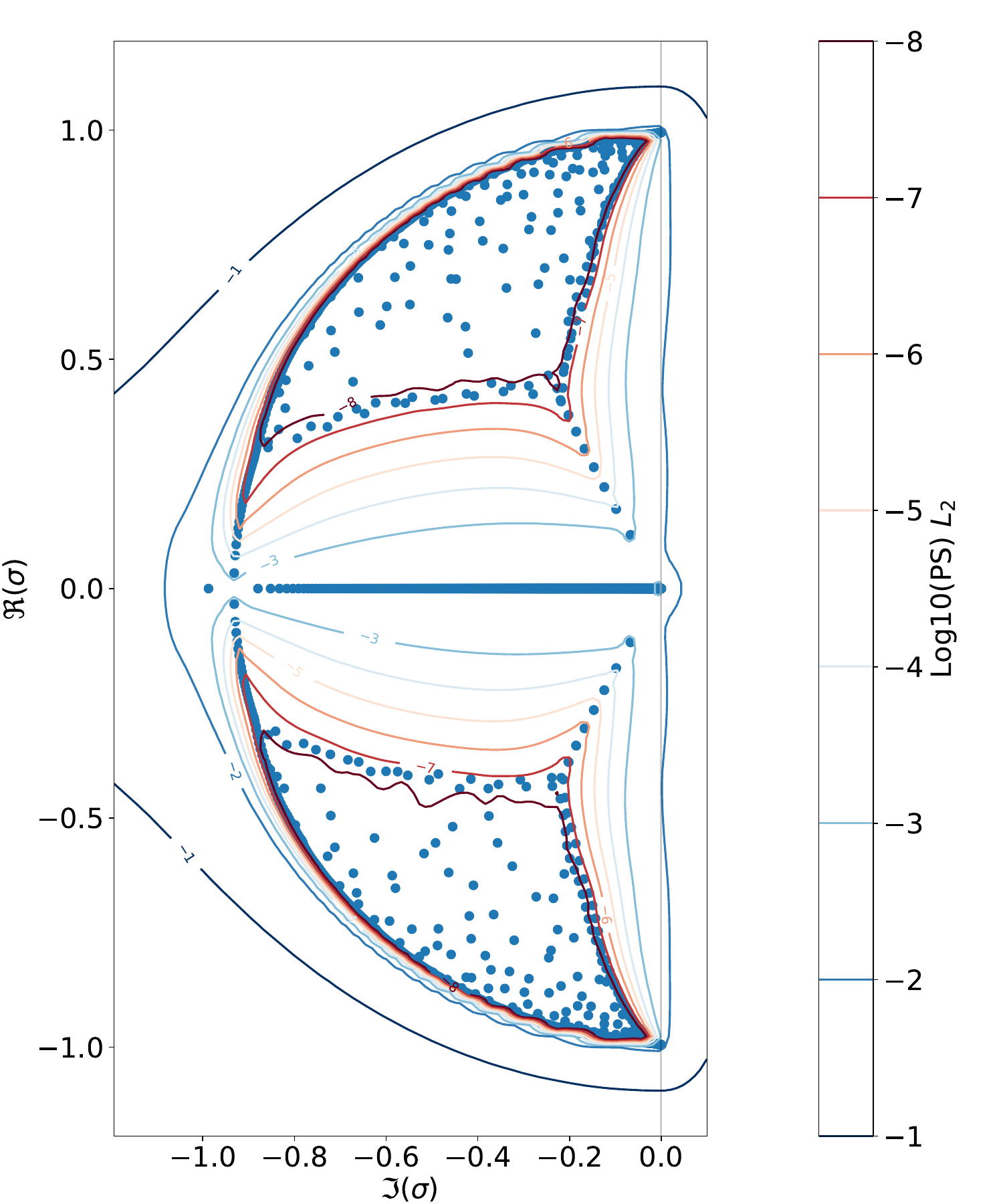}
  \end{minipage}
  \caption{Pseudospectra at $S=10^4$ and $k_x = 0.995$, which means there is no unstable tearing mode, as this wavenumber is slightly past the numerical bound for marginal stability. A grid resolution of $N=2000$ was used for both computations.}
  \label{fig:PS_s1e4_2000}
\end{figure}

Let us examine the pseudospectra of a wavenumber corresponding to a barely stable case ($k_x=0.995$), at a low Lundquist number of $S=10^3$, using once the $L_2$ norm, and once the energy norm, as seen in Fig. \ref{fig:PS_s1e3_1024}. All pseudospectra were computed using the iterative algorithm described in Sec. \ref{sec:PS_ALGO}. We can see nicely how the connection points of the branches of the spectrum follow our predictions. While the two pseudospectra might seem similar at first glance, closer examination of the $\epsilon=2$ curve reveals that the energy norm shows a smaller crossover into the unstable part of the complex plane, with only a very small, pronounced salient being visible around $z=0$ and $z=\pm k_x$. This indicates that a higher transient amplification is possible when measuring growth using the $L_2$ norm, according to the bounds described in Sec. \ref{sec:transient_theory}. The same set of parameters, but using a Lundquist number of $S=10^4$ shows the connecting branch moving closer to the line $\Im(z)=0$ in Fig. \ref{fig:PS_s1e4_1600}. This leads to the $\epsilon = -2$ contour moving further to the right, indicating a higher potential for transient growth. While both norms show the same qualitative behavior, the $L_2$ norm still indicates a higher growth potential. The $\epsilon=-8$ contour, which represents an area of the spectrum that is highly sensitive, becomes much larger with increasing $S$, indicating that this portion of the spectrum will be nearly impossible to numerically resolve since any numerical error on the order of $10^{-8}$ will move the eigenvalues by an order of $0.1$. Looking at the same plot computed at a higher resolution in Fig. \ref{fig:PS_s1e4_2000} we can see that the eigenvalues inside this contour have moved significantly. 

\begin{figure}[htbp]
  \centering
  \begin{minipage}[b]{0.49\textwidth}
    \centering
    \includegraphics[width=\textwidth]{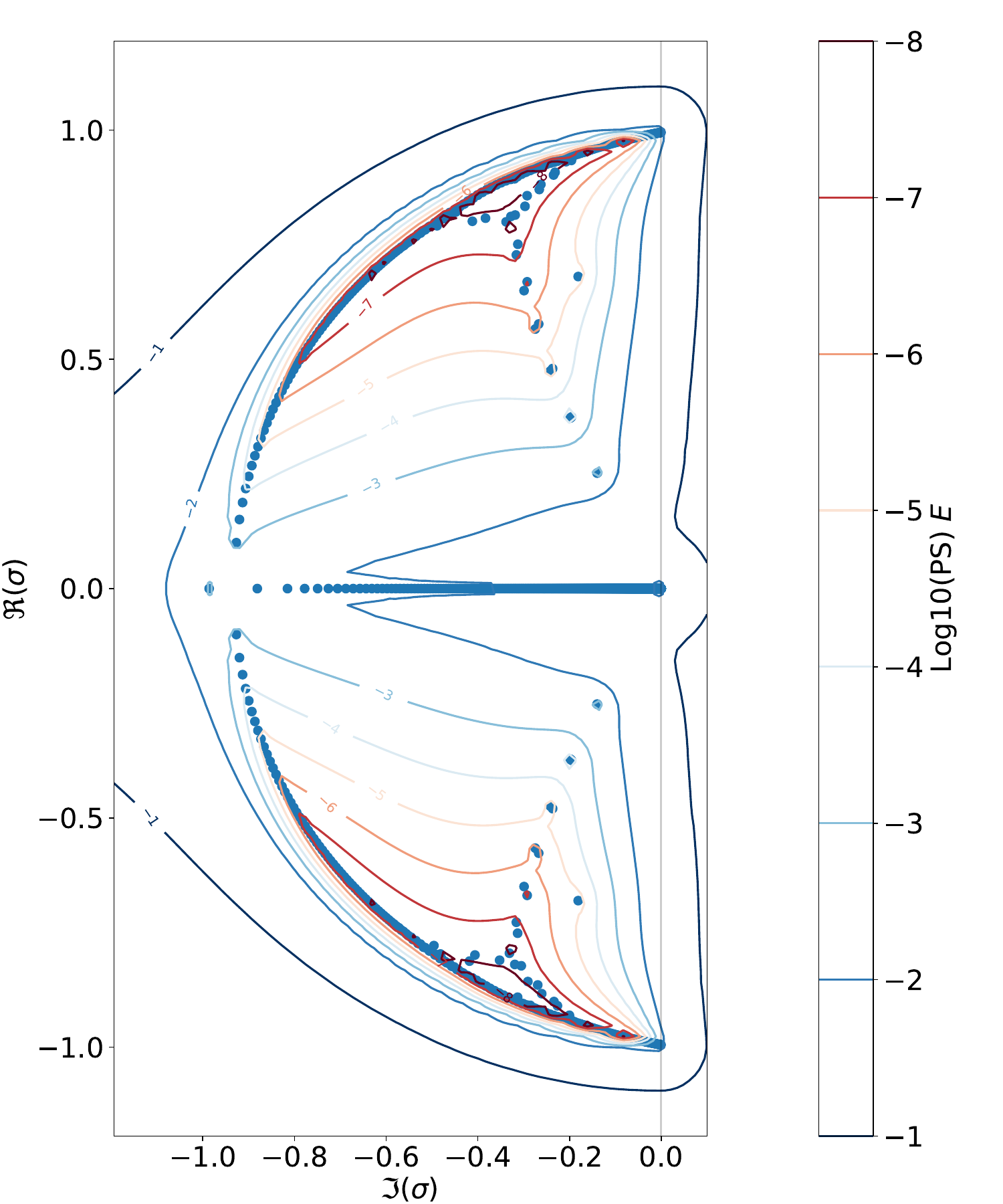}
  \end{minipage}
  \hfill
  \begin{minipage}[b]{0.49\textwidth}
    \centering
    \includegraphics[width=\textwidth]{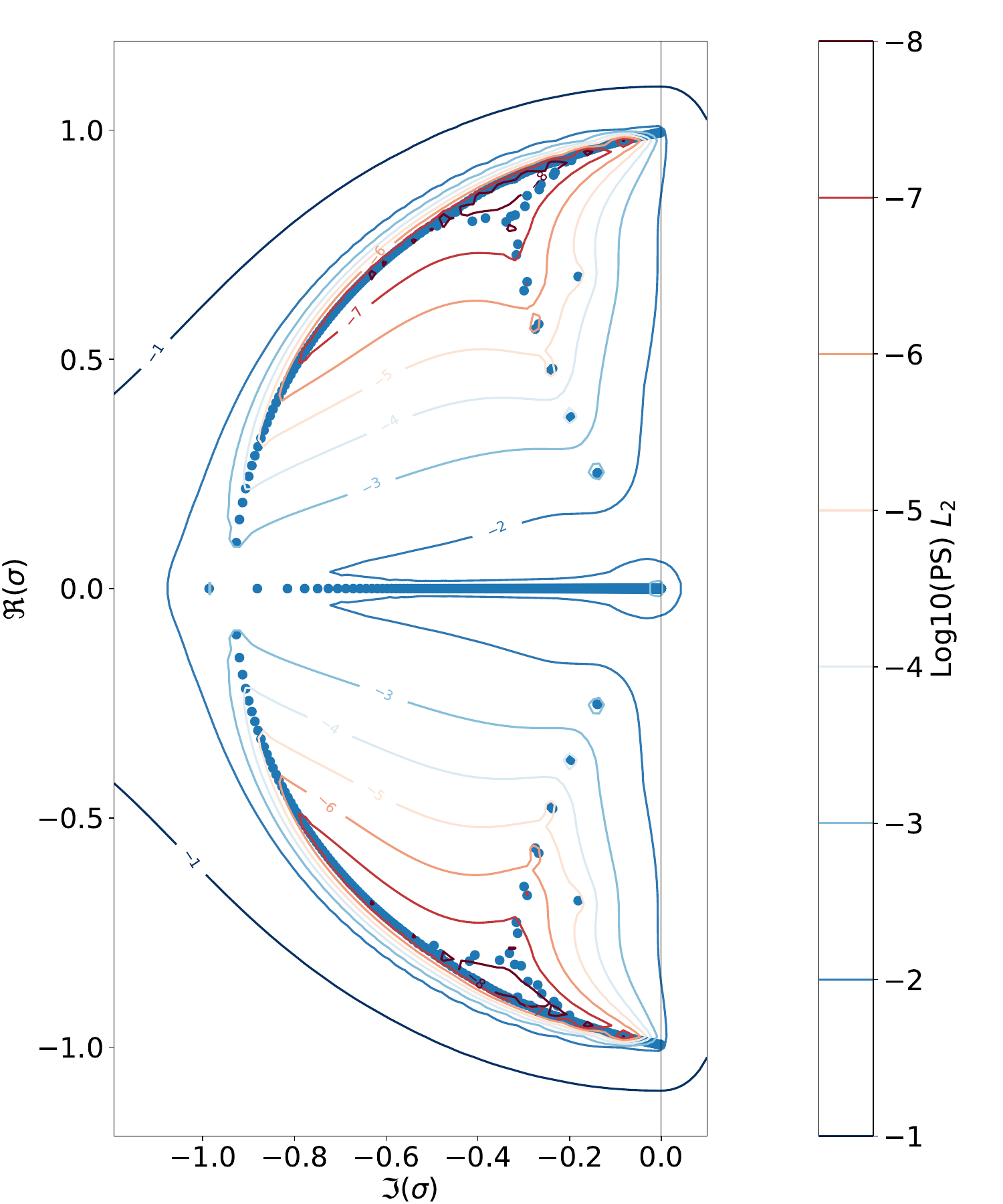}
  \end{minipage}
  \caption{Pseudospectra at $S=10^3$ and $k_x = 0.995$, which means there is no unstable tearing mode, as this wavenumber is slightly past the numerical bound for marginal stability. A grid resolution of $N=1600$ was used for both computations.}
  \label{fig:PS_s1e3_1600}
\end{figure}

It shall be noted that for all wavenumbers, the contour corresponding to disturbances of order $\mathcal{O}(10^{-2})$ protrudes into the right half of the complex plane. This means that there is an initial condition that leads to transient growth. In other words, for all Lundquist numbers considered in this thesis, wavenumbers corresponding to a stable system have an initial condition that will lead to transient growth.

\FloatBarrier
\subsubsection{Transient growth in linearized system}

\begin{figure}[htbp]
  \centering
  \begin{minipage}[b]{0.49\textwidth}
    \centering
    \includegraphics[width=\textwidth]{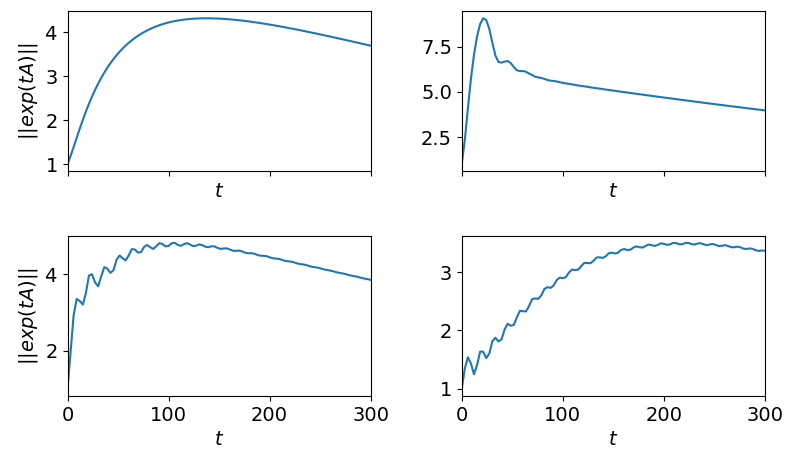}
    \label{fig:fig4}
  \end{minipage}
  \hfill
  \begin{minipage}[b]{0.49\textwidth}
    \centering
    \includegraphics[width=\textwidth]{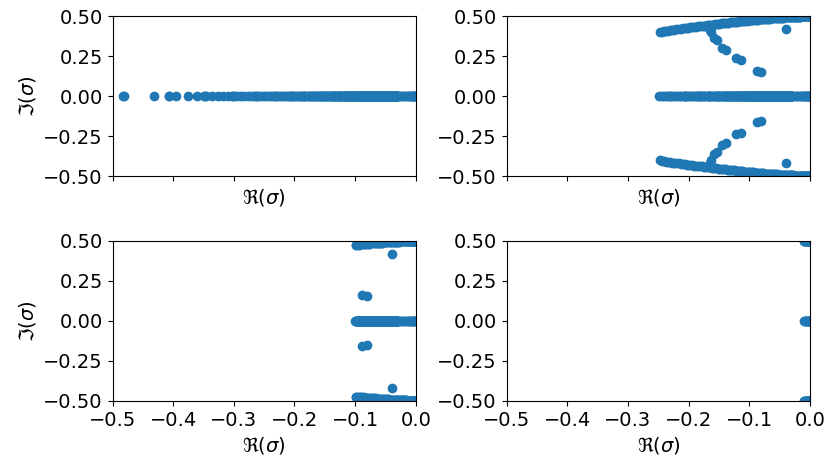}
    \label{fig:fig_4_traj}
  \end{minipage}
  \caption{Subsets of the spectrum considered for orthonormal projection (left) and the resulting upper bounds on transient growth (right). Including the branches connecting the imaginary and mostly real parts of the spectrum leads to significantly higher transient growth. The same computation was carried out, including viscosity, using our algorithm, and we exactly reproduce the results from \cite{mactaggart_2018}.}
  \label{fig:fig_4_comb}
\end{figure}

Since the pseudospectra indicate that Eq.\eqref{eq:reduced_MHD_2d} will exhibit transient growth, we will now closely investigate how large this growth can be, which initial conditions produce it, how it depends on the choice of $\mathcal{S}^N$, and, by consequence, which portions of the spectrum are responsible for this phenomenon. In order to do this, we utilize the concept of orthonormal projection, described earlier in sec. \ref{sec:PS_ALGO} or, alternatively, by selecting the contents of $\mathcal{S}^N$. This allows us to select eigenvalues of interest and examine the dynamics restricted to the set of corresponding eigenvectors. We performed this orthonormal projection for the partitions of the spectrum seen in Fig. \ref{fig:fig_4_comb}, along with the corresponding upper bounds on the $L_2$ norm of the projected system. We can see that the two complex valued branches (A, C in Fig. \ref{fig:sepctrum_labeled}) play an important role in the transient growth. More specifically, the connection point of branch with the central branch seem to be very important. Including this connection point in $\mathcal{S}^N$ leads to almost twice the amplification, in $L_2$ norm compared to other choices of $\mathcal{S}^N$, which do not include this connection point. Since this connection is located close to $\Im(z)=0.5 \cdot k_x$, including all eigenfunctions corresponding to eigenvalues $\Im(\sigma) > -k_x$ in $\mathcal{S}^N$ should allow us to observe a good representation of the transient dynamics, while simultaneously ignoring spurious eigenfunctions associated with numerical eigenvalues, which typically have very large eigenvalues in magnitude.

\begin{figure}[htbp]
  \centering
  \begin{minipage}[b]{0.49\textwidth}
    \centering
    \includegraphics[width=\textwidth]{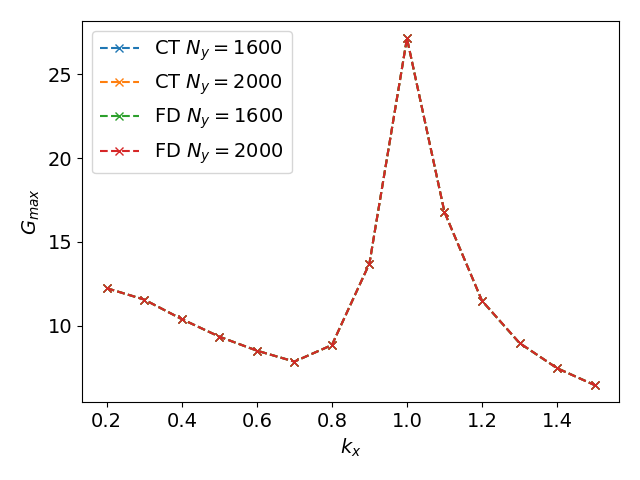}
  \end{minipage}
  \hfill
  \begin{minipage}[b]{0.49\textwidth}
    \centering
    \includegraphics[width=\textwidth]{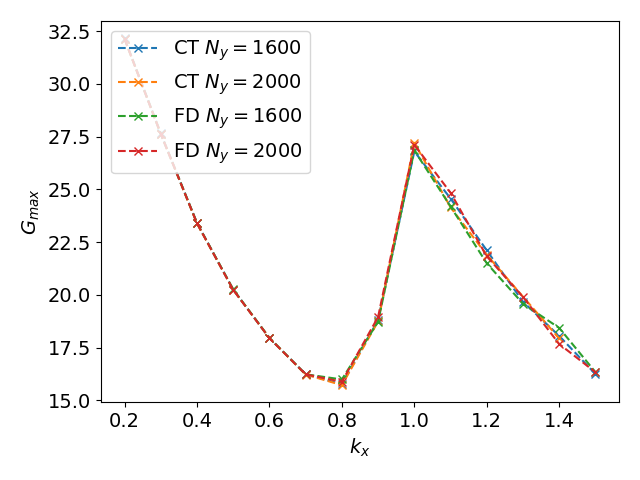}
  \end{minipage}
  \caption{Maximum of $G(t)$ for $t \in [0, 300]$ without viscosity at $S=10^3$ (left) and $S=10^4$ (right), using the $L_2$ norm.}
  \label{fig:truncated}
\end{figure}

In order to get a good sense of how nonmodal effects compare to the modal dispersion relation computed earlier, we first removed the most unstable eigenfunction from the system via orthogonal projection, and found the maximum growth achieved by the remaining, damped eigenfunctions. The time horizon for this optimization was $t=300$. Results computed in the $L_2$ norm can be seen in Fig. \ref{fig:truncated} This computation was carried out for wavenumbers $k_x \in [0.2, 1.5]$. Increasing $S$ by a factor of 10 allows for higher transient growth at unstable wavenumbers. This scaling was observed by Mactaggart \cite{MacTaggart2020}, who also remarks that it does not carry over to unstable wavenumbers. Our results indicate that stable wavenumbers are much less influenced by changing $S$, compared to unstable ones. All results were computed using both compact finite difference matrices and Chebyshev differentiation matrices (denoted FD and CT in plots).

\begin{figure}[htbp]
  \centering
  \begin{minipage}[b]{0.49\textwidth}
    \centering
    \includegraphics[width=\textwidth]{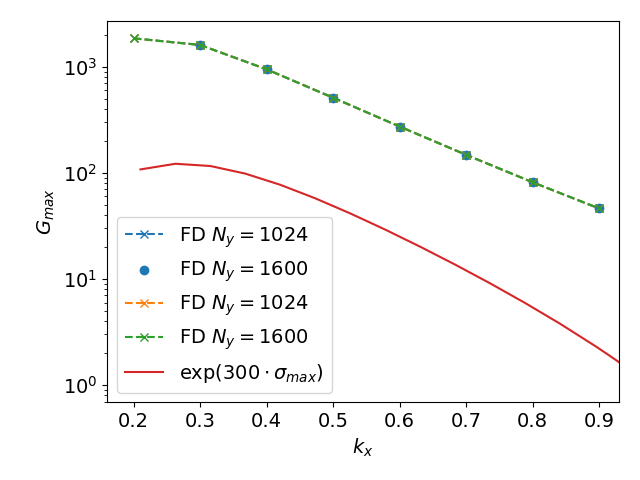}
  \end{minipage}
  \hfill
  \begin{minipage}[b]{0.49\textwidth}
    \centering
    \includegraphics[width=\textwidth]{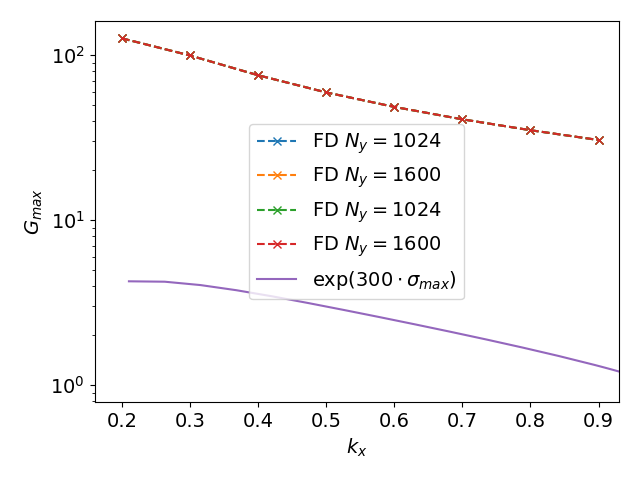}
  \end{minipage}
  \caption{Value of $G(300)$ for $t = 300$ without viscosity at $S=10^3$ (left) and $S=10^4$ (right), using the $L_2$ norm. For comparison, the modal growth at $t=300$ is included as well.}
  \label{fig:full_t_300}
\end{figure}

To emphasize the effect of nonmodal growth, we directly compare the modal growth, achieved by the tearing mode alone, with the highest possible linear transient growth at unstable wavenumbers. For this comparison, both quantities are compared at $t=300$. The results can be seen in Fig. \ref{fig:full_t_300}. Including nonmodal effects leads to significantly larger growth, and one can clearly see that increasing $S$ leads to lower modal growth. Again, all results were computed both with Chebyshev and finite difference matrices.

\begin{figure}[htbp]
  \centering
  \begin{minipage}[b]{0.49\textwidth}
    \centering
    \includegraphics[width=\textwidth]{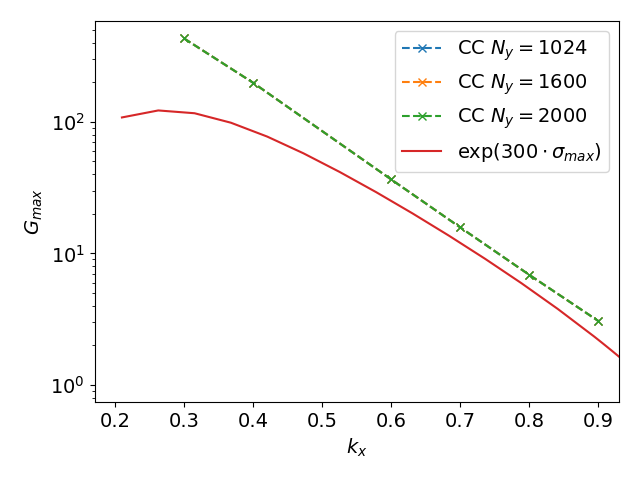}
  \end{minipage}
  \hfill
  \begin{minipage}[b]{0.49\textwidth}
    \centering
    \includegraphics[width=\textwidth]{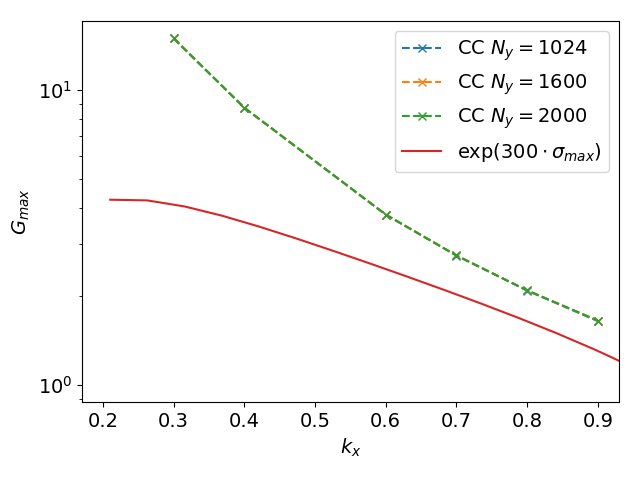}
  \end{minipage}
  \caption{Value of $G(300)$ for $t = 300$ without viscosity at $S=10^3$ (left) and $S=10^4$ (right), using the energy norm. For comparison, the modal growth at $t=300$ is included as well.}
  \label{fig:full_t_300_e}
\end{figure}

The same computation can be carried out using the energy norm. Looking at unstable wavenumbers again, we again see significantly higher growth potential compared to purely modal growth. The growth potential is significantly smaller compared to the $L_2$ norm. The results can be seen in Fig. \ref{fig:full_t_300_e}. This computation was only carried out using the system in Chebyshev coefficient form (denoted CC in plot), as this was the only formulation the energy norm was implemented in for this thesis.

To get a full picture of the maximum possible transient growth, we will numerically compute the envelope given by Eq. \ref{eq:matrix_exponential}. We recall that this represents a bound that is obtained by optimizing over all admissible initial conditions and selecting the one that will lead to the amplification, represented by the value of $G(t)$, at $t=t_{opt}$. At any other value of $t$, we have no information about the dynamics of said initial condition.

\begin{figure}[htbp]
  \centering
  \begin{minipage}[b]{0.49\textwidth}
    \centering
    \includegraphics[width=\textwidth]{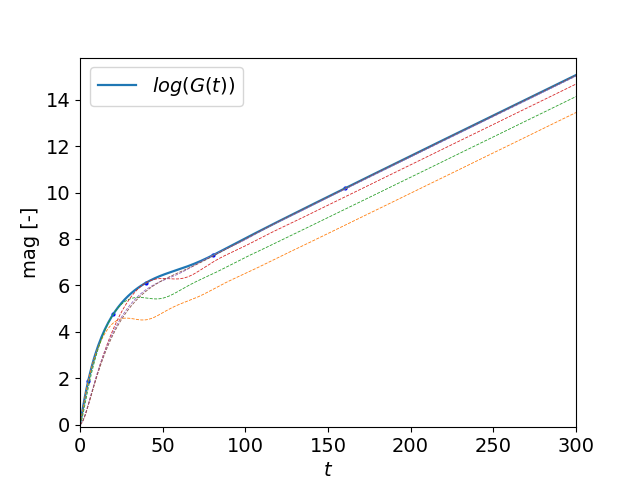}
  \end{minipage}
  \hfill
  \begin{minipage}[b]{0.49\textwidth}
    \centering
    \includegraphics[width=\textwidth]{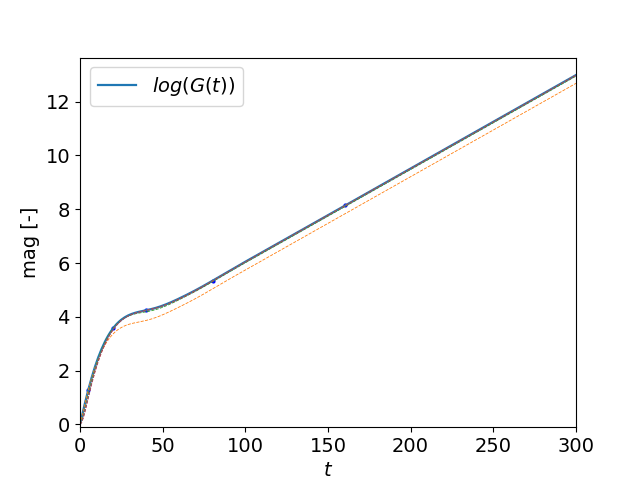}
  \end{minipage}
  \caption{Upper bounds on transient growth using two different norms. The left plot shows results using the $L_2$ norm of the state vector, the plot on the right shows results using the perturbation energy as the norm. $S=10^3$, $k_x=0.2$, $N=1000$. Dashed lines correspond to the evolution of optimal initial conditions optimized for the time at which they meet the growth bound, marked with a dot.}
  \label{fig:ev_vs_opt_s1e3_02}
\end{figure}

First, we consider a low Lundquist number of $S=10^3$ and an unstable wavenumber of $k_x=0.2$. All transient growth computations were performed up to a time horizon of 300 time units. For moderate Lundquist numbers below $10^5$, a significantly shorter time horizon would suffice, since it has been shown that modal effects start to dominate the behavior at a time of around $S^{\frac{1}{2}}$. Fig. \ref{fig:ev_vs_opt_s1e3_02} shows the upper bound, defined by the matrix exponential, along with the linear evolution of initial conditions, leading to the highest possible transient growth. It can be seen that the sampled initial conditions touch the upper bound at exactly the time they were optimized for, with the behavior at different times not necessarily being related. As can be seen in a time horizon of around $t=100$, all sampled initial conditions grow at an exponential rate, corresponding to the modal growth rate, although at a significantly larger amplitude than possible via modal growth. The upper bound computed with the $L_2$ norm is significantly higher than the one computed with the energy norm. 

\begin{figure}[htbp]
  \centering
  \begin{minipage}[b]{0.49\textwidth}
    \centering
    \includegraphics[width=\textwidth]{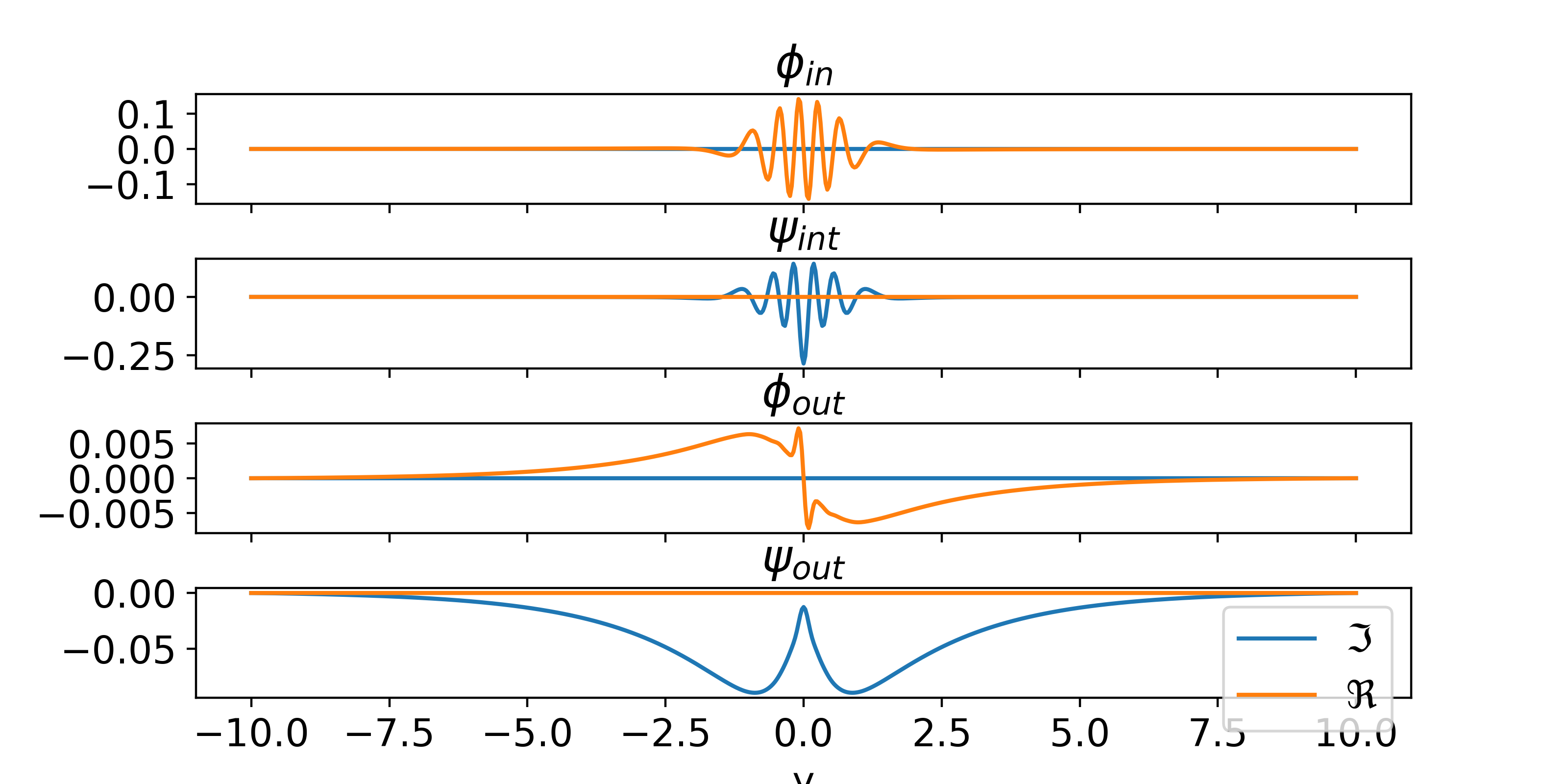}
  \end{minipage}
  \hfill
  \begin{minipage}[b]{0.49\textwidth}
    \centering
    \includegraphics[width=\textwidth]{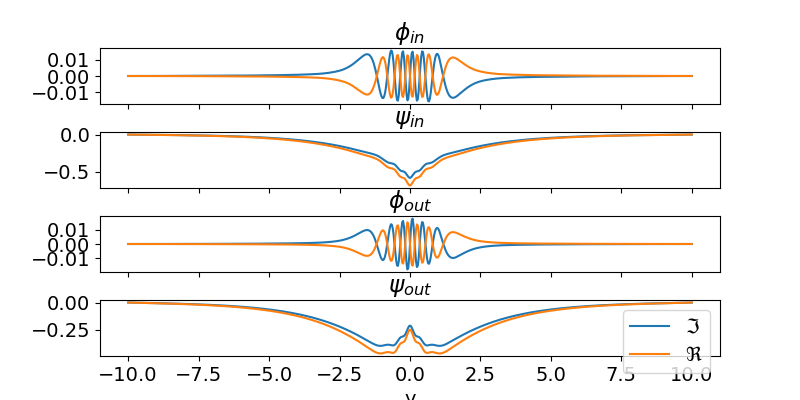 }
  \end{minipage}
  \caption{Pseudomodes that lead to maximum transient growth in $L_2$ norm (left) and energy norm (right). $S=10^5$, $k_x=0.5$, $N=1000$, $Re=0$, $t_{opt} = 40$. The pseudomode computed with the $L_2$ criterion clearly exhibits strong oscillations in both field components, whereas the one computed with the energy norm has a much more tame mode corresponding to the magnetic vector potential, which does not have a single oscillation.}
  \label{fig:pseudomes_1e5_05}
\end{figure}

\begin{figure}[htbp]
  \centering
  \begin{minipage}[b]{0.49\textwidth}
    \centering
    \includegraphics[width=\textwidth]{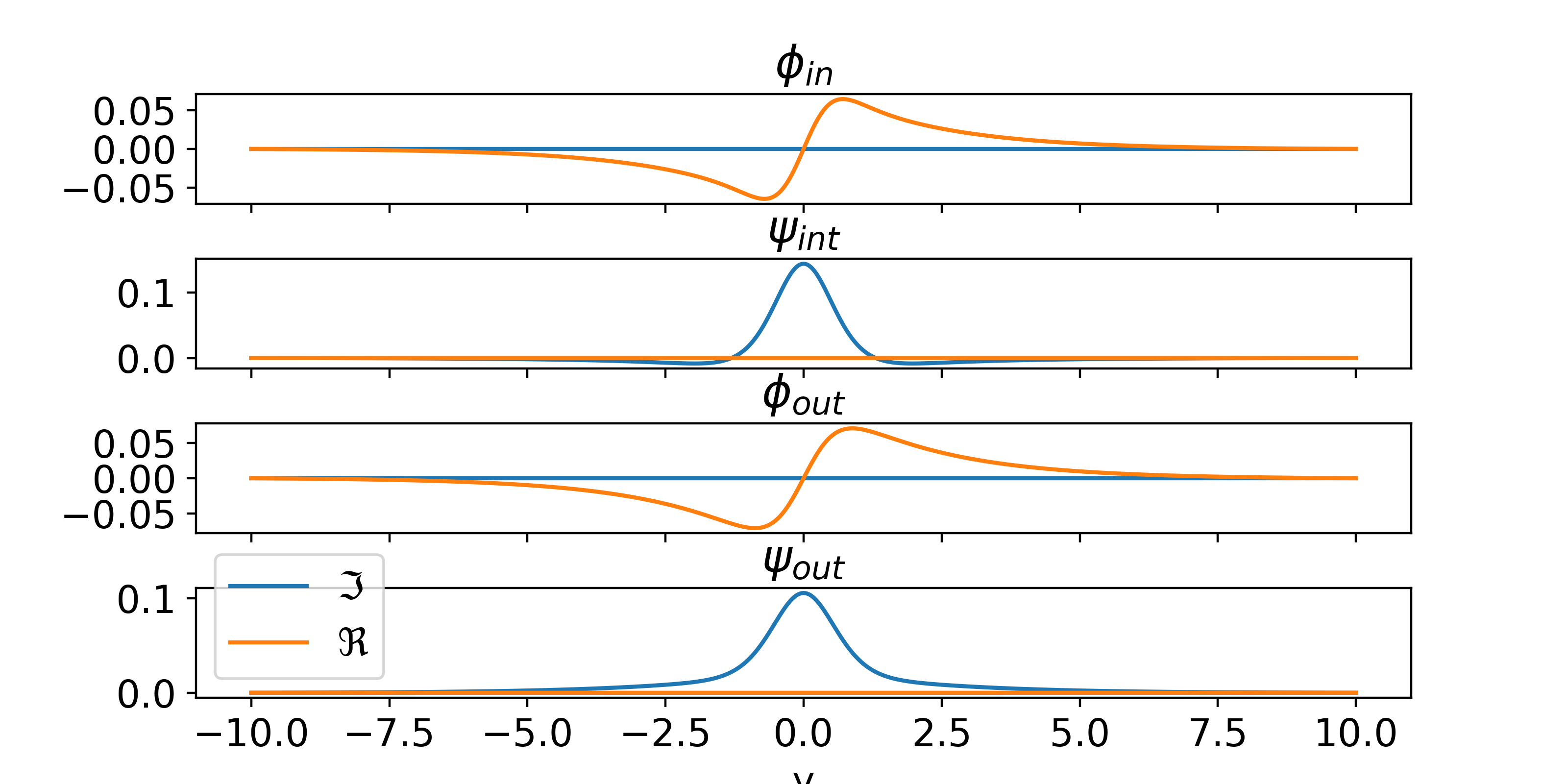}
  \end{minipage}
  \hfill
  \begin{minipage}[b]{0.49\textwidth}
    \centering
    \includegraphics[width=\textwidth]{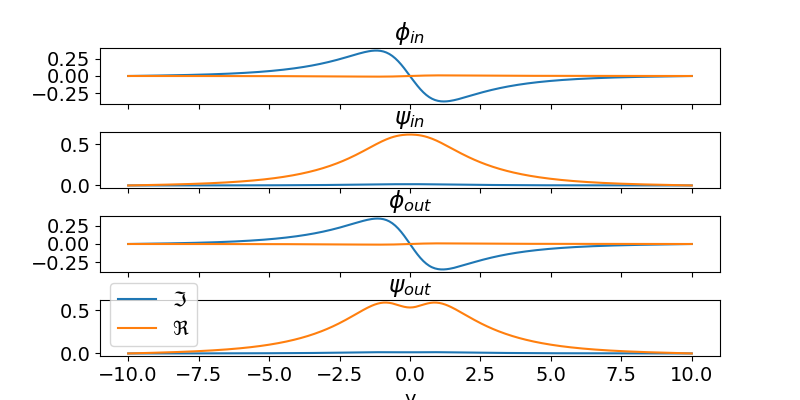}
  \end{minipage}
  \caption{Pseudomodes that lead to maximum transient growth in $L_2$ norm (left) and energy norm (right). $S=10^3$, $k_x=0.5$, $N=1024$, $Re=0$, $t_{opt} = 1$. While the difference between the two modes is not as striking as in previous examples, there is still a noticeably higher degree of localization associated with the $L_2$ norm.}
  \label{fig:pseudomodes_1e3_05}
\end{figure}

\begin{figure}[htbp]
  \centering
  \begin{minipage}[b]{0.49\textwidth}
    \centering
    \includegraphics[width=\textwidth]{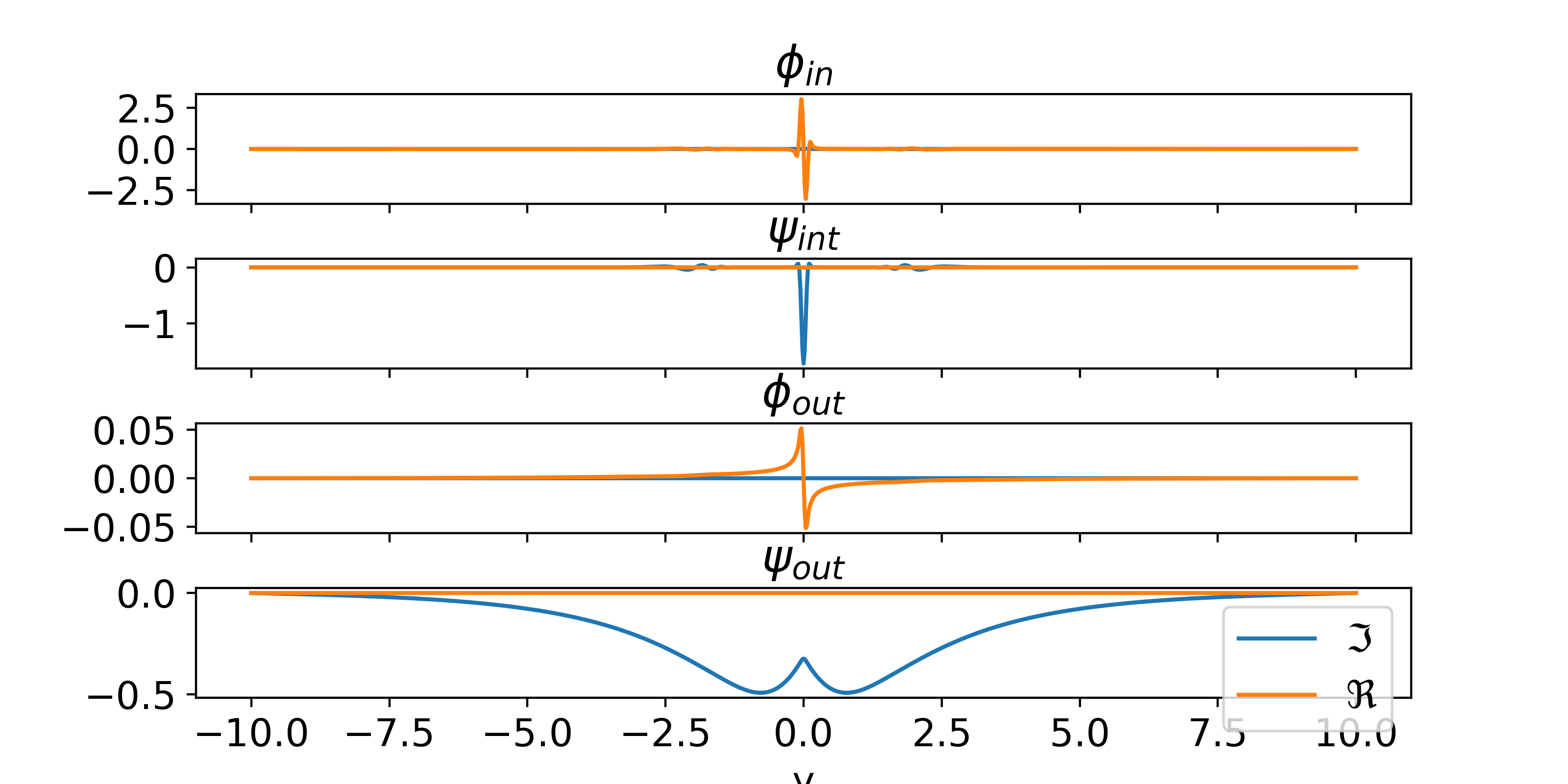}
  \end{minipage}
  \hfill
  \begin{minipage}[b]{0.49\textwidth}
    \centering
    \includegraphics[width=\textwidth]{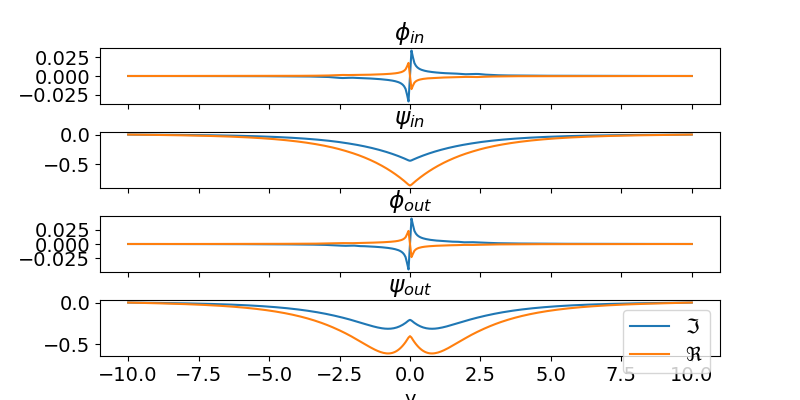}
  \end{minipage}
  \caption{Pseudomodes that lead to maximum transient growth in $L_2$ norm (left) and energy norm (right). $S=10^4$, $k_x=0.2$, $N=1000$, $Re=0$, $t_{opt} = 300$. In both norms, after transient growth until $t_{opt}$, the pseudomodes continue to grow at the modal growth rate obtained by an eigenvector.}
  \label{fig:looks_like_eigenvector}
\end{figure}

Representative examples of initial conditions leading to the upper bound in transient growth can be seen in Figures \ref{fig:pseudomes_1e5_05}, \ref{fig:pseudomodes_1e3_05} and \ref{fig:looks_like_eigenvector}. The first two figures were computed at an unstable wavenumber $k_x=0.5$, for $S=10^3$ and $S=10^5$. Plotted are both the initial conditions and the state at which maximum amplification is achieved. The modes denoted by $\phi_{in}, \psi_{in}$ are the initial conditions that get propagated to $\phi_{out}, \psi_{out}$ at $t_{opt}$, amplified by the nonmodal growth factor $G(t_{opt})$. Fig. \ref{fig:looks_like_eigenvector} was computed at $s=10^4, k_x=0.2$ at a very long time horizon $t_{opt} = 300$. Comparing pseudomode initial conditions computed with the $L_2$ norm with results computed with the energy norm, we can see that the initial conditions that lead to growth in the $L_2$ norm have a much more oscillatory character. This has a simple explanation: the $L_2$ norm does not include the derivative of the modes in the optimization that we solve via SVD. The energy norm does, and therefore finds a mode with less oscillations. This effect is especially obvious in the magnetic field. In the component corresponding to $\phi$, oscillations can also be found using the energy norm, although this leads to $\phi$ being an order of magnitude smaller relative to $\psi$. This might be an explanation for the higher bounds on transient growth for the $L_2$ norm; in the physical system, energy stored in the derivative of the initial conditions could be converted into energy stored in the magnitude of the stream functions, albeit at a specific time. Using our interpretation of the state vector as a stream function, this would be equivalent to energy being transferred from the x-component of the magnetic and velocity field being transferred to the y-component. Interpreting the state vector as perturbation fields in y-direction along with the solenoidal field constraint leads us to the exact same conclusion. In general, a trend can be seen that with increasing time horizon for optimization, $t_{opt}$, one computes maximal growth initial conditions with more oscillations. The pseudomodes that optimize transient growth at $t_{opt} = 300$ show another very interesting result: for very long time horizons, where modal growth overweighs, as seen for $t>100$ in Fig. \ref{fig:ev_vs_opt_s1e3_02} we expect to get $\phi_{out}, \psi_{out}$ that are very similar to the eigenvectors of the system, as these are the structures that grow at the modal growth rate. We can see that this is the case. Interestingly, the initial conditions do not necessarily bear resemblance to the eigenvectors of the linear system. This is more rigorously derived in \cite{Farrell_1996}.

\begin{figure}[htbp]
  \centering
  \begin{minipage}[b]{0.49\textwidth}
    \centering
    \includegraphics[width=\textwidth]{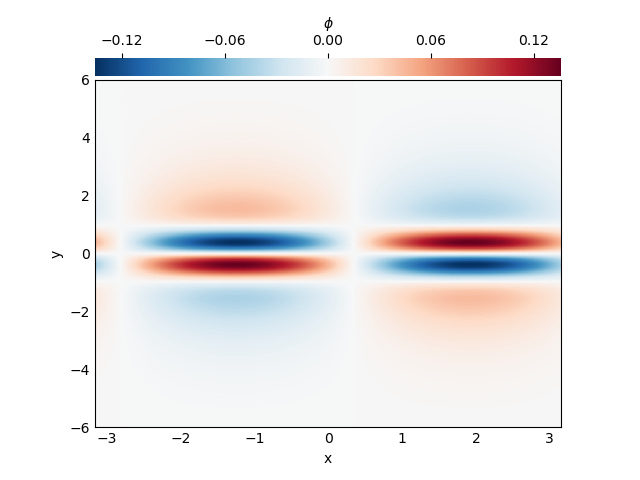}
  \end{minipage}
  \hfill
  \begin{minipage}[b]{0.49\textwidth}
    \centering
    \includegraphics[width=\textwidth]{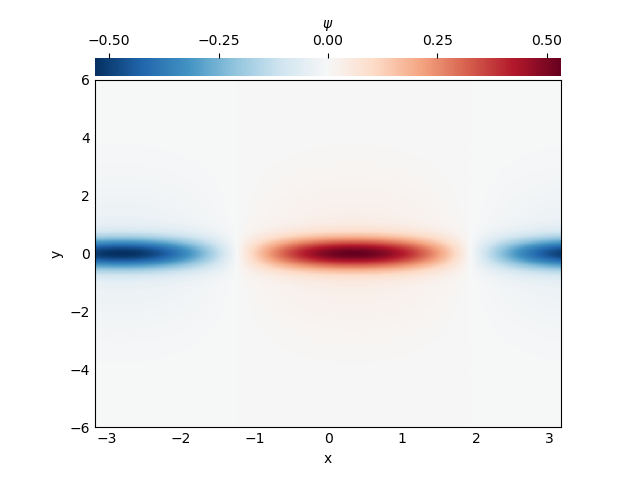}
  \end{minipage}
  \caption{Pseudomode initial conditions that lead to maximum transient growth in energy norm. $S=10^5$, $k_x=0.995$, $N=1000$, $Re=0$, $t_{opt} = 5$. The modes were projected onto a two-dimensional domain using a wavenumber of unity, normalized with respect to the computational domain.}
  \label{fig:init_2d}
\end{figure}

\begin{figure}[htbp]
  \centering
  \begin{minipage}[b]{0.49\textwidth}
    \centering
    \includegraphics[width=\textwidth]{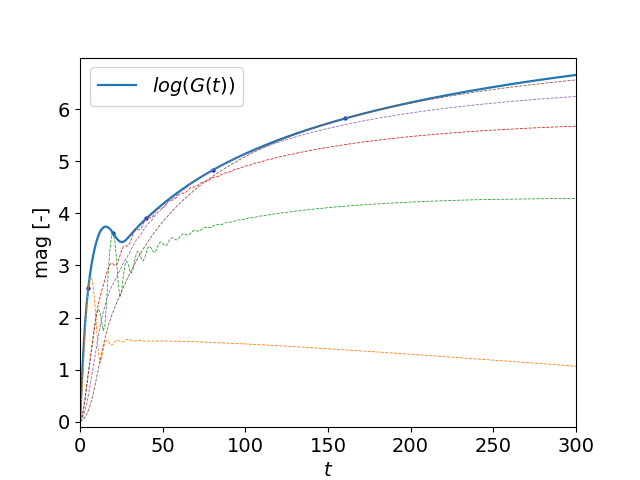}
  \end{minipage}
  \hfill
  \begin{minipage}[b]{0.49\textwidth}
    \centering
    \includegraphics[width=\textwidth]{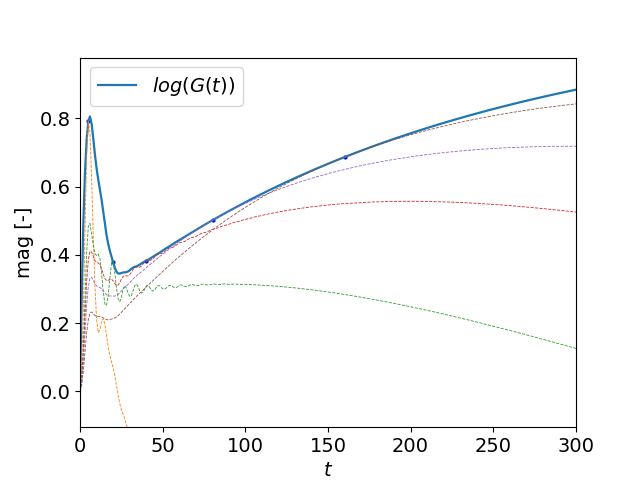}
  \end{minipage}
  \caption{Upper bounds on transient growth using two different norms. The left plot shows results using the $L_2$ norm of the state vector, the plot on the right shows results using the perturbation energy as the norm. $S=10^3$, $k_x=0.995$, $N=1600$.}
  \label{fig:ev_vs_opt_lin_s1e3}
\end{figure}

\begin{figure}[htbp]
  \centering
  \begin{minipage}[b]{0.49\textwidth}
    \centering
    \includegraphics[width=\textwidth]{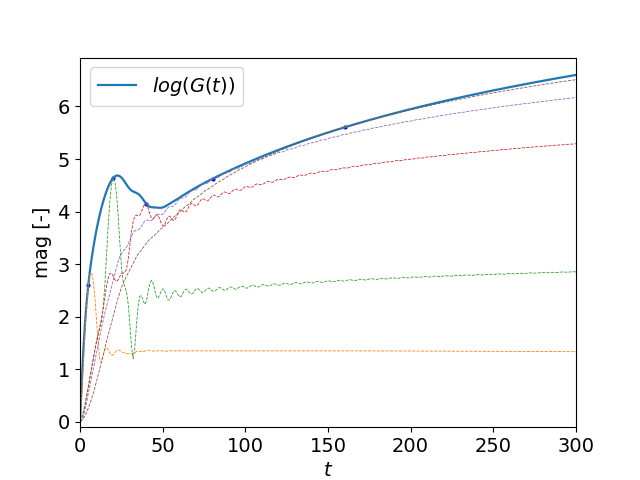}
  \end{minipage}
  \hfill
  \begin{minipage}[b]{0.49\textwidth}
    \centering
    \includegraphics[width=\textwidth]{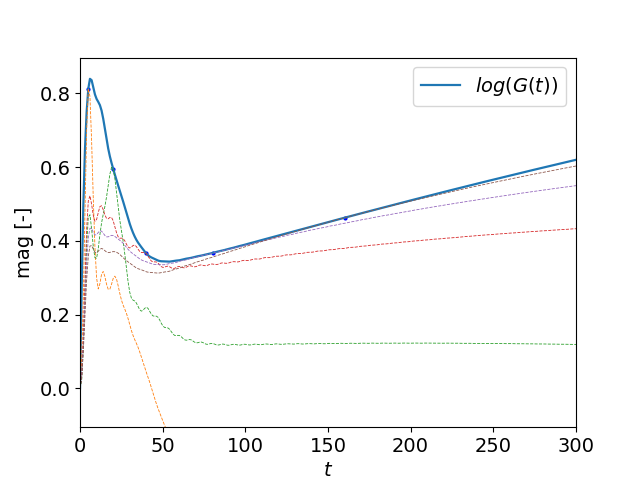}
  \end{minipage}
  \caption{Upper bounds on transient growth using two different norms. The left plot shows results using the $L_2$ norm of the state vector, the plot on the right shows results using the perturbation energy as the norm. $S=10^4$, $k_x=0.995$, $N=1600$.}
  \label{fig:ev_vs_opt_lin_1e4}
\end{figure}

\begin{figure}[htbp]
  \centering
  \begin{minipage}[b]{0.49\textwidth}
    \centering
    \includegraphics[width=\textwidth]{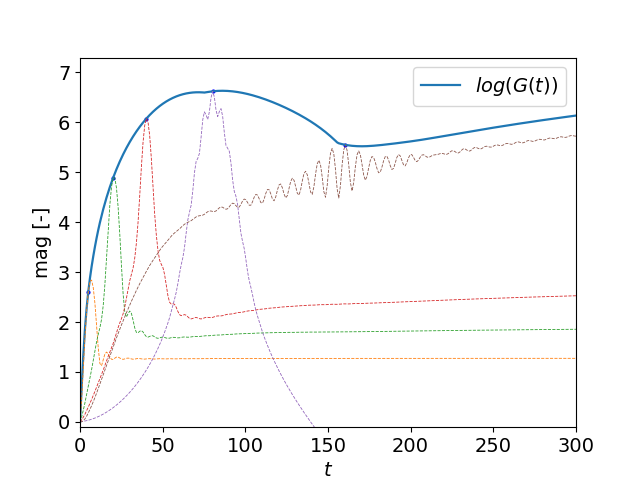}
  \end{minipage}
  \hfill
  \begin{minipage}[b]{0.49\textwidth}
    \centering
    \includegraphics[width=\textwidth]{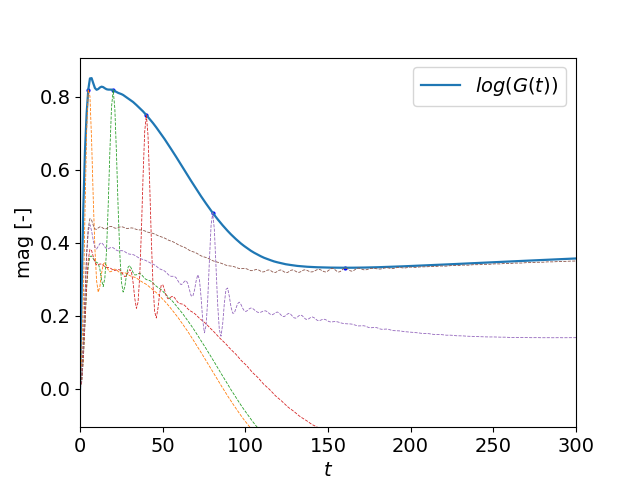}
  \end{minipage}
  \caption{Upper bounds on transient growth using two different norms. The left plot shows results using the $L_2$ norm of the state vector, the plot on the right shows results using the perturbation energy as the norm. $S=10^5$, $k_x=0.995$, $N=1600$.}
  \label{fig:ev_vs_opt_lin_1e5}
\end{figure}

Looking at the same bounds, but computed for a wavenumber $k_x=0.995$ very close to marginal stability, we can see a few differences compared to the unstable wavenumber case. In the energy norm, lower amplifications are achieved. In the $L_2$ norm, this is not necessarily the case, as seen in Fig. \ref{fig:truncated}. The separation between growth computed in the energy norm and the $L_2$ norm is more pronounced,  with the energy norm only allowing significantly lower amplification of initial conditions. This can be seen to be true for three different Lundquist numbers in Figs. \ref{fig:ev_vs_opt_lin_s1e3} \ref{fig:ev_vs_opt_lin_1e4} \ref{fig:ev_vs_opt_lin_1e5} The sampled linear trajectories for optimum initial conditions now also reveal much more clearly that they only exhibit optimal growth at a short instance of time around $t_{opt}$, with initial conditions corresponding to an early $t_{opt}$ decaying rapidly after reaching the growth bound. This happens in both norms, although it is more pronounced in the energy norm. Pseudomode initial conditions can be seen in Fig. \ref{fig:init_2d}, after projection onto a sinusoidal perturbation in x-direction. This plot displays the relative phase between the two components of the solution vector.

\subsubsection{\label{sec:resolvent_results} Computation of resolvent modes}

\begin{figure}[htbp]
  \centering
  \begin{minipage}[b]{0.49\textwidth}
    \centering
    \includegraphics[width=\textwidth]{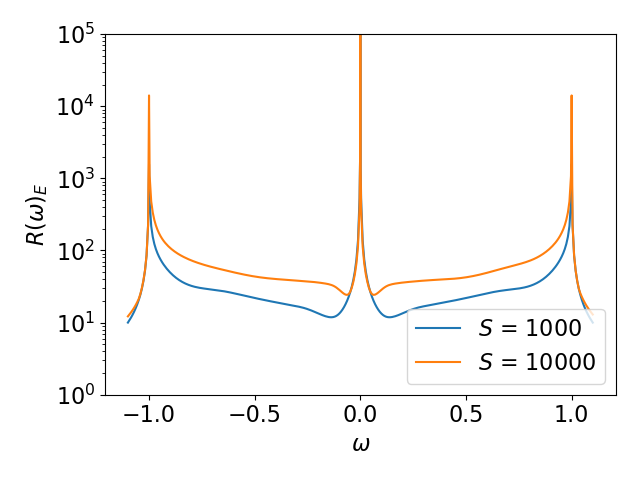}
  \end{minipage}
  \hfill
  \begin{minipage}[b]{0.49\textwidth}
    \centering
    \includegraphics[width=\textwidth]{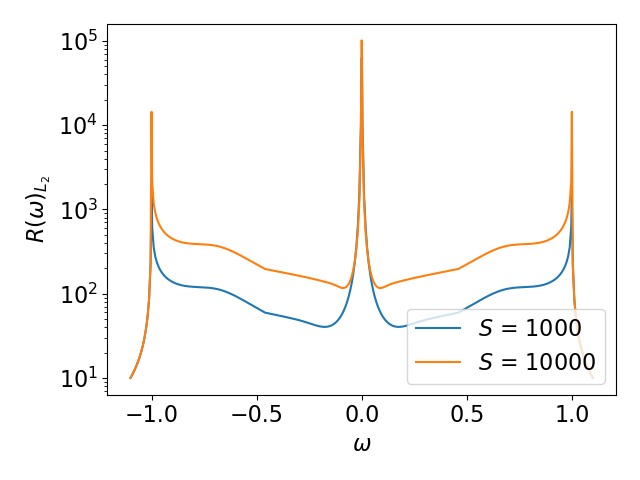}
  \end{minipage}
  \caption{Receptivity curves computed both in the energy norm (left) and the $L_2$ norm (right). The curves were computed at a wavenumber of $k_x=1.0$ using $N=2000$ grid points.}
  \label{fig:receptivity}
\end{figure}

Apart from transient growth, the possibility of pseudoresonance is an interesting property of non-normal systems. The curves characterizing the maximum steady-state response to harmonic forcing were computed for a similar parameter range as the transient growth computations. The curves can be seen in Fig. \ref{fig:receptivity}. Each curve has three distinct peaks; these peaks correspond to frequencies $\omega$ close to an eigenvalue of the system. As we saw earlier in Fig. \ref{fig:sepctrum_labeled}, there are three points where branches of the spectrum come close to $\Im(\omega) = 0$: at $\Re(\omega)\pm k_x$ and $\Re(\omega)=0$. At exactly these frequencies peaks can be observed. This would also be expected for a normal system. What is a clear indication of non-normal behavior however, is the fact that the resolvent norm remains fairly large in between these peaks (see Sec. \ref{sec:non_normal_theory}). At frequencies around $\omega \pm k_{x}$ the resolvent norm decays in an asymmetric manner, with fairly high amplification factors being achieved for $\vert \omega \vert < 1$ as opposed to $\vert \omega \vert > 1$. Increasing $S$ leads to higher responsiveness to forcing, similar to the increased transient growth observed.

\begin{figure}[h]
    \centering
    \includegraphics[width=0.9\textwidth]{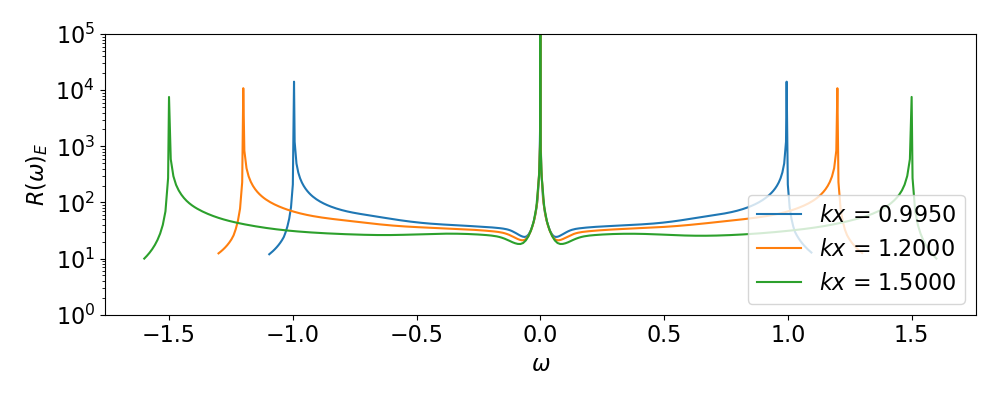}
    \caption{Receptivity curve at $S=10^4$ for different, spectrally stable, values of $k_x$}
    \label{fig:recept_1}
\end{figure}

Changing the wavenumber $k_x$ to have a larger stability margin, i.e., higher, we can see that in general we get smaller amplification factors, although the resolvent norm still decays slower than it would in a normal system. The corresponding response curves can be seen in Fig. \ref{fig:recept_1}

\begin{figure}[htbp]
  \centering
  \begin{minipage}[b]{0.49\textwidth}
    \centering
    \includegraphics[width=\textwidth]{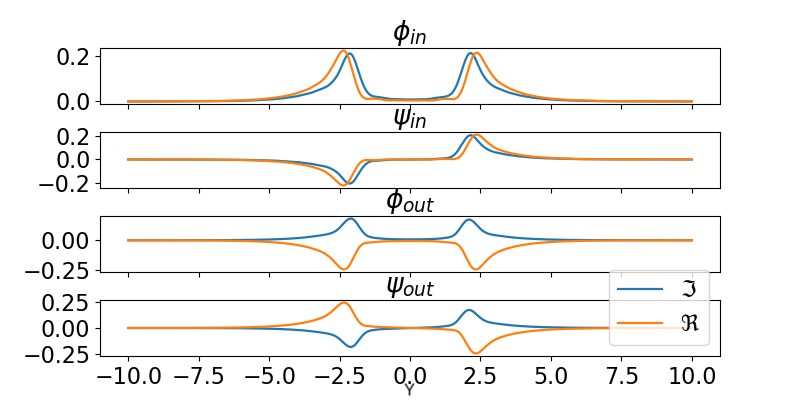}
  \end{minipage}
  \hfill
  \begin{minipage}[b]{0.49\textwidth}
    \centering
    \includegraphics[width=\textwidth]{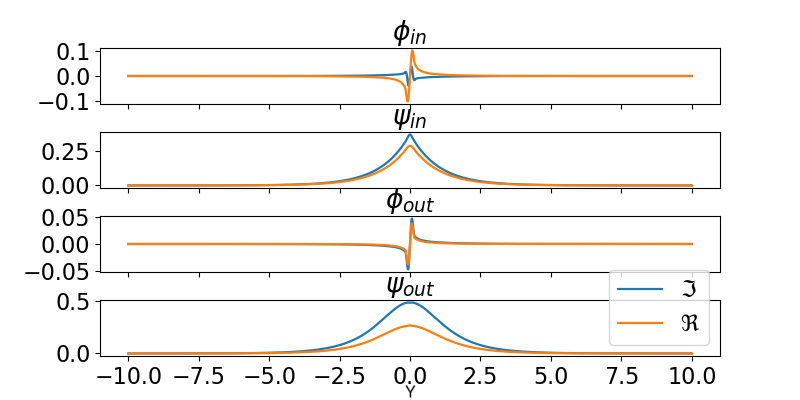}
  \end{minipage}
  \caption{Input and output modes corresponding to modes of maximum amplification in the energy norm when used as harmonic forcing with frequency $\omega$. The left mode was computed for $\vert\omega\vert=0.968$, the right one for $\vert\omega\vert=0.011$. Both modes were computed using $N=2000$ grid points at a wavenumber of $k_x = 1.0$, a Lundquist number of $S=10^3$.}
  \label{fig:resolvent_modes}
\end{figure}

Looking at the modes that lead to this optimum amplification of harmonic forcing, we can see that their shape changes with the forcing frequency $\omega$. In Fig. \ref{fig:resolvent_modes}, the input and output modes of the resolvent operator were computed for a frequency close to $\omega = k_x$ and for another close to zero. Clearly, the mode close to zero is much more localized in the center of the domain. The pseudomode closer to $\omega=k_x$ has two distinct peaks and seems to be close to zero inside the magnetic shear layer, located at $y=0$.

\subsubsection{\label{sec:numerical_instability} Sensitivity of energy norm results}

\begin{figure}[htbp]
  \centering
  \begin{minipage}[b]{0.49\textwidth}
    \centering
    \includegraphics[width=\textwidth]{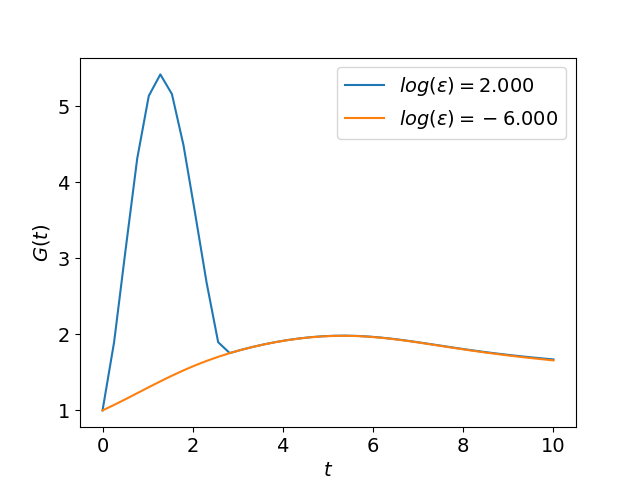}
  \end{minipage}
  \hfill
  \begin{minipage}[b]{0.49\textwidth}
    \centering
    \includegraphics[width=\textwidth]{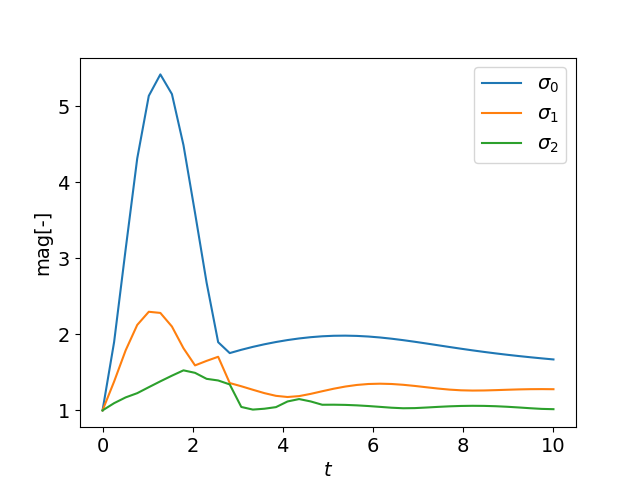}
  \end{minipage}
  \caption{Sensitivity of growth bound with respect to pseudoinverse tolerance (left). Magnitude of the first three SVD coefficients over time (right).}
  \label{fig:numerics_energy_norm}
\end{figure}

A strong dependence of transient growth bounds was noticed both with respect to the resolution used and the technique used to invert the matrix $\bm F$. This is demonstrated in Fig. \ref{fig:numerics_energy_norm}. The left plot shows the upper bound on transient growth, defined by the leading singular value as a function of time. Each curve is computed by using a different value $\epsilon$ for the pseudoinverse of $\bm F$. The growth bounds differ drastically. Using no cutoff at all for the pseudoinverse, a growth bound is computed that includes a very large spike at early time horizons. This spike completely vanishes when using a truncated pseudoinverse, with a tolerance of around $\epsilon = 10^{-6}$. If we look at the first three singular values of the matrix exponential, computed without truncating the pseudoinverse, i.e. the amplification factors of the three most amplified initial conditions, we can see that until $t \approx 2.5 $, there are two initial conditions with relatively high amplification factors, whereas after $t \approx 2.5$, a mode with an amplification factor of around two persists as the optimum initial condition. The modes leading to the high initial amplification have very high-frequency components, and are visually hard to distinguish from numerical artifacts. 

\FloatBarrier
\subsection{Nonlinear System}

After computing various quantities from the linear system, the question of how these results carry over to the full nonlinear system arises. For this purpose, the spectral PDE solver Dedalus \cite{dedalus} was used. We focus on reproducing the transient growth seen in the linear set of equations in the fully nonlinear form of the system.

\subsubsection{Scaling study on the ETH Euler cluster}
\begin{figure}[h]
    \centering
    \includegraphics[width=0.5\textwidth]{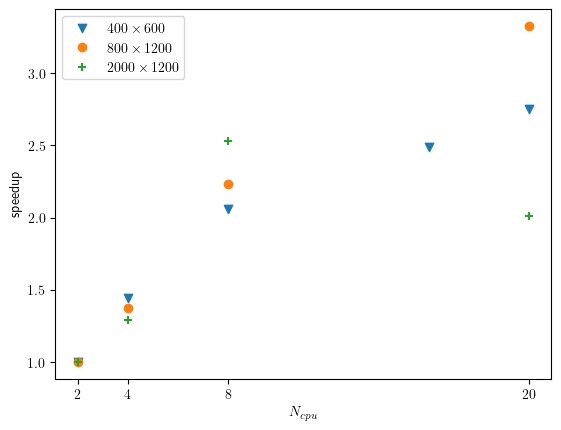}
    \caption{Scaling with number of cores on the cluster. Test problem chosen was the nonlinear resistive tearing instability, with different problem sizes.}
    \label{fig:scaling_1}
\end{figure}

In order to evaluate the performance of the spectral code Dedalus on distributed systems, a test problem was solved with different configurations. The test problem used was the nonlinear evolution of the tearing instability for 10 Alfvén timescales $\tau_{A}$. A modest resolution of $N_{x} = 400$ and $N_{y} = 600$ was first chosen, as this would represent the typical problem size that is still manageable with a normal computer, with the problem typically being solved in a matter of hours. An initial condition was used that leads to representative transient dynamics in order to best represent future computations. A conservative timestep of $\Delta t = 0.001 \cdot \tau_{A}$ was chosen to prevent excessive CFL adjustments. The simulation used between 4GB and 6GB of RAM, depending on the number of cores, with a higher core count using slightly more memory. As can be seen in Fig. \ref{fig:scaling_1}, the wall time scales almost linearly up to 8 cores. Adding more cores leads to reduced returns in wall time. The CPU time increases linearly with the core count. The suboptimal scaling at high core counts could be attributed to the relatively small problem size. Since Dedalus parallelizes along the Fourier direction, the resulting sub-problems at this resolution might not be large enough to fully leverage the low-level FFT-based operations. Another possible cause for suboptimal scaling could be the increased number of writing operations to the hard drive, as every sub-problem dumps data into a separate file before combining the sub-dumps into a large dump. These results led to the conclusion that the most efficient way to use the resources at hand was to run a higher number of simulations at the same time using a single core, instead of a lower number of simulations on many cores. 

Repeating the same experiment, but with a larger problem size of $N_{x} = 800$ and $N_{y} = 1200$ along with a lower timestep of $\Delta t = 0.5 \cdot 0.001 \cdot \tau_{A}$ to account for the change in CFL condition with the finer grid, we see slightly better scaling up to 8 cores, with a dropoff in scaling performance for the largest problem.

\subsubsection{Nonlinear Simulations with Dedalus}

The tau-spectral code Dedalus \cite{dedalus} was used to simulate the nonlinear transient dynamics of system Eq. \ref{eq:reduced_MHD_2d}. The initial conditions used were of three main categories: The ones used for the GEM challenge, designed to benchmark nonlinear MHD solvers, eigenvector perturbations, and pseudo modes. The pseudomodes were used to excite strong transient growth, and are introduced as initial conditions added to the background field. Lastly, eigenmodes are a good way to verify that linear growth rates are at least somewhat reproduced by the nonlinear solver, although nonlinear effects and dissipation of the background field quickly make simulations deviate from linear growth rates. This is a well documented phenomenon and numerous approaches exist to attenuate it, such as removing dissipation of the background fields \cite{Landi}. We expect slightly slower growth rates, and we expect the growth rates to drop below an exponential one as the background field diffuses \cite{Zanna_2016}. The influence of the diffusing background field should be more pronounced the lower the Lundquist number is \cite{Steinolfson1983NonlinearEO}. The values we are dealing with in this work, around $10^3 - 10^5$ are considered fairly low/moderate, astrophysical flows would have Lundquist numbers of order $10^{8}$.

For time stepping, a RK443 (see Ascher \cite{ASCHER1997151} sec. 2.8) scheme was used, which is a 3rd-order 4-stage Runge-Kutta scheme. It is an implicit-explicit multistep scheme, which is known to be fairly robust and specifically developed with convection-diffusion problems as a focus. In general, it is recommended to use some form of anti-aliasing when using a spectral method based on Fourier transforms. The so-called ``2/3-rule'' is often recommended as a good starting point \cite{boyd2013chebyshev} and was used for all nonlinear simulations, for both the Fourier and Chebyshev grids.

\subsection{Extracting the linear tearing instability from nonlinear simulations}
To verify that our nonlinear solver was set up correctly and is able to reproduce the dynamics of the linearized system, as a first step, the growth of eigenfunction initial conditions was examined. The growth was measured using kinetic energy, magnetic energy and reconnected flux. All three are considered valid quantities of interest, with reconnected flux being the most widely used by plasma practitioners. They are defined as follows:

\begin{align}
    \Delta \Psi &= \int_{0}^{L_{x}/2} \vert B_{y}(x, y=0, t) \vert dx \label{eq:reconnected_flux} \\
    ME_{y} &= \int_{-L_{x}/2}^{L_{x}/2} \int_{-d}^{d} (\partial_{y}\psi)^{2} dx dy \\
    KE &= \int_{-L_{x}/2}^{L_{x}/2} \int_{-d}^{d} \left((\partial_{x}\phi)^{2} + (\partial_{y}\phi)^{2}\right) dx dy
\end{align}

The contribution of the magnetic field in the x-direction was not considered, as it contains the background field, which is many times larger than the perturbations and would obscure any dynamics. This is a possible source of discrepancy when comparing nonlinear simulation results to linear simulation results. It would be possible to simulate only the perturbations, including the background field as a non-constant coefficient, i.e., simulating Eq. \eqref{eq:perturbation_equation_full} instead of Eq. \eqref{eq:reduced_MHD_2d}.

The integrals are evaluated numerically, except for Eq. \ref{eq:reconnected_flux} which can be analytically integrated using the stream function, although the analytic formula only works if the magnetic island is perfectly aligned with the computational domain, making it less valuable in practice. Another quantity one might be interested in is the projection of the solution (denoted here $\mathbf{y}$) onto another vector (eg. initial condition, eigen- or pseudomode, etc., denoted here $\mathbf{u} \in V$). This is done via the projection operator:

\begin{align}
    \mathrm{proj}_{V} \mathbf{y} = \frac{\mathbf{y} \cdot \mathbf{u}}{\mathbf{u} \cdot \mathbf{u}} \mathbf{u}.
\end{align}

The dot product in this case is the $L_{2}$ inner product of two continuous functions, approximated by Fourier and Chebyshev basis functions. This approximation can be done in coefficient space or by using vector representations of the sampled solutions combined with an appropriate weighting matrix.

Exponential growth rates can be extracted from these quantities. For the magnetic energy and kinetic energy, this is done using finite differences. The formula used is the following:

\begin{align}
    \gamma = \frac{1}{\Delta \Psi} \partial_{t} (\Delta \Psi) = \frac{\Delta \Psi^{t} - \Delta \Psi^{t-1}}{\Delta t} \frac{1}{\Delta \Psi^{t}}.\label{eq:numerical_growth_rate}
\end{align}

One quickly notices that inserting an exponential into Eq. \ref{eq:numerical_growth_rate} leads to a constant $\gamma$. The growth rate for the reconnected flux can be obtained analytically from the vector potential $\phi$, after combining Faraday's law and Ohm's law, by evaluating $\partial_{t} (\Delta \Psi)= \frac{1}{S} (J_{z}(0, 0) - J_{z}(0.5L_{x}, 0)) $ \cite{goldston2020introduction}.

This quantity measures the magnetic flux being trapped inside the forming magnetic island. In order to benchmark computational routines for plasma flows, a commonly used benchmark is the GEM challenge. It consists of a perturbed Harris current sheet. The perturbation is typically chosen to be fairly large so as to excite the nonlinear phase of the instability as quickly as possible. The initial conditions used are as follows:

\begin{align}
\label{eq:GEM_init}
    \psi(x, y, 0) = -\mathrm{log}(\mathrm{cosh}(y)) + \mathrm{log}(\mathrm{cosh}(L_{y}/2)) + \psi_{0}\mathrm{cos}(2 \pi x / L_{x}) \mathrm{cos}(\pi y / L_{y})
\end{align}

\begin{figure}[htbp]
\centering
\begin{minipage}[b]{0.49\textwidth}
  \centering
    \includegraphics[width=\textwidth]{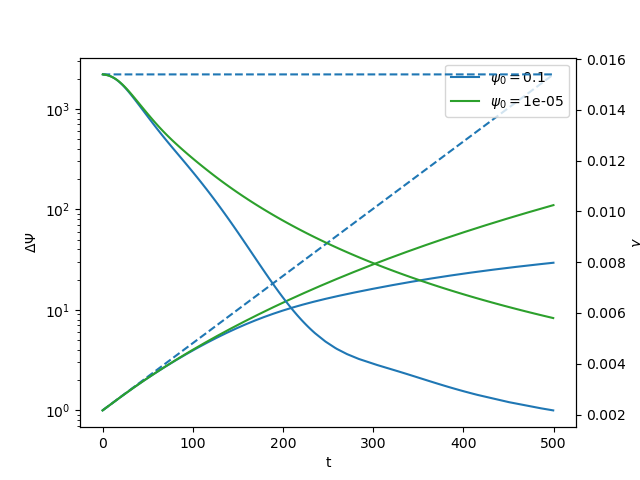}
    \caption{Growth rates obtained from nonlinear transient simulations of an eigenvector. The growth rates drop to sub-exponential levels quickly due to the diffusion of the background field. The dashed lines represent the time evolution of a perfectly linear system response. The parameters used for the simulations were: $k_x=0.2, S=10^3$.}
    \label{fig:psi_0_sacling}
\end{minipage}
\hfill
\begin{minipage}[b]{0.49\textwidth}
  \includegraphics[width=\textwidth]{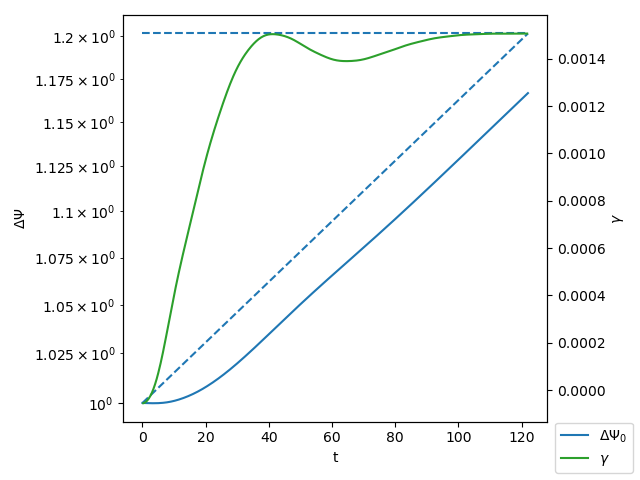}
    \caption{Growth rates obtained from nonlinear transient simulations, using only a perturbation in the magnetic field as defined in Eq. \eqref{eq:GEM_init}. This simulation uses the parameters $S=10^4$ and $k_X=0.7$. The dashed lines represent the time evolution of a perfectly linear system response. The higher lundquist number leads to linear growth being sustained by a less diffused background field for a longer time horizon. The first 20 time units are spent building up a velocity field, using energy of the magnetic field.}
    \label{fig:gem_0_sacling}
\end{minipage}
\end{figure}

The magnetic energy of the background field is computed: $\int_{-L_x/2}^{L_x/2}\int_{-d}^{d} \vert \mathrm{tanh(y)} \vert^{2} \mathrm{dy}\mathrm{dx} = \\ \frac{(L_y -2)e^{L_y} + L_y + 2}{e^{L_y}+1} = 10.00002457 \cdot L_{x}$. Perturbation energies are also readily computed, assuming a perturbation $\mathrm{p}(y)$ wavenumber of unity: $\int_{-L_x/2}^{L_x/2}\int_{-d}^{d} \vert \mathrm{p(y)} \vert^{2} \mathrm{dy}\mathrm{dx} = \frac{L_x}{2} \int_{-d}^{d} \vert \mathrm{p(y)} \vert^{2} \mathrm{dy}$. This value will be used to measure approximately how large perturbations are with respect to the background equilibrium, as follows: 
\begin{align}
r = \frac{\int_{-d}^{d} \vert \mathrm{p(y)} \vert^{2} \mathrm{dy}}{2 \cdot 10.00002457}.
\end{align} 
In order to compute dispersion relations, eigenvectors of the linearized system were deemed most appropriate for use as initial conditions for the nonlinear solver. So as to rule out any transient growth effects on the obtained growth rates. The eigenvectors were scaled to be fairly small, in order to rule out strong nonlinear effects, at least at the very beginning of the simulations. First, we shall show the evolution of the reconnected flux as a function of time for a periodic wavenumber of $k_x=0.2$ and $S=10^3$ with different scaling factors for the eigenvectors. This shall demonstrate the influence of the perturbation size on the validity of the linearized system. We would expect smaller perturbations to grow linearly for a longer time. On the other hand, the background field diffuses irrespective of perturbation size, so this will always lead to a shrinking of the growth rate over time. The result can be seen in Fig. \ref{fig:psi_0_sacling}. The same computations were carried out for different values of $\psi_0$, although for all values lower than $\psi_0=0.1$, the results were nearly identical, implying that the deviation from the exponential growth was due to the background field diffusing, rather than nonlinear interactions. The growth was nearly perfectly linear for around 10 time units. This test also confirmed that linear growth is observed for small perturbations. The results obtained are similar to the ones in \cite{Steinolfson1983NonlinearEO}. Repeating a similar computation with an initial condition corresponding to a small GEM-type perturbation, we get the results seen in Fig. \ref{fig:gem_0_sacling}. Exponential growth is achieved after a short phase of attenuated growth. This phenomenon was the main motivation for choosing eigenvectors as initial conditions for obtaining dispersion relations.

\begin{figure}[h]
    \centering
    \includegraphics[width=0.75\textwidth]{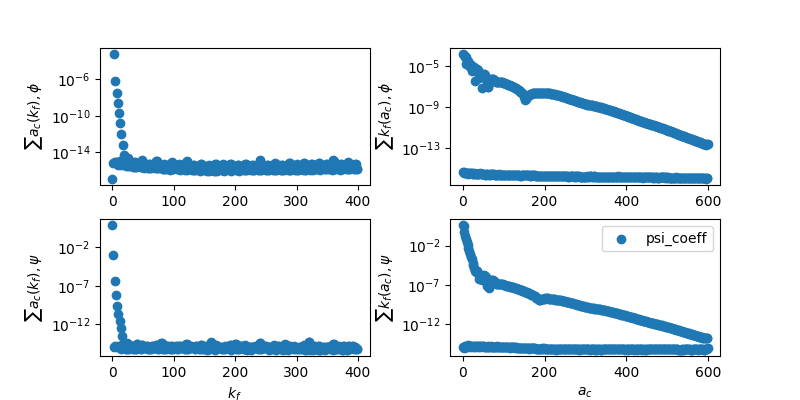}
    \caption{Coefficients of solution in periodic and spanwise direction at a time $t=120$. In the periodic direction, the coefficients very quickly approach machine precision, and therefore do not need to be considered for the simulation. In spanwise direction, the coefficients of the Chebyshev polynomials drop off quickly, but seem to be contributing to the dynamics and therefore a higher resolution was retained. This is due to the gradients in x-direction being much less steep than in y-direction. Similar observations are made in the literature \cite{RDP}.}
    \label{fig:coeffs}
\end{figure}

\begin{figure}[htbp]
\centering
\begin{minipage}[b]{0.49\textwidth}
  \centering
    \includegraphics[width=\textwidth]
    {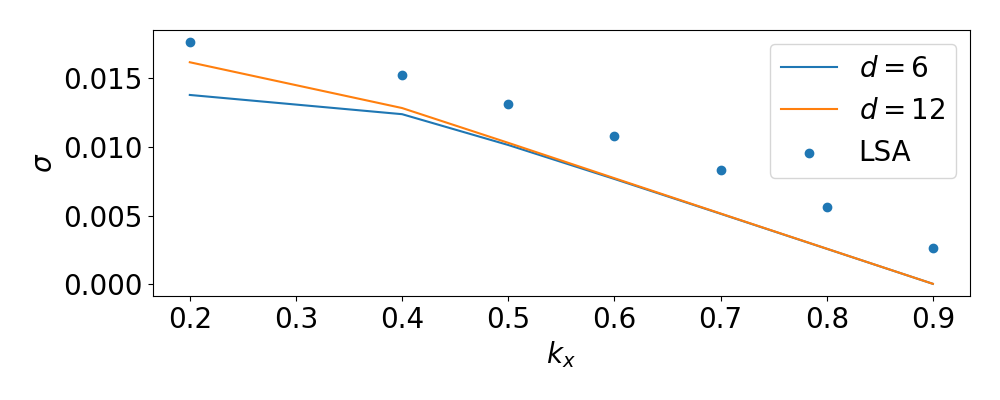}
    \caption{Dispersion relation extracted from nonlinear simulations using only a sinusoidal perturbation of $\psi$ (Eq. \eqref{eq:GEM_init}) as initial conditions. The maximum measured growth rate of the reconnected flux at time $t=0$ was chosen for this plot. The plot shows the results of two runs of simulations, with the total domain size in y-direction being doubled, to better understand the influence of the wall distance on growth rates at $S=10^3$.}
    \label{fig:gem_disp_wall}
\end{minipage}
\hfill
\begin{minipage}[b]{0.49\textwidth}
  \centering
    \includegraphics[width=\textwidth]{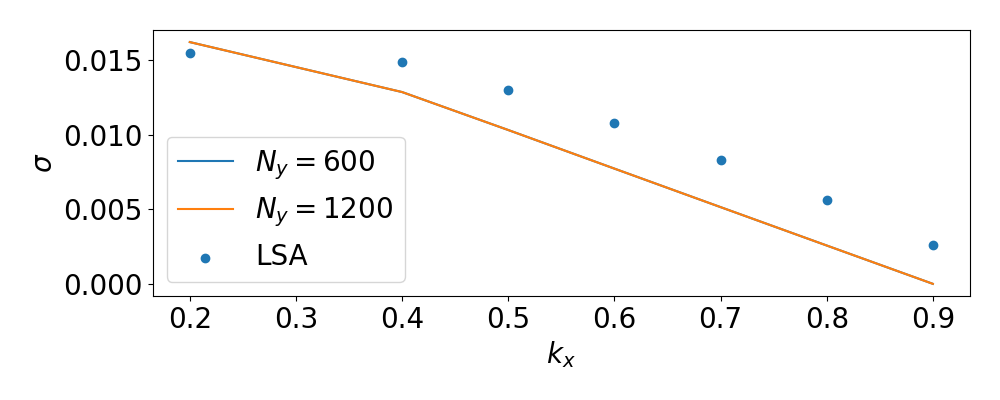}
   
          \caption{Dispersion relations extracted from nonlinear simulations using only a sinusoidal perturbation of $\psi$ (Eq. \eqref{eq:GEM_init}) as initial conditions. The maximum measured growth rate of the reconnected flux at time $t=0$ was chosen for this plot. The plot shows the growth rates obtained at $S=10^4$, with a doubling of the grid resolution in $y$-direction.}
    \label{fig:gem_disp_ny}
\end{minipage}
\end{figure}

\begin{figure}[htbp]
\centering
\begin{minipage}[b]{0.49\textwidth}
  \centering
    \includegraphics[width=\textwidth]{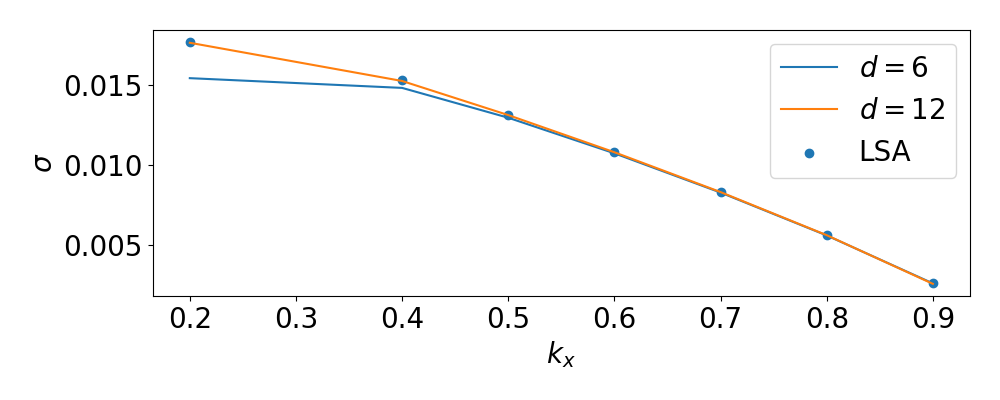}
    \caption{Dispersion relation extracted from nonlinear simulations using eigenvectors as initial conditions. The growth rate of the reconnected flux at time $t=0$ was chosen for the plots. The plot shows the results of two runs of simulations, with the total domain size in y-direction being doubled, to better understand the influence of the walls on growth rates at $S=10^3$. At low wavenumbers, the walls lead to decreased growth. Similar results were reported by \cite{RDP}.}
    \label{fig:eig_disp_ly}
\end{minipage}
\hfill
\begin{minipage}[b]{0.49\textwidth}
  \includegraphics[width=\textwidth]{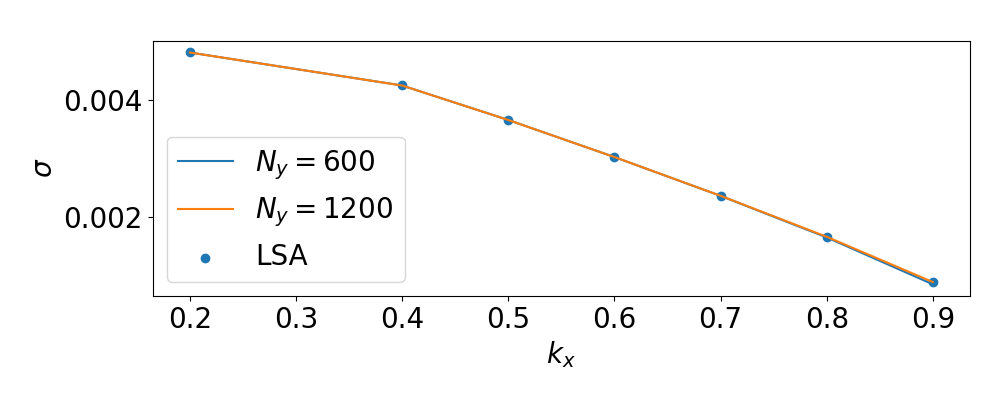}
  \caption{Dispersion relations extracted from nonlinear simulations using eigenvectors as initial conditions. The growth rate of the reconnected flux at time $t=0$ was chosen for the plots. The plot shows the growth rates obtained at $S=10^4$, with a doubling of the grid resolution in $y$-direction. Almost no difference can be seen.}
  \label{fig:eig_disp_s}
\end{minipage}
\end{figure}

Apart from computing a dispersion relation from nonlinear simulations, some sensitivity analysis was carried out with respect to wall distance $L_y \in [12, 24]$, spanwise resolution $N_y \in [600, 1200]$ and Lundquist number $S \in [10^3, 10^4, 10^5]$. The parameters were varied independently of each other, with some promising combinations of parameters also tested simultaneously (Lundquist number and spanwise resolution). A time step of around $10^{-3}$ was chosen. This is a relatively conservative time step that proved to be small enough for most parameter regimes. Experiments were carried out with a CFL-based adaptive time step, but this was scrapped due to it not being robust enough to be fully reliable and also requiring significantly more memory to be allocated to each simulation. The resulting growth rates for wavenumbers $k_x \in [0.2, 0.4, 0.5, 0.6, 0.7, 0.8, 0.9]$ are compared to the results using a linear stability analysis with the same parameters. The experiments showed that varying the wall distance $L_y$ has a minor effect on growth rates at lower wavenumbers, albeit without changing the fundamental dynamics of the instability, as can be seen in Fig. \ref{fig:gem_disp_wall} and Fig. \ref{fig:eig_disp_ly}. The lower growth rates expected with higher Lundquist numbers were accurately reproduced in Fig. \ref{fig:eig_disp_s}. A resolution of $N_y = 600$ was seen to be sufficient for simulations initiated with eigenvectors or GEM-type perturbations, as was verified with experiments; see Fig. \ref{fig:gem_disp_ny}. Inspecting the magnitude of the Fourier series coefficients, it was found that as little as 20 coefficients can resolve the dynamics in the periodic direction sufficiently, assuming we are using small perturbations to examine linear effects. Fig. \ref{fig:coeffs} shows the coefficients of the spectral expansion at time $t=120$ of a simulation with a small perturbation.

Using the perturbation from the GEM challenge, which only perturbs the magnetic field, a distinct phase of linear growth can also be seen. It shall be noted that the instability only starts growing exponentially after a velocity field is established by transferring energy from the magnetic field to the velocity field. By the time this transfer has happened, the background field has already diffused somewhat, leading to slightly lower growth rates than predicted by linear stability analysis. This can be seen in Figs \ref{fig:gem_0_sacling}\ref{fig:gem_disp_ny}\ref{fig:gem_disp_wall}.

\subsection{Nonmodal linear effects in nonlinear simulations}
Next, it was verified whether the bounds on transient growth derived and computed earlier (Sec. \ref{sec:transient_theory}) for the linear system held significance for the nonlinear case. This was done by initializing the nonlinear modes with wave packets corresponding to optimal initial disturbances at different times. These are the pseudomodes that give the highest possible linear transient growth at a time $t_{opt}$, and are computed via SVD. We chose the times $t_{opt} \in [1, 300]$. Since these modes were computed using a linear system, it is expected that they have only limited validity when used with the nonlinear solver. We shall also focus on the two norms defined in Sec. \ref{sec:energynorm} to quantify growth in these cases, since these are the norms we used to compute the wavepackets in the first place. The simulations were carried out at $k_{x} = 0.995$, as this was found to be the wavenumber with marginal stability for the grid used by the nonlinear solver.

\subsubsection{Infinitesimal perturbations}

The transient evolution of optimal initial conditions with a very small magnitude was computed, in order to rule out strong nonlinear effects. The influence of nonlinear effects was excluded by verifying that changing the perturbation amplitude did not change the observed growth, given it was normalized by the initial value, similar to the previous section where we computed modal growth rates. Since we used a very small perturbation, we only needed to use a resolution in x-direction of $40$ Fourier modes, with the vast majority of these 40 modes only being utilized by the perturbations at $t_{opt} > 100$. It was verified that the smallest coefficients were on the order of machine precision at all times of the simulations, leaving no doubt about convergence.

\begin{figure}[htbp]
\centering
\begin{minipage}[b]{0.49\textwidth}
  \centering
    \includegraphics[width=\textwidth]{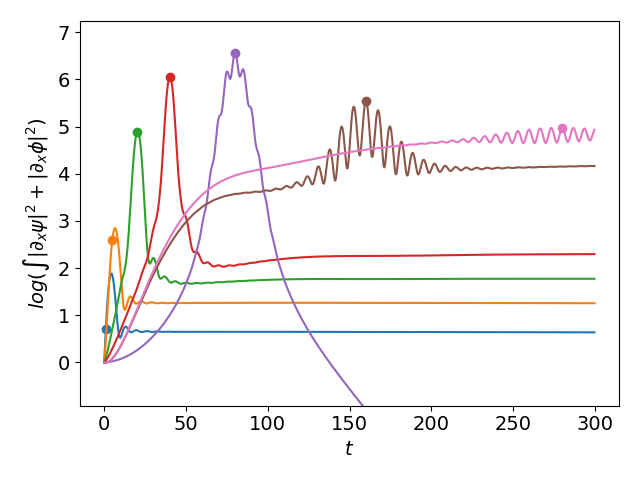}
\end{minipage}
\hfill
\begin{minipage}[b]{0.49\textwidth}
  \includegraphics[width=\textwidth]{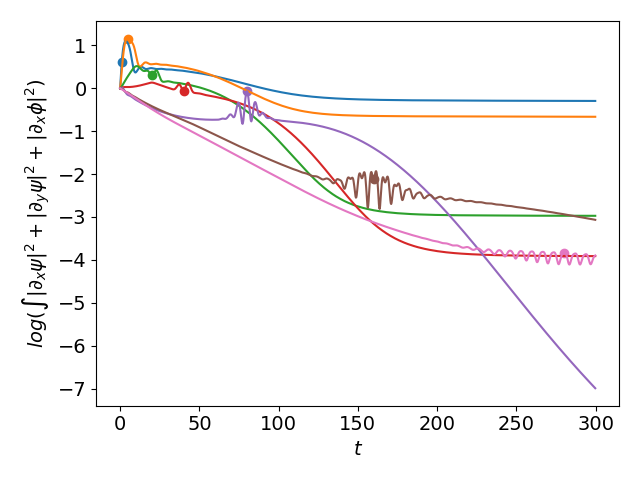}
\end{minipage}
\caption{Nonlinear evolution of the $L_2$ norm of a small perturbation. The initial condition was constructed to increase the $L_2$ norm, and it can be seen that the linear predictions are verified in this norm (left). Including the x-component of the velocity in the postprocessing of the same run shows a much lower growth (right), and the initial conditions do not necessarily lead to the highest possible growth at the times of optimization. The parameters used for the simulation were: $S=10^5$, $k_x=0.995$, $N_y=1600$, $N_x=40$.} \label{fig:L2_s_1e5}
\end{figure}

First, simulations were run with initial conditions derived using the $L_2$ norm. The results for $S=10^5$ and $k_x=0.995$ can be seen in Fig. \ref{fig:L2_s_1e5}. The left plot shows the evolution of the $L_2$ norm of said simulation, and the right plot shows the evolution of the energy norm of the same simulation. While the growth in the $L_2$ norm looks very similar to the predictions made by the linear computations, it is remarkable that even the individual trajectories of the nonlinear simulations look extremely similar to the linear predictions. Further, it should be noted that the transient growth in $L_2$ norm is very large. Including the x-component of the velocity demonstrates that this huge amplification does not carry over into the energy norm. Many of the initial conditions even start decaying instantly in the energy norm. At very short time horizons, there is comparable energy growth to initial conditions derived using the energy norm. 

\begin{figure}[htbp]
\centering
\begin{minipage}[b]{0.49\textwidth}
  \centering
    \includegraphics[width=\textwidth]{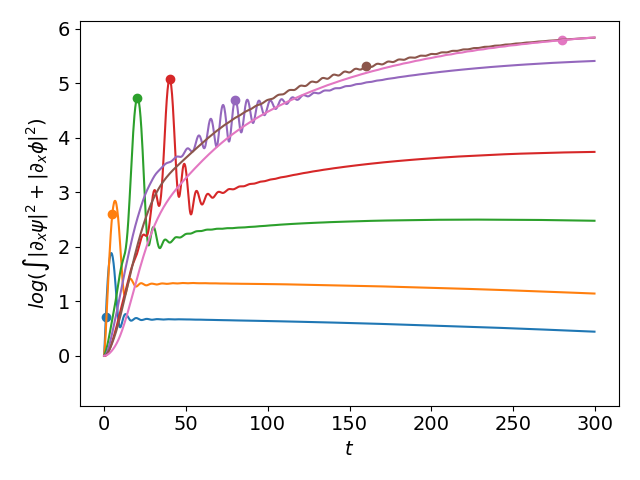}
\end{minipage}
\hfill
\begin{minipage}[b]{0.49\textwidth}
  \includegraphics[width=\textwidth]{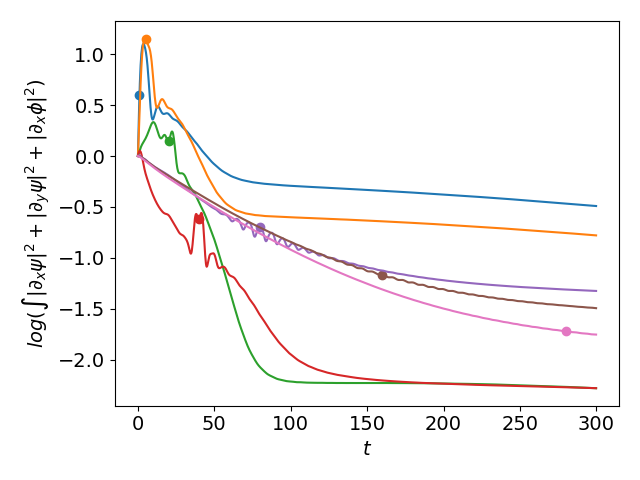}
\end{minipage}
\caption{Nonlinear evolution of the $L_2$ norm of a small perturbation. The initial condition was constructed to increase the $L_2$ norm, and it can be seen that the linear predictions are verified in this norm (left). Including the x-component of the velocity in the postprocessing of the same run shows a much lower growth (right), and the initial conditions do not necessarily lead to the highest possible growth at the times of optimization. The parameters used for the simulation were: $S=10^4$, $k_x=0.995$, $N_y=1600$, $N_x=40$.} \label{fig:L2_s_1e4}
\end{figure}

\begin{figure}[htbp]
\centering
\begin{minipage}[b]{0.49\textwidth}
  \centering
    \includegraphics[width=\textwidth]{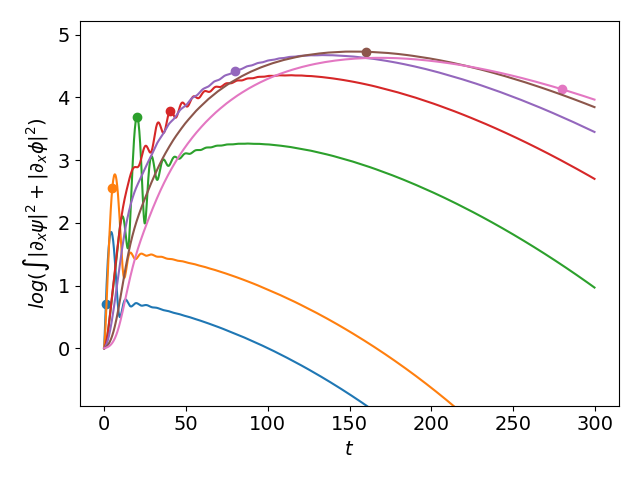}
\end{minipage}
\hfill
\begin{minipage}[b]{0.49\textwidth}
  \includegraphics[width=\textwidth]{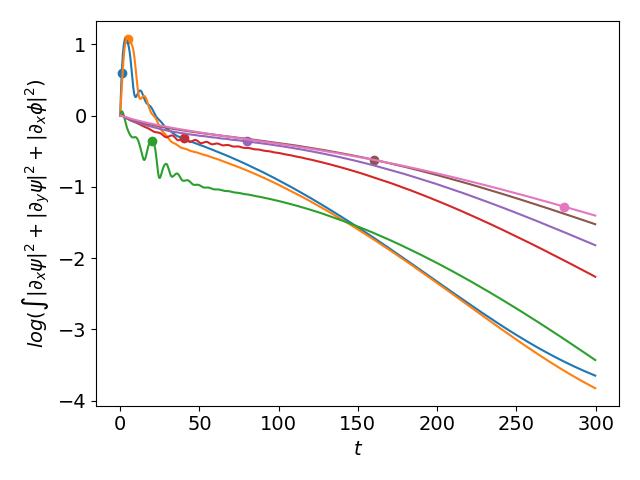}
\end{minipage}
\caption{Nonlinear evolution of the $L_2$ norm of a small perturbation. The initial condition was constructed to increase the $L_2$ norm, and it can be seen that the linear predictions are verified in this norm (left). Including the x-component of the velocity in the postprocessing of the same run shows a much lower growth (right), and the initial conditions do not necessarily lead to the highest possible growth at the times of optimization. The parameters used for the simulation were: $S=10^3$, $k_x=0.995$, $N_y=1600$, $N_x=40$.} \label{fig:L2_s_1e3}
\end{figure}

The same simulations were run for lower Lundquist numbers of $S=10^4$ and $S=10^3$ (Fig. \ref{fig:L2_s_1e4}, \ref{fig:L2_s_1e3}). The effect of the background field diffusing can be seen, with the trajectories for higher $t_{opt}$ not reaching the linear upper bound computed earlier. Again, transient growth in the $L_2$ norm is very large and does not seem to imply large growth in the energy norm.

\begin{figure}[htbp]
\centering
\begin{minipage}[b]{0.49\textwidth}
  \centering
    \includegraphics[width=1.0\textwidth]{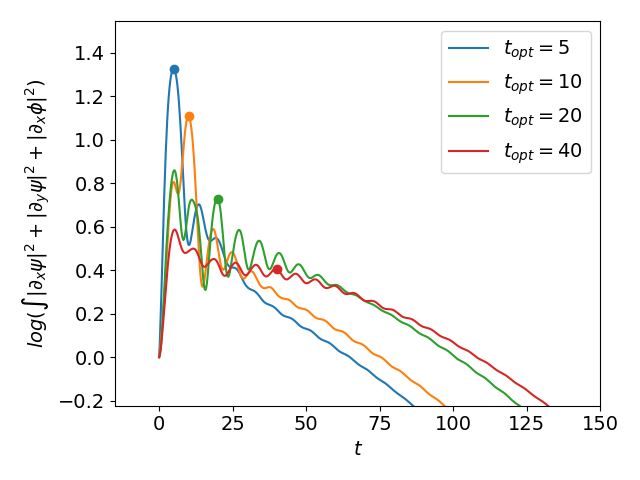}
    \caption{Energy norm evolution of a small perturbation computed using the energy norm at $S=10^3$, $k_x=0.995$, $Re=10^6$, with various time horizons chosen for the optimization.}
    \label{fig:inf_s1e3}
\end{minipage}
\hfill
\begin{minipage}[b]{0.49\textwidth}
  \centering
    \includegraphics[width=1.0\textwidth]{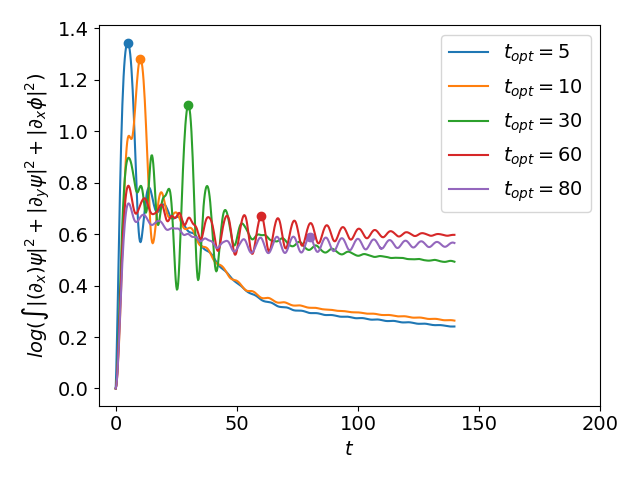}
    \caption{Energy norm evolution of a small perturbation computed using the energy norm at $S=10^4$, $k_x=0.995$, $Re=10^6$, with various time horizons chosen for the optimization.}
    \label{fig:inf_s1e4}
\end{minipage}
\end{figure}

For initial conditions derived using the energy norm, we can see linear predictions being verified by the nonlinear solver as well. Compared to the growth of more than an order of magnitude observed for perturbations optimized using the $L_2$ norm, the growth is more modest, with amplifications of less than $10$ being observed. As expected, at lower Lundquist numbers, the diffusion of the background field significantly influences the trajectories, as can be seen in Fig. \ref{fig:inf_s1e3}, with linear predictions only being verified up to a time horizon of $t\approx40$. For a Lundquist number of $S=10^4$ (Fig. \ref{fig:inf_s1e4}), the effect becomes less pronounced and linear predictions hold up to $t\approx60$. Finally, the highest Lundquist number used in this thesis showed that all computed initial conditions behaved as predicted by the linear computation.

\begin{figure}[h]
    \centering
    \includegraphics[width=0.5\textwidth]{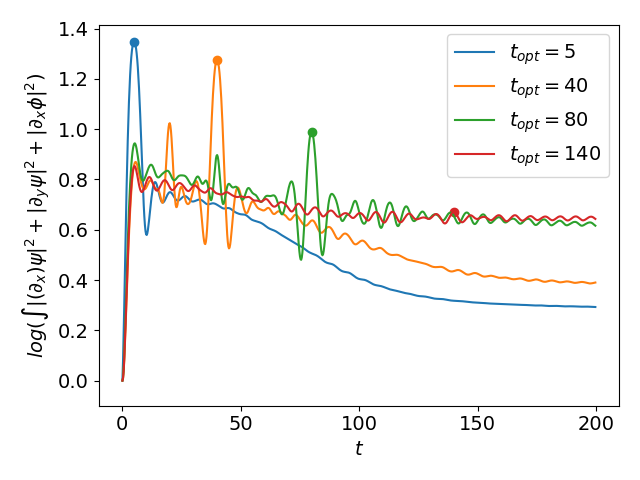}
    \caption{Energy norm evolution of a small perturbation computed using the energy norm at $S=10^5$, $k_x=0.995$, $Re=10^6$, with various time horizons chosen for the optimization.}
    \label{fig:inf_s1e5}
\end{figure}

\FloatBarrier

\subsubsection{Finite perturbations}

Next, the perturbation size was increased to explore how triggering nonlinear effects changes the dynamics of the transient growth. In particular, a question of interest was whether specific initial conditions can trigger enough transient growth to change the stability behavior at marginal stability, i.e. effectively shift the value of marginal stability and potentially trigger the full nonlinear tearing instability. Again, the wavenumber we focused on was $k_x=0.995$, which is very close to marginal stability and initial conditions comuted using the energy norm. The perturbation strengths tested were up to 0.25\% of the background field, which would already be considered fairly large. For all simulations, a resolution of $N_y=1600$ and $N_x=400$ was used. A second run was set up with $N_x=800$ to verify that this did not change the results. We focused on initial conditions computed with the energy norm.  

\begin{figure}[htbp]
\centering
\begin{minipage}[b]{0.49\textwidth}
  \centering
    \includegraphics[width=\textwidth]{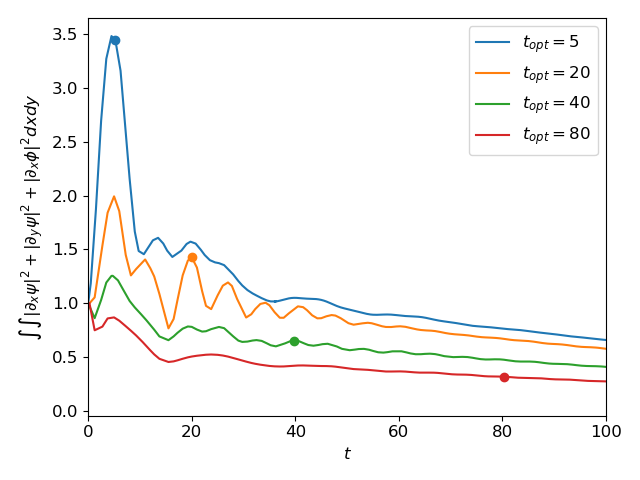}
\end{minipage}
\hfill
\begin{minipage}[b]{0.49\textwidth}
  \includegraphics[width=\textwidth]{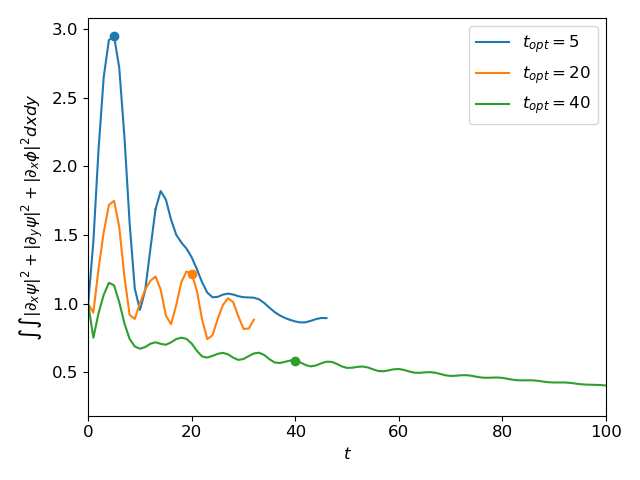}
\end{minipage}
\caption{Energy growth of large perturbations at $S=10^3$. The initial conditions were computed using the energy norm, similar to Fig. \ref{fig:inf_s1e3}. The magnitude was around 0.05\% of the background flow (left) and 0.31\% (right) in the energy norm. The parameters used for the simulation were: $S=10^4$, $k_x = 0.995$ and $Re=10^6$.} \label{fig:E_finite_s_1e3}
\end{figure}

The results for Lundquist number $S=10^3$ can be seen in Fig. \ref{fig:E_finite_s_1e3}. We can see that increasing the perturbation size from $0.05$\% to $0.31$\% of the background field leads to lower transient growth. Also, for both magnitudes, the optimum initial condition that should achieve maximum linear growth at $t_{opt} = 20$ is surpassed by a different run. This was not the case for a very small perturbation (see Fig. \ref{fig:inf_s1e3}), indicating that nonlinear effects are the cause and not diffusion of the background field.

\begin{figure}[htbp]
\centering
\begin{minipage}[b]{0.49\textwidth}
  \centering
    \includegraphics[width=\textwidth]
{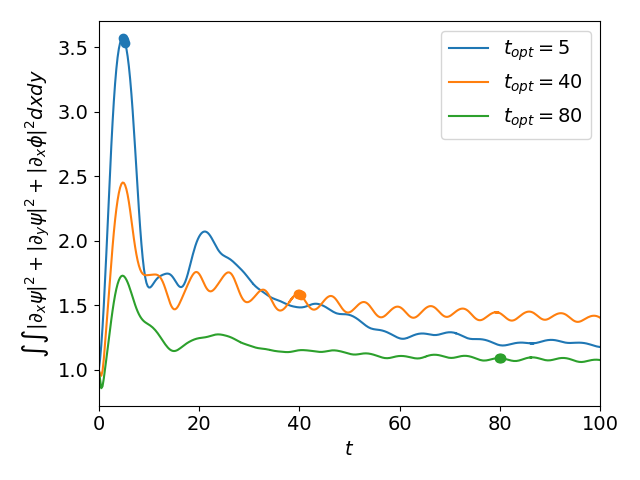}
\end{minipage}
\hfill
\begin{minipage}[b]{0.49\textwidth}
  \includegraphics[width=\textwidth]{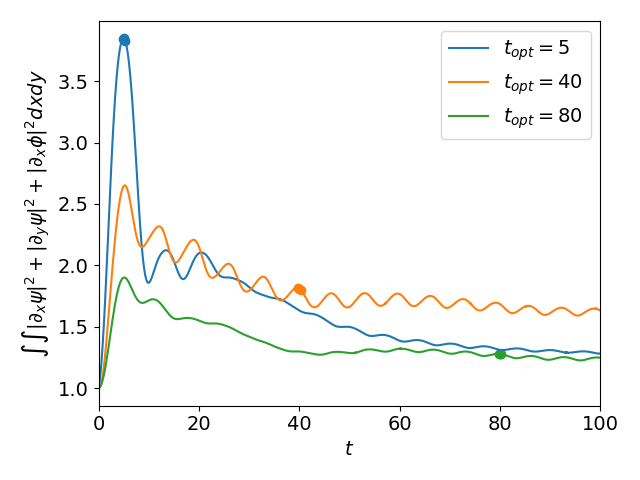}
\end{minipage}
    \caption{Energy growth of large perturbations at $S=10^4$. The initial conditions were computed using the energy norm, similar to Fig. \ref{fig:inf_s1e4}. The magnitude was around 0.05\% of the background flow (left) and 0.01\% (right), in the energy norm. The parameters used for the simulation were: $S=10^4$, $k_x = 0.995$ and $Re=10^6$.} \label{fig:E_finite_s_1e4}
\end{figure}

Increasing the Lundquist number to $S=10^4$, we can see somewhat similar results (Fig. \ref{fig:E_finite_s_1e4}). Increasing the perturbation size slightly lowers the highest growth achieved by our simulations, and the oscillating energy norm evolution seen for longer time horizons $t_{opt}$ was reduced, presumably by nonlinear effects, as can be seen by comparing the growth achieved by initial conditions optimized for $t_{opt} = 80$ with the infinitesimal limit in Fig. \ref{fig:inf_s1e4}.

\begin{figure}[h]
    \centering
    \includegraphics[width=0.5\textwidth]{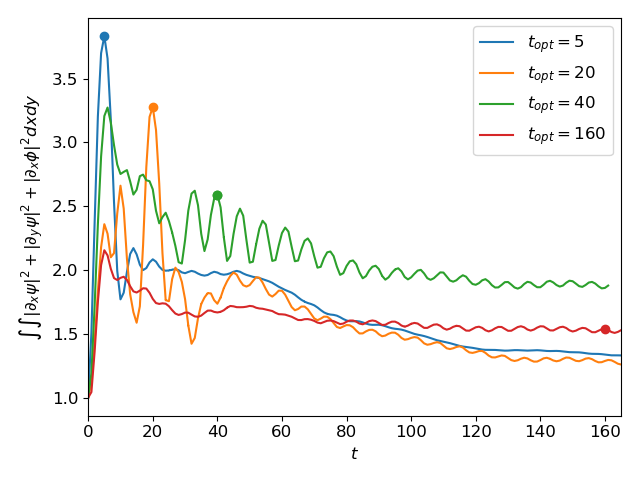}
    \caption{Growth of a finite perturbation at $S=10^5$, $k_x=0.995$, $Re=10^6$, with various time horizons chosen for the optimization. The initial conditions were computed using the energy norm, similar to Fig. \ref{fig:inf_s1e5}. The magnitude of the perturbation was around 0.001 \% of background field. The parameters used for the simulation were: $S=10^4$, $k_x = 0.995$ and $Re=10^6$.}
    \label{fig:E_finite_s1e5}
\end{figure}

While we were not able to use the same magnitude of perturbations at $S=10^5$, we can still see some nonlinear effects changing the energy growth evolution when comparing finite perturbations in Fig. \ref{fig:E_finite_s1e5} with small ones in Fig. \ref{fig:inf_s1e5}. Again, high values of $t_{opt}$ are seen not to lead to optimum energy growth for larger perturbations.

\section{Discussion}

The first result, that we also consider to be the most interesting, is the large difference the choice of norm makes for both linear and nonlinear transient growth calculations. Across all wavenumbers, much larger transient growth is possible in the $L_2$ norm compared to the energy norm. In particular, close to marginal stability, our results indicate a difference of an order of magnitude. The choice of norm significantly changing the character of pseudomodes and growth bounds has been previously demonstrated in other, related work \cite{Zhuravlev_2014}. We can explain to some part where this difference comes from. If we optimize using the $L_2$ norm, we are only penalizing the magnitude of the stream function, which is directly proportional to $u_y, B_y$ when only a single wavenumber is considered. We note that no constraint is imposed on $u_x, B_x$, allowing for the highly oscillating wavepackets we reported to be formed. While these wavepackets certainly lead to growth in the $L_2$ norm, it is clear how this occurs: The energy stored in the fields $u_x, B_x$ is released in the form of growth in $u_y, B_y$. Looking at the energy norm of the simulations that led to large transient growth in the $L_2$ norm, we saw that the transient growth was either significantly weaker or did not occur at all. This means the energy is mainly transferred between states of the perturbation, not taken from the background field. For many of these pseudomodes computed in the $L_2$ norm, their initial norm would be significantly larger if measured using the energy norm, again supporting the theory that they store a lot of energy in the $u_x$ and $B_x$ fields. 

The reported growth in the energy norm is much lower, and the pseudomodes are much less oscillatory, with smaller derivatives in general. While the growth in the energy norm is much lower, it is still remarkable that this growth occurs even at wavenumbers beyond the stability boundary.

Transient growth both in the $L_2$ norm and the energy norm was verified with a full nonlinear solver, using both infinitesimal and finite-sized perturbations. Especially for short time horizons, the transient growth showed up at all Lundquist numbers and perturbation sizes considered in this work. The transient simulations mainly focused on a wavenumber of $k_x=0.995$, which is slightly higher than the value for marginal stability for the discretization used, meaning that the system is spectrally stable and does not allow for modal growth. None of the nonlinear simulations run in this thesis resulted in a full tearing instability being triggered by pseudomode initial conditions. 

Our linear computations in the energy norm were limited by numerical ill-conditioning as soon as high numbers of grid points were needed. This was especially true at high Lundquist numbers and wavenumbers close to marginal stability. We believe that a more conclusive picture of magnitude and scaling laws for transient growth in the energy norm can be obtained with a more robust linear solver, potentially based on finite differences or finite elements. In \cite{Squire_2014}, a similar pseudospectral approach to this thesis is used. The authors mention an extreme sensitivity of the growth bounds and pseudomode shaped with respect to: 1. the number of Chebyshev modes used, 2. the round-off error of the Cholesky factorization of $\bm F$ and 3. the exact composition of $\mathcal{S}^N$. They indicate that a high-precision datatype was used for the Cholesky factorization. This is not the route taken in this thesis, where the factorization was instead performed with an SVD. Using the $L_2$ norm, which we formulated without the need for matrix factorizations, directly in physical space, we experienced fewer issues, leading us to believe that either of the three factors mentioned in \cite{Squire_2014} could be the issue. The hydrodynamic stability community generally uses Chebyshev polynomials due to the ease of deriving the matrix $\bm F$ and the near-perfect grid distribution for wall-bounded flows, allowing the use of a very small number of polynomials. The small number of polynomials used in these hydrodynamic examples avoids many, if not all, numerical issues identified in this thesis. It can be said with certainty that extending the present calculations to higher Lundquist numbers will require a different approach than the one outlined in this work. The most interesting parameter regime includes high Lundquist numbers of the order of $10^8$ and higher, where solar flares and fusion reactors reside. This is \textit{far} beyond the capabilities of the framework of this thesis, both for the linear solver and the nonlinear solver. The finite difference scheme proved to be a good, robust alternative to the Chebyshev differentiation matrices for computations in the $L_2$ norm, motivating further work into adapting it for use with different norms.

For very large perturbations, our nonlinear simulations, especially the highly oscillating ones corresponding to large $t_{opt}$, or simulations with high Lundquist numbers, were limited by the fact that sufficiently large initial conditions led to an immediate excitation of nearly all Fourier modes of the nonlinear solver, preventing convergence form being achieved with a reasonable amount of computing power. Running simulations with perturbation strengths of multiple percent of the background field would likely take a significant amount of computation time. Our simulations used up to $800 \times 1600$ modes, which used a minimum of 24GB of RAM and required 24 hours on 8 CPU cores to advance around 20 time units. This was the limit of what was feasible given the time and resource constraints at hand. As mentioned, the Chebyshev grid distribution was not well suited to the problem at hand, given the narrow boundary layer in the center of the domain. This could be aided with a higher node density in the center of the domain, as mentioned in the discussion of the linear solver. Another possibility would be an adaptive grid, although that would be a significantly more involved solution. Both of these solutions are currently not available in Dedalus.

We computed the leading resolvent modes for the tearing instability, along with response curves indicating how harmonic forcing with different frequencies is amplified. Again, we saw that the choice of norm is significant; higher amplification factors are possible in the $L_2$ norm.

\subsection{Future research}

The optimum initial conditions for growth in the energy norm should be computed using a well-conditioned discretization scheme to verify whether some of the high-frequency initial conditions leading to large energy growth (Sec. \ref{sec:numerical_instability}) found by the spectral collocation scheme are numerical artifacts or evidence of an ill-resolved but physically relevant mode. This could be done with either a finite difference or finite element method. For our computations in the $L_2$ norm, finite differences have proven to be robust. 

While two popular choices of norm were covered in this work, there might be a norm choice better suited to the problem. Typically, in simulations of the tearing instability, the diagnostic quantity of interest is the reconnected flux, not the energy norm or the $L_2$ norm of the state vector. It is possible that choosing a norm closely related to the reconnected flux allows for the computation of modes that are highly effective at triggering the tearing instability. 

Future work should be carried out to verify how sensitive transient growth is with respect to the exact initial conditions used. In \cite{Squire_2014}, random initial conditions are generated to show that the probability of noise leading to transient growth is very high. Showing that not only specifically constructed initial conditions can lead to transient growth would make research into nonmodal effects even more relevant.

This thesis has focused on using transient growth to change the stability properties of the current sheet, i.e, trigger an instability in a spectrally stable system. A systematic study on how transient growth affects the triggering of a secondary instability could also be carried out. There exists a wavenumber $k_{crit}$ below which secondary islands are generated, and optimum initial conditions might be able to significantly shift that wavenumber.

Finally, if non-modal effects prove to be relevant in resistive MHD, and more specifically the plasmoid instability, resolvent-based modelling and control could be feasible, with huge potential implications for coherent structure formation, energy transfer between structures and gaining a better understanding of self-sustaining processes in MHD turbulence. This could also prove to be a valuable tool for the control of instabilities in fusion reactors.

\section{Conclusion}

The tearing mode in slab geometry is one of the simplest resistive instabilities. Nonetheless, relatively little is known about the onset of this instability. The modal growth rate of the tearing instability decreases with higher Lundquist numbers and is too low to explain the plasmoid instability, which has generally been explained as a fully nonlinear phenomenon. Similar to wall-bounded turbulence, transient growth and nonmodal effects could play a role in the linear onset of the tearing mode. Such transient growth, if proven to be substantial enough, could help explain the rapid onset of the plasmoid instability.

This thesis adapted nonmodal stability tools from hydrodynamics for use with resistive, incompressible MHD. The importance of norm choice was highlighted, with the most obvious choice, the $L_2$ norm of the state vector, producing somewhat misleading results. The problem at hand required a higher number of basis functions/grid points compared to most hydrodynamic problems and proved to be challenging numerically. The spectral collocation approach used extensively in hydrodynamics and in this thesis is limited to modest Lundquist numbers, with numerical ill-conditioning preventing robust results at many parameter combinations. Finite differences were shown to be a viable alternative for computations in the $L_2$ norm, although further work remains in order to reformulate computations in the energy norm  using finite differences or finite elements. Transient growth was observed at stable and unstable wavenumbers, both in the $L_2$ norm and the energy norm. This growth was verified with a nonlinear spectral solver, both for infinitesimal and finite perturbation sizes. None of the nonlinear simulations performed exhibited sufficient growth to generate a full tearing instability for a spectrally stable current sheet.

\newpage
\bibliographystyle{plain}
\bibliography{MHD}

\begin{appendices}
\section{\label{subsec::SVD} The singular value decomposition}

The singular value decomposition (SVD) is one of the most important tools in linear algebra. It shows up in nearly every scientific discipline and has been the subject of a lot of research - making it a fairly well-understood tool. Since it is also at the heart of our work, we will briefly review the properties and implications of the SVD, in particular when applied to a matrix representing a linear differential operator.

For any complex-valued matrix $\bm M \in \mathbb{C}^{m \times n}$, the following decomposition always exists:

\begin{align}
\bm M = \bm U \boldsymbol \Sigma \bm V^{H}
\end{align}    

Where $\bm U \in \mathbb{C}^{m \times m} $, $\bm V \in \mathbb{C}^{n \times n} $ and $\boldsymbol \Sigma \in \mathbb{R}^{m \times n}$
is a diagonal matrix containing the diagonal elements referred to as the singular values. The matrices $\bm U$ and $\bm V$ are unitary, i.e. $\bm U \bm U^{H}=\bm U^{H}U=\bm I_{m}$ and $\bm V \bm V^{H}=\bm V^{H} \bm V=\bm I_{n}$, where the $H$ superscript denotes complex-conjugation. The decomposition is unique up to a constant multiplicative complex factor, meaning if $\bm U \boldsymbol \Sigma \bm V^{H}$ is a SVD, so is $(e^{i\theta} \bm U) \boldsymbol \Sigma( \bm V^{H}e^{-i\theta})$. Applying a unitary transform to the left of $\bm M$ does not change the singular values. In fact, if $\bm M= \bm U \boldsymbol \Sigma \bm V^{H}$ is a SVD and $\bm A$ is unitary, $\bm A \bm M=\bm A \bm U \boldsymbol \Sigma \bm V^H$ holds, in the same way $\bm M \bm A = \bm U \boldsymbol \Sigma \bm V^{H} \bm A$ does. In general, the SVD can be formulated by the following two eigenvalue problems: $\bm M^H \bm M = \bm V( \boldsymbol \Sigma^H \boldsymbol \Sigma) \bm V^H$ and $\bm M \bm M^H= \bm U(\boldsymbol \Sigma \boldsymbol \Sigma^H) \bm U^H$.

If one permutes the system such that the singular values are of decreasing magnitude, an approximation of the matrix $M$ can be built in a successive manner: 
\begin{align}
\label{eq:SVD_decomp}
    \bm M = \bm U \boldsymbol{\Sigma} \bm V^{H} = \sum_{i=1}^{m} \sigma_i u_i v^{H}_i
\end{align}
Since we can interpret $M$ as an operator acting on a domain, spanned by the basis $V$ and mapping to the range, the action of $M$ on any element of  its domain can be written, using the fact that $V$ is unitary, and therefore $v^{H}_i v_j = \delta_{ij}$ as such:

\begin{align}
    \bm M v_j &= \sum_{i=1}^{m}\sigma u_i v^{H}_i v_j = \sigma_j u_j,\\ \notag
     \mathbf{a} &\in \mathrm{span}\{\bm V\} = v_i c_i, \\ \notag 
    \bm M \mathbf{a} &= \sum_{i=1}^{m} u_i \sigma_i v^{*}_i a = \sum_{i=1}^{m} u_i \sigma_i c_i. \notag
\end{align}

A natural question to ask is the following: How many singular modes and values are needed to approximate $M$ or  its action well? Luckily, there exist very precise bounds on this approximation error. Given $M_r = \sum_{i=1}^{r}u_i \sigma_i v_{i}^{*}$, the approximation error is the following:

\begin{align}
    \bm M \mathbf{a} - \bm M_{r} \mathbf{a} &= \sum_{i=r+1}^{m} u_i \sigma_{i} v_{i}^{H} a, \\
    \vert\vert \bm M \mathbf{a} - \bm M_{r} \mathbf{a} \vert\vert &\leq \sigma_{r+1} \vert\vert \mathbf{a} \vert\vert. \notag
\end{align}

In summary, the SVD provides us with a tailored basis for an arbitrary linear mapping that allows for a hierarchical ordering of orthonormal input and output directions, no matter what kind of matrix is considered \cite{bisgard2021analysis}. Often times, with the help of this ordering, a low-rank representation of a matrix can be found, or a better understanding of the behavior of the matrix can be obtained. The SVD is one of the most fundamental tools to study the input-output behavior of a system, precisely due to the hierarchical ordering of the modes, and has been extensively used in all aspects of control theory \cite{skogestad_mimo}.

\end{appendices}
\end{document}